\documentclass[prd,showpacs,nofootinbib,preprint, 10 pt]{revtex4}
\usepackage[a4paper, total={7.0in, 10in}]{geometry}
\usepackage{amsmath,graphicx,xcolor,epsfig,slashed}
\usepackage{appendix}
\usepackage{tabularx}
\usepackage{float}
\usepackage{pdfpages}
\usepackage{epstopdf}
\usepackage{subfigure}
\usepackage{caption}
\usepackage{multirow}
\usepackage{booktabs,siunitx}
\usepackage{hyperref}
\usepackage[normalem]{ulem}

\begin{document}

\preprint{ALBERTA-THY-9-24}

\vspace*{0.2cm}

\begin{center}
{\large \bf Angular observables of the four-fold \boldmath $B \to K_{1}(1270,1400)(\to V P) \ell^{+}\ell^{-}$ decays in and beyond the Standard Model}\\[5mm]
\renewcommand{\thefootnote}{\fnsymbol{footnote}}
\setlength {\baselineskip}{0.3in}
{Faisal Munir Bhutta$^{1}$\footnote[2]{faisal.munir@sns.nust.edu.pk}, Abdur Rehman$^{2, 3}$, M. Jamil Aslam$^{4}$, Ishtiaq Ahmed$^{5}$, Saadi Ishaq$^{1}$}\\[4mm]
$^1$~{\it School of Natural Sciences, Department of 
			Physics, National University of Sciences and 
			Technology (NUST), \\ \vspace{-0.25cm} Sector H-12, Islamabad, Pakistan}\\\vspace{-0.25cm}
$^2$~{\it Physics Department, University of Alberta, Edmonton, T6G 2E1, Alberta, Canada}\\
\vspace{-0.25cm}
$^3$~{\it Department of Environmental and Physical Sciences, Faculty of Science, Concordia University of Edmonton,\\ \vspace{-0.25cm} T5B 4E4, Alberta, Canada}\\
\vspace{-0.25cm}
$^4$~{\it Physics Department, Quaid-i-Azam University, 45320, Islamabad, Pakistan}\\ 
\vspace{-0.25cm}
$^5$~{\it National Centre for Physics, Quaid-i-Azam University Campus, 45320, Islamabad, Pakistan}
\\[5mm]
\end{center}
{\bf Abstract}\\[5mm]
Recent measurements of the lepton flavor universality ratios $R^{\mu e}_{K}$ and $R^{\mu e}_{K^*}$ in $B\to \left(K,K^*\right)\mu^{+}\mu^{-}\left(e^+e^-\right)$ at LHCb align with the Standard Model predictions, necessitating search for the complementary decay modes. In this context, we derive the angular decay distributions of the four-fold $B \to K_{1}(1270,1400)(\to V P) \mu^{+}\mu^{-}$ decays, where $K_1 $ is an axial-vector meson, $V=\rho, K^{\ast}$, and $P=K, \pi$. Considering the weak effective Hamiltonian with vector and axial-vector new physics operators, and employing the helicity formalism, we obtain the four-dimensional differential decay distributions and extract various physical observables. These include differential branching ratios, lepton forward-backward asymmetry, forward-backward asymmetry for transversely polarized $K_1$ meson, longitudinal and transverse polarization fractions of the $K_1$ meson, and the normalized angular coefficients with longitudinally and transversely polarized final state vector meson $V$, in the cascade decay $K_1\to VP$. Based on the latest global fit analysis data for all $b\to s$ transitions, we predict the results of these observables in various bins of the square of the momentum transfer, $q^2$. Our findings indicate that these physical observables are not only sensitive to new physics but can also differentiate between various new physics scenarios in certain kinematical regions. The precise measurements of these observables in $B \to K_{1}(1270,1400)(\to V P) \mu^{+}\mu^{-}$ decays at current and future experiments will provide opportunities for the complementary searches for physics beyond the Standard Model in $b\to s\ell^+\ell^-$ decays.

\maketitle

\section{Introduction}\label{sec1}
The studies of inclusive and exclusive $B$-meson decays are pertinent for improving our understanding of the dynamics of quantum chromodynamics (QCD). Of all the $B$-meson decays, the theoretical description of the semileptonic decays is relatively straightforward. These decays can occur through the flavor-changing-charged-current (FCCC) or the flavor-changing-neutral-current (FCNC) transitions. In the Standard
Model (SM), FCCC decays proceed through tree-level diagrams in the weak interaction, where as the FCNC decays are occurred via electroweak penguin and box diagrams.
Among the FCNC processes, the $b \to s \ell^{+}\ell^{-}$ transitions have been at the spotlight of both theoretical and experimental studies over the past decade or so.  These transitions are both Cabibbo-Kobayashi-Maskawa (CKM) and loop suppressed in the SM, and therefore quite sensitive to new physics (NP), \textit{i.e.,} the physics beyond SM. Experimentally, several observables of the exclusive processes involving quark level $b \to s \mu^{+}\mu^{-}$ transitions, such as the branching ratios of $B \to K^{(*)}\mu^{+}\mu^{-}$ \cite{Aaij:2014pli, Aaij:2013iag, Aaij:2016flj}, $B_s \to \phi \mu^{+}\mu^{-}$ \cite{Aaij:2015esa}, and the optimized observables of $B\to K^{*}\mu^{+}\mu^{-}$ \cite{Matias:2012xw, Descotes-Genon:2013vna, Aaij:2013aln, ATLAS:2017dlm, CMS:2017ivg, Aaij:2020nrf, Abdesselam:2016llu} have been measured, showing deviations from the SM results.  
Initially, the measurements of lepton-flavor-universality (LFU) ratio $R_{K^{(\ast)}}$, which is the ratio of $B \to K^{(*)} \mu^{+}\mu^{-}$ to $B \to K^{(*)} e^{+}e^{-}$, in different bins of the square of the momentum transfer, \textit{i.e.,} $q^2$, mismatched with its SM predictions, and hence provided the signatures of the NP \cite{Aaij:2014ora, Abdesselam:2019lab, Aaij:2017vbb}. Consequently, a number of theoretical approaches (see, for example, \cite{Alguero:2019ptt, Descotes-Genon:2015uva,Capdevila:2017bsm,Descotes-Genon:2013wba,Ahmed:2017vsr,Bhattacharya:2019dot,Aebischer:2019mlg, Ciuchini:2019usw, Alok:2019ufo,Kumar:2019qbv,Biswas:2020uaq, Arbey:2019duh,Bifani:2018zmi,Alguero:2019pjc}),
were developed to explain these deviations \cite{Hiller:2003js, Bordone:2016gaq}. Nevertheless, the recent measurements of these LFU ratios in the low- and central-$q^2$ bins at the LHCb, based on $9$ fb$^{-1}$ integrated luminosity \cite{LHCb:2022qnv, LHCb:2022vje}, are aligned with the SM predictions, which provide the indications for the universal lepton couplings. To attribute NP to any deviations in $b\to s\ell^+\ell^-$ decays, it is crucial to explore complementary channels, such as $B\to K_{1}\ell^{+}\ell^{-}$.

In this work, we investigate the four-fold $B \to K_{1}(1270,\; 1400)\left(\to VP\right) \mu^{+}\mu^{-}$ decays, where $K_1(1270,\; 1400)$ is an axial-vector meson, and in the cascade decay $VP = \rho K,\; K^*\pi$ are the dominant decay modes. 
It is established that the axial-vector physical states $K_1(1270)$ and $K_1(1400)$ are mixture of the $^{3}P_{1}$ and $^{1}P_{1}$ states, $K_{1A}$ and $K_{1B}$, respectively, such that:
\begin{eqnarray}
|K_{1}(1200)\rangle &=& |K_{1A}\rangle \sin\theta_{K_{1}}+|K_{1B}\rangle\cos\theta_{K_{1}}\label{mix1},\\
|K_{1}(1400)\rangle &=& |K_{1A}\rangle \cos\theta_{K_{1}}-|K_{1B}\rangle\sin\theta_{K_{1}},\label{mix2}
\end{eqnarray}
where $\theta_{K_{1}}$ is the mixing angle, and currently, there is no consensus on its value. For instance, in Ref. \cite{Suzuki:1993yc}, an analysis of the widths and masses of the $K_1\left(1270\right)$ and $K_1\left(1400\right)$ states suggests $\theta_{K_{1}}$ to be $ 33^{\circ}$ or $57^{\circ}$. However, phenomenological analyses of the data from $B\to K_{1}(1270)\gamma$ and $\tau \to K_{1}(1270)\nu_{\tau}$ decays, as presented in \cite{Cheng:2003bn, Li:2009tx}, indicate values of $37^{\circ}$ or $58^{\circ}$, while another study reports a value of $-\left(34\pm13\right)^\circ$ \cite{Hatanaka:2008xj}. By studying the correlation between the mixing angle of $f_1(1285)$ - $f_1(1420)$ and $\theta_{K_{1}}$, the possible range is further refined to $31.7^\circ$ or $56.3^\circ$ \cite{Close:1997nm}. Additionally,  incorporating $h_1(1170)$- $h_1(1380)$ with these $f$-states, the range shrinks to $28^\circ<\theta_{K_{1}}<30^\circ$ \cite{Cheng:2013cwa}. Besides these phenomenological analyses, the values obtained in the non-relativistic constituent quark model \cite{Burakovsky:1997dd} and QCD sum rules \cite{Dag:2012zz} are $34^\circ<\theta_{K_{1}}<55^\circ$ and $\theta_{K_{1}} = 39^\circ\pm4^\circ$, respectively. The QCD sum rules analysis is recently revisited in \cite{Shi:2023kiy}, where the calculation of the quark-gluon dynamics, based on the operator product expansion up to dimension-5 condensates, gives $\theta_{K_{1}} = 22^\circ \pm 7^\circ$ or $68^\circ\pm 7^\circ$. Motivated by this, considering both the mixed and unmixed $K_{1A}$ and $K_{1B}$ states, the semileptonic $B\to K_{1}(1270)\mu^{+}\mu^{-}$ decay has been extensively studied in and beyond the SM in literature, see, \textit{e.g.,} \cite{Hatanaka:2008xj,Hatanaka:2008gu,Paracha:2007yx,Ahmed:2008ti,Ahmed:2010tt,Ahmed:2011vr, Ishaq:2013toa,Munir:2015gsp, MunirBhutta:2020ber, Li:2011nf,Huang:2018rys}.  The symmetry breaking corrections arsing due to the vertex renormalization and hard-spectator scattering in the form factors of $B\to K_1 \mu^{+}\mu^{-}$ decay along with their impact on various physical observables were calculated in Ref. \cite{Sikandar:2019qyb}.  


In 2007, Choudhary $\textsl{et al.}$ \cite{Choudhury:2007kx} analyzed the angular correlations for the cascade decay process $B \to K_{1}(1270)(\to \rho K)\mu^{+}\mu^{-}$ by using form factors estimated with large energy effective theory (LEET). Later, Li $\textsl{et al.}$ \cite{Li:2009rc} did more detailed calculations of the branching ratios, forward-backward asymmetry, and different angular distributions of $B\to K_{1}\left(\to K\pi\pi\right)\ell^{+}\ell^{-}$ decay where $\ell=e,\mu$. The angular analysis of different FCCC and FCNC $B$ decay modes involving cascade decays such as, $V\to PP$ and $V\to P\gamma$, have been extensively studied in the literature. For example, in $b\to (u, c) \ell \bar{\nu_{\ell}}$ FCCC transitions, $\bar{B}\to D^{\ast}(\to D\gamma, D\pi)\ell^-\bar{\nu_{\ell}}$ \cite{Colangelo:2018cnj}, $\bar{B}\to D^{\ast}(\to D\pi)\ell\bar{\nu}$ \cite{Becirevic:2019tpx, Alguero:2020ukk, Mandal:2020htr, Colangelo:2024mxe}, $B_s\to D_s^{\ast}(\to D_s\gamma, D_s\pi)\tau \nu$ \cite{Das:2021lws}, and $\bar{B}\to \rho (\to \pi\pi)\ell^-\bar{\nu_{\ell}}$ \cite{Colangelo:2019axi} have been investigated. Similarly, in $b\to (d, s) \ell \bar{\ell}$ FCNC transitions, angular analysis of various decay channels, such as $B \to K^{\ast}(\to K \pi) \mu^{+}\mu^{-}$ \cite{Altmannshofer:2008dz}, $B_c \to D^{\ast}(\to D \pi) \ell\bar{\ell}$ \cite{Faessler:2002ut}, $B_{c}\to D_{s}^{\ast}(\to D_{s}\pi)\ell^{+}\ell^{-}$ \cite{Li:2023mrj}, $B_{c}\to D_{s}^{\ast}(\to D_{s}\gamma,(D_{s}\pi))\ell^{+}\ell^{-}$ \cite{Zaki:2023mcw}, and $B\to\rho (\to\pi\pi)\mu^{+}\mu^{-}$ \cite{Farooq:2024owx}, have been explored. However, only a few angular studies have been devoted to FCCC and FCNC decay modes that involve the cascade decay of the final state axial-vector meson $\left(A\to VP\right)$, \textit{e.g.,} the angular analysis of the FCCC decay $B \to a_1(\to\rho \pi)\ell\bar{\nu_{\ell}}$ was performed in \cite{Colangelo:2019axi, Mohapatra:2024knf} . Likewise, the angular observables of the FCNC decay $B \to a_1(\to\rho \pi)\mu^+\mu^-$, were studied in \cite{Farooq:2024owx}.


In this work, we focus on the complete four-fold angular analysis of the $B \to K_{1}(\to VP)\mu^{+}\mu^{-}$ decay, where $V =\rho, K^{*}$, and $P = K,\pi$. Particularly, for the decay modes involving the $K_1 \to K\pi\pi$ and $K_1\to VP$, the full differential decay distribution involves numerous angular observables, that are worth studying within the SM and in its extensions. In addition, the decay $K_1\to VP$ has a peculiar feature that the final state vector mason $(\rho, K^{*})$ can be longitudinally $\left(\|\right)$ and transversely $\left(\perp\right)$ polarized, increasing the number of experimentally accessible observables. Utilizing these features, we present the analytical results of the differential branching ratios $\left(d\mathcal{B}/{dq^2}\right)$, the lepton forward-backward (FB) asymmetry $\left(\mathcal{A}_{\text{FB}}^{K_1}\right)$, the FB asymmetry for transversely polarized $K_1$ meson $\left(\mathcal{A}_{\text{FB}}^{K_{1T}}\right)$, the longitudinal and transverse polarization fractions of $K_1$ meson, $f_{L}^{K_1}$ and $f_{T}^{K_1}$, respectively, in terms of different angular coefficients.

To incorporate the NP, we follow the recent updates on the LFU ratios $R_K$ and $R_{K^*}$ from the LHCb, and the $B_s\to \mu^{+}\mu^{-}$ branching ratio where the possibility of NP in $b\to s\ell^{+}\ell^{-}$ decay is discussed \cite{Alguero:2023jeh}. The global fit analysis was performed using three different frameworks: (i) an update limited to experimental results only, (ii) a full update considering both experimental and theoretical frameworks, and (iii) by performing the analysis without including the LHCb results on electron modes. In this work, we use the constraints obtained from the second possibility to see their impact on the various observables mentioned above for $B\to K_{1}\left(1270,\; 1400\right)\left(\to VP\right)\mu^{+}\mu^{-}$ decays.  


We have organized this study as follows. In Sec. \ref{framework}, we introduce the  weak effective Hamiltonian (WEH) for $b\to s \ell^{+}\ell^{-}$ transition both in the SM and for the NP scenarios. To construct the full differential decay distributions for $B\to K_1(\to V_{\|(\perp)}P)\ell^{+}\ell^{-}$ decay, we provide the complete helicity framework and the corresponding helicity amplitudes. In Sec. \ref{observables}, we present the expressions for the four-fold decay distributions of $B\to K_1(\to V_{\|(\perp)}P)\mu^{+}\mu^{-}$, in terms of the longitudinally and transversely polarized angular coefficients describing the state of final vector meson in the cascade decay of $K_{1}\left(1270,1400\right)\to VP$. The expressions of these angular coefficients in terms of the helicity amplitudes are also included in this section. In Sec. \ref{physObs}, we write the expression of physical observables in the form of these angular coefficients and analyze them numerically in Sec. \ref{num-analysis}. Furthermore, by studying the impact of various NP scenarios, we discuss the discriminatory power of the normalized angular coefficients. We conclude our findings in Sec. \ref{conclusions}. This work is supplemented with six appendices. The Appendix \ref{append} and \ref{appendHME} contain the expressions of the Wilson coefficients and matrix elements of $B\to K_1$ transitions, respectively. The leptonic tensors calculated in the dilepton rest frame are given in Appendix \ref{leptcal}. The detailed calculation of the cascade decay $K_1\to VP$ for parallel and perpendicular polarization of the vector meson is presented in Appendix \ref{Casdecay2}. In the end, the calculation of the four-fold decay distribution and the expressions of different angular coefficients in terms of the transversality amplitudes are presented in Appendix \ref{diffcal} and \ref{transversality-amp}, respectively. 

\section{Theoretical Framework}\label{framework}

In this section, we present the WEH for $b\to s \ell^{+}\ell^{-}$ transition and compute the four-fold transition amplitude for the exclusive $B\to K_{1}\left(\to V_{\|(\perp)} P\right)\ell^{+}\ell^{-}$ decay. By parameterizing the matrix elements of $B\to K_1$ in terms of form factors, and using the helicity framework, we calculate the different helicity amplitudes for the four-fold decay distribution.

\subsection{Effective Hamiltonian and decay amplitude}\label{effH}
The SM Hamiltonian for $b\to s\ell^{+}\ell^{-}$ transition can be given as
\begin{align}\label{H1}
\mathcal{H}_{\text{eff}}^{\text{SM}}=-\frac{4 G_{F}}{\sqrt{2}}V_{tb}V^{\ast}_{ts}&\Bigg[C_{7}^{\text{eff}}(\mu)O_{7}(\mu)+C_{9}^{\text{eff}}(\mu)O_{9}(\mu)+C_{10}(\mu)O_{10}(\mu)\Bigg],
\end{align}
where the effective operators $\left(O_i\right)$ are defined as:
\begin{align}\label{op1}
O_{7}(\mu) &=\frac{e}{16\pi ^{2}}m_{b}\left( \bar{s}\sigma _{\mu \nu }P_{R}b\right) F^{\mu \nu },
& O_{9}(\mu) &=\frac{e^{2}}{16\pi ^{2}}(\bar{s}\gamma _{\mu }P_{L}b)(\bar{\ell}\gamma^{\mu }\ell),
& O_{10}(\mu) &=\frac{e^{2}}{16\pi ^{2}}(\bar{s}\gamma _{\mu }P_{L}b)(\bar{\ell} \gamma ^{\mu }\gamma _{5} \ell).
\end{align}
Here, $e$ is the electric charge, $P_{L/R} = \frac{1\mp \gamma_5}{2}$ are the chiral projection operators, and $\mu$ is the renormalization scale. 
The $b$-quark mass appearing in the operator $O_7$ is treated as the running quark mass in the modified minimal-subtraction $\overline{\text{MS}}$ scheme. The contributions of the
factorizable quark-loop corrections to current-current and
penguin operators are absorbed in the effective Wilson coefficients $C_{7,9}^{\text{eff}}(q^2)$ \cite{Bobeth:1999mk, Beneke:2001at, Asatrian:2001de, Asatryan:2001zw, Greub:2008cy, Du:2015tda}. Their explicit expressions are summarized in Appendix \ref{append}. The effective Hamiltonian characterizing the existence of NP through the vector and axial-vector operators is given as
\begin{align}\label{H2}
\mathcal{H}_{\text{eff}}^{\text{NP}}=-\frac{4 G_{F}}{\sqrt{2}}V_{tb}V^{\ast}_{ts}&\Bigg[C_{9\ell}^{\text{NP}}O_{9}+C_{10\ell}^{\text{NP}}O_{10}+C_{9^\prime\ell}^{\text{NP}}O_{9^{\prime}}+C_{10^\prime\ell}^{\text{NP}}O_{10^{\prime}}\Bigg],
\end{align}
where
\begin{eqnarray}\label{H2tr}
O_{9^{\prime}} =\frac{e^{2}}{16\pi ^{2}}(\bar{s}\gamma _{\mu }P_{R}b)(\bar{\ell}\gamma^{\mu }\ell), \qquad
 O_{10^{\prime}} =\frac{e^{2}}{16\pi ^{2}}(\bar{s}\gamma _{\mu }P_{R}b)(\bar{\ell} \gamma ^{\mu }\gamma _{5} \ell).    
\end{eqnarray}

Using the Hamiltonian given in Eqs. (\ref{H1}) and (\ref{H2}), we can write the decay amplitude for the process $B\to K_1\ell^+\ell^-$ as:
\begin{eqnarray}
\mathcal{M}_1^{\lambda_{l^+},\lambda_{l^-}}\left(B\to K_1\ell^+\ell^-\right)=\frac{G_{F}\alpha}{2\sqrt{2}\pi}V_{tb}V^{\ast}_{ts}\,g^{\mu^{\prime}\mu}\Big\{T^{1}_{\mu}L_{V,\mu^{\prime}}^{{\lambda_{l^+},\lambda_{l^-}}}
+T^{2}_{\mu}L_{A,\mu^{\prime}}^{{\lambda_{l^+},\lambda_{l^-}}}\Big\},\label{Amp1ag}
\end{eqnarray}
where
\begin{eqnarray}
L_{V,\mu^{\prime}}^{{\lambda_{l^+},\lambda_{l^-}}}&=&\left\langle l^+\left(p_{l^+},\lambda_{l^+}\right)l^-\left(p_{l^-},\lambda_{l^-}\right)\left|\bar{\ell}\gamma_{\mu^{\prime}}{\ell}\right|0\right\rangle = \bar u \left(\vec{p}_{l^-},\lambda_{l^-}\right)\gamma_{\mu^{\prime}}\,v\left(\vec{p}_{l^+},\lambda_{l^+}\right),\label{Amp1alep}
\\
L_{A,\mu^{\prime}}^{{\lambda_{l^+},\lambda_{l^-}}}&=&\left\langle l^+\left(p_{l^+},\lambda_{l^+}\right)l^-\left(p_{l^-},\lambda_{l^-}\right)\left|\bar{\ell}\gamma_{\mu^{\prime}} \gamma_{5}{\ell}\right|0\right\rangle=\bar u \left(\vec{p}_{l^-},\lambda_{l^-}\right)\gamma_{\mu^{\prime}}\gamma_{5} \,v\left(\vec{p}_{l^+},\lambda_{l^+}\right),\label{Amp1blep}
\end{eqnarray}
with $\lambda_l$ specifying the helicity states, which correspond to the orientation of lepton's spin relative to its momentum. The contributions to $T^{1}_{\mu}$ and $T^{2}_{\mu}$ are given by
\begin{eqnarray}
T^{1}_{\mu}(r)&=&(C_{9}^{\text{eff}}+C_{9\ell}^{\text{NP}})\Big\langle K_1\left(k,\epsilon_{K_1}(r)\right)|\bar s\gamma_{\mu}(1-\gamma_{5})b|B(p)\Big\rangle
+C_{9^\prime\ell}^{\text{NP}}\Big\langle K_1\left(k,\epsilon_{K_1}(r)\right)|\bar s\gamma_{\mu}(1+\gamma_{5})b|B(p)\Big\rangle\notag\\
&-&\frac{2m_{b}}{q^{2}}C_{7}^{\text{eff}}
\Big\langle K_1\left(k,\epsilon_{K_1}(r)\right)|\bar s i\sigma_{\mu\nu}q^{\nu}(1+\gamma_{5})b|B(p)\Big\rangle\notag\\
&=&\epsilon_{K_1}^{\ast\alpha}(r)T^{1}_{\mu\alpha},\label{Amp1a}
\\
T^{2}_{\mu}(r)&=&(C_{10}+C_{10\ell}^{\text{NP}})\Big\langle K_1\left(k,\epsilon_{K_1}(r)\right)|\bar s\gamma_{\mu}(1-\gamma_{5})b|B(p)\Big\rangle
+C_{10^\prime\ell}^{\text{NP}}\Big\langle K_1\left(k,\epsilon_{K_1}(r)\right)|\bar s\gamma_{\mu}(1+\gamma_{5})b|B(p)\Big\rangle\notag\\
&=&\epsilon_{K_1}^{\ast\alpha}(r)T^{2}_{\mu\alpha}.\label{Amp1b}
\end{eqnarray}

In Eqs. (\ref{Amp1a}), and (\ref{Amp1b}), $p$ and $k$ denote the momenta of $B$ and $K_1$ mesons, respectively, whereas $\epsilon_{K_1}(r)$ represents the polarization vector of the final state $K_1$ meson with $r=+,-,0$. The matrix elements for $B\to K_1$ that appear in $T^{i=1,2}_\mu$ encapsulate the non-perturbative effects, which can be parameterized in terms of the transition form factors as given in Appendix \ref{appendHME}.

\begin{figure}[b!]
\centering
\includegraphics[scale=0.4]{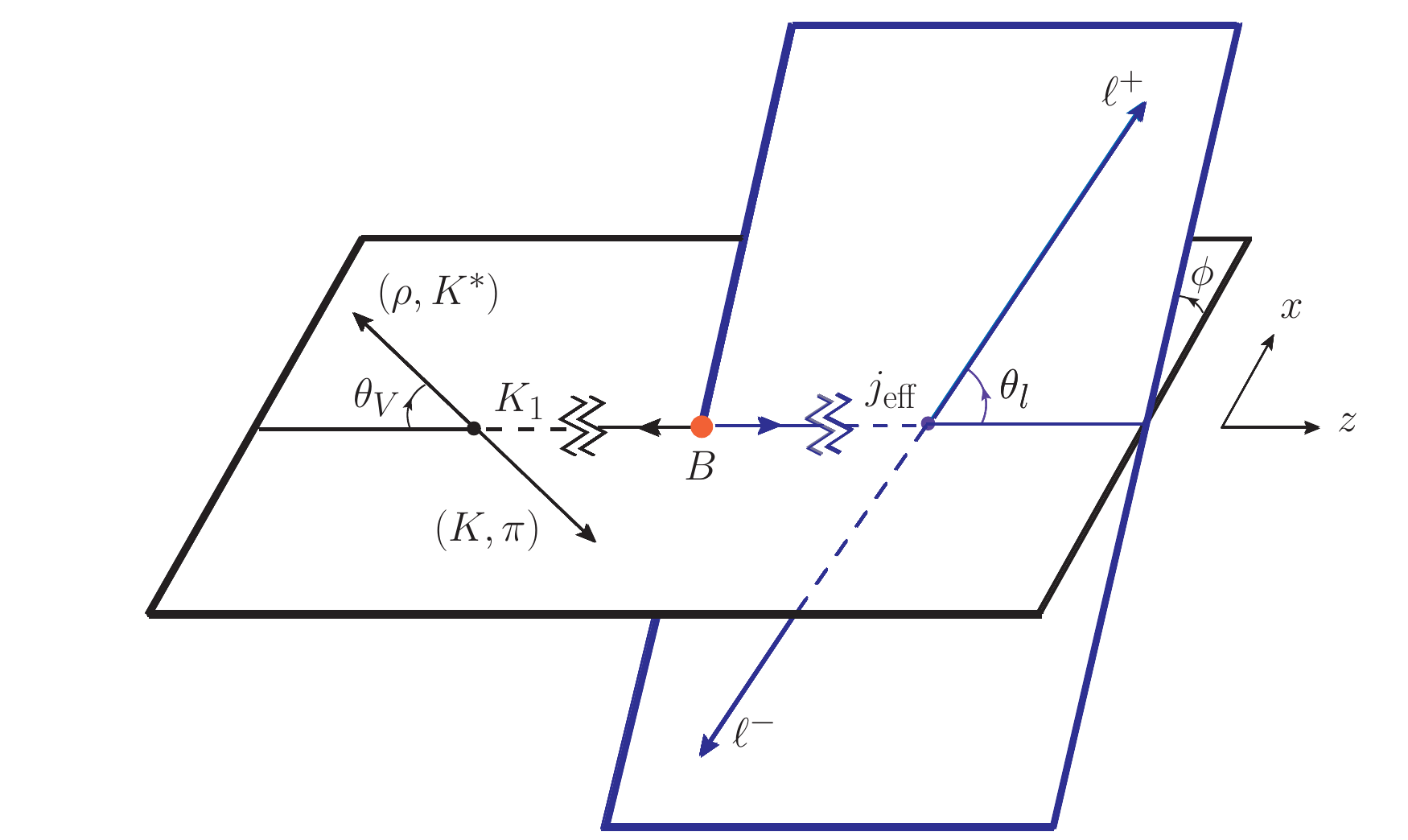}
\caption{The kinematics of $B \to K_{1}(1270,1400)(\to V P) \ell^{+}\ell^{-}$ decays, where $V=\rho, K^*$ and  $P=K, \pi$.}
\label{figfourfold}
\end{figure}

\subsection{Helicity formalism and helicity amplitudes}\label{secHelicity}

To write the decay amplitudes for semileptonic $B\to V$ decays using the helicity formalism, an elegant approach is presented in \cite{Faessler:2002ut}. In this approach, the orthonormality and the completeness properties of helicity basis $\varepsilon^{\alpha}(n=t, +, -, 0)$, with three
spin-1 components orthogonal to momentum transfer, \textit{i.e.,} $q\cdot\varepsilon(\pm)=q\cdot\varepsilon(0)=0$ are used. These properties can be expressed as follows:
\begin{eqnarray}
\varepsilon^{\ast\mu}(m)\varepsilon_{\mu}(m^{\prime})=g_{mm^{\prime}}, \qquad\quad \sum_{m, m^{\prime}=t, +, -, 0}\varepsilon^{\mu^{\prime}}(m)\varepsilon^{\ast\mu}(m^\prime)g_{mm^{\prime}}=g^{\mu^{\prime}\mu}.\label{C22}
\end{eqnarray}
Here, $0,\; t,\; \pm$ denote the longitudinal, time-like and transverse components of the polarization vector corresponding to $j_{\text{eff}}^\mu$ (see Fig. \ref{figfourfold}), with $g_{mm^{\prime}}=\text{diag}(+, -, -, -)$.


Using Eqs. (\ref{Amp1a}) and (\ref{Amp1b}), along with the completeness property given in Eq. (\ref{C22}), we can write the amplitude in Eq. (\ref{Amp1ag}) as 
\begin{eqnarray}
\mathcal{M}_1^{\lambda_{l^+},\lambda_{l^-}}&=&N_1\,
\sum_{m=t, +, -, 0}g_{mm}\Big\{\varepsilon^{\ast\mu}(m)\epsilon_{K_1}^{\ast\alpha}(r)T^{1}_{\mu\alpha}\,\varepsilon^{\mu^{\prime}}(m)L_{V,\mu^{\prime}}^{{\lambda_{l^+},\lambda_{l^-}}}
+\varepsilon^{\ast\mu}(m)\epsilon_{K_1}^{\ast\alpha}(r)T^{2}_{\mu\alpha}\varepsilon^{\mu^{\prime}}(m)L_{A,\mu^{\prime}}^{{\lambda_{l^+},\lambda_{l^-}}}\Big\}\notag\\
&=&N_1\,
\sum_{m=t, +, -, 0}g_{mm}\Big\{H_m^1\,L_{V,m}^{{\lambda_{l^+},\lambda_{l^-}}}
+H_m^2\,L_{A,m}^{{\lambda_{l^+},\lambda_{l^-}}}\Big\},\label{Amp2ag}
\end{eqnarray}
where, $N_1\equiv\frac{G_{F}\alpha}{2\sqrt{2}\pi}V_{tb}V^{\ast}_{ts}$. Also, $H_m^i=\varepsilon^{\ast\mu}(m)\epsilon_{K_1}^{\ast\alpha}(r)T^{i=1,2}_{\mu\alpha}$ and $L_{V(A),m}^{{\lambda_{l^+},\lambda_{l^-}}}=\varepsilon^{\mu^{\prime}}(m)L_{V(A),\mu^{\prime}}^{{\lambda_{l^+},\lambda_{l^-}}}$ are the hadronic and leptonic helicity amplitudes, respectively. The polarization states of $K_1$ are related to that of the $j_{\text{eff}}^\mu$ current through angular momentum conservation (c.f. Fig. \ref{figfourfold}), such that $r=m$ for $m=\pm,0$ and $r=0$ for $m=t$.

\subsection{Helicity amplitudes for \boldmath $B\to K_1\ell^+\ell^-$ decays}
The helicity amplitude for $B\to K_1$ transition is defined as \cite{Faessler:2002ut}
\begin{eqnarray}
H^{i}_{m} &=&\varepsilon^{\ast\mu}(m)\epsilon_{K_1}^{\ast\alpha}(r)T^{i}_{\mu\alpha}.\label{HA7}
\end{eqnarray} 
Using the same conventions for the polarization vectors as in Ref. \cite{Faessler:2002ut}, and the matrix elements written provided in Appendix \ref{appendHME}, we find the following analytical expressions of helicity amplitudes in terms of the form factors for $B\to K_1\ell^+\ell^-$ decays,
\begin{align}
H^{1}_t&=-\sqrt{\frac{\lambda}{q^2}}(C_{9}^{\text{eff}}+C_{9\ell}^{\text{NP}}+C_{9^{\prime}\ell}^{\text{NP}})V^{K_1}_0\left(q^2\right),\quad\quad\quad
H^{2}_t=-\sqrt{\frac{\lambda}{q^2}}(C_{10}+C_{10\ell}^{\text{NP}}+C_{10^{\prime}\ell}^{\text{NP}})V^{K_1}_0\left(q^2\right),\notag\\
H^{1}_{\pm}&=-\left(m^2_{B}-m^2_{K_1}\right)\Big[(C_{9}^{\text{eff}}+C_{9\ell}^{\text{NP}}+C_{9^{\prime}\ell}^{\text{NP}})
\frac{V_{1}^{K_1}\left(q^2\right)}{m_{B}-m_{K_1}}+\frac{2m_{b}}{q^{2}}C_{7}^{\text{eff}}T_{2}^{K_1}\left(q^2\right)\Big]\notag
\\
&\pm \sqrt{\lambda}\Big[(C_{9}^{\text{eff}}+C_{9\ell}^{\text{NP}}-C_{9^{\prime}\ell}^{\text{NP}})
\frac{A^{K_1}\left(q^2\right)}{m_{B}+m_{K_1}}+\frac{2m_{b}}{q^{2}}C_{7}^{\text{eff}}T_{1}^{K_1}\left(q^2\right)\Big],\notag
\\
H^{2}_{\pm}&=-(C_{10}+C_{10\ell}^{\text{NP}}+C_{10^{\prime}\ell}^{\text{NP}})\left(m_{B}+m_{K_1}\right)
V_{1}^{K_1}\left(q^2\right)\pm \sqrt{\lambda}(C_{10}+C_{10\ell}^{\text{NP}}-C_{10^{\prime}\ell}^{\text{NP}})
\frac{A^{K_1}\left(q^2\right)}{m_{B}+m_{K_1}},\notag
\\
H^{1}_0&=-\frac{1}{2m_{K_1}\sqrt{q^2}}\Bigg[(C_{9}^{\text{eff}}+C_{9\ell}^{\text{NP}}+C_{9^{\prime}\ell}^{\text{NP}})
\Big\{(m^2_{B}-m^2_{K_1}-q^2)\left(m_{B}+m_{K_1}\right)V_{1}^{K_1}\left(q^2\right)\notag
\\
&-\frac{\lambda}{m_{B}+m_{K_1}}V_{2}^{K_1}\left(q^2\right)\Big\}+2m_b C_{7}^{\text{eff}}\Big\{(m^2_{B}+3m^2_{K_1}-q^2)T_{2}^{K_1}\left(q^2\right)
-\frac{\lambda}{m^2_{B}-m^2_{K_1}}T_{3}^{K_1}\left(q^2\right)\Big\}
\Bigg],\notag
\\
H^{2}_0&=-\frac{1}{2m_{K_1}\sqrt{q^2}}(C_{10}+C_{10\ell}^{\text{NP}}+C_{10^{\prime}\ell}^{\text{NP}})
\Bigg[(m^2_{B}-m^2_{K_1}-q^2)\left(m_{B}+m_{K_1}\right)V_{1}^{K_1}\left(q^2\right)\notag
\\
&-\frac{\lambda}{m_{B}+m_{K_1}}V_{2}^{K_1}\left(q^2\right)\Bigg].\label{HA8}
\end{align}

\subsection{Propagation of \boldmath $K_1$ and inclusion of $K_1 \to V P$ decay}
Including the propagation and decay of $K_1$ into the three-body decay, $B\to K_1\ell^+\ell^-$, the total four-body amplitude for $B\to K_1\,(\to V P)\ell^+\ell^-$ decay can be written as
\begin{eqnarray}
\mathcal{M}^{\lambda_{l^+},\lambda_{l^-}}&=&
\sum_{m=t, +, -, 0}N_1\,g_{mm}\Big\{L_{V,m}^{{\lambda_{l^+},\lambda_{l^-}}}\varepsilon^{\ast\mu}(m)T^{1}_{\mu\alpha}
+L_{A,m}^{{\lambda_{l^+},\lambda_{l^-}}}\varepsilon^{\ast\mu}(m)T^{2}_{\mu\alpha}\Big\}\frac{S^{\alpha\alpha^{\prime}}(k)}{k^2-m_{K_1}^2+im_{K_1}\Gamma_{K_1}}\notag\\
&\times&\frac{2\lambda_{K_1 VP}}{m_{K_1}m_V} \Big\{(k\cdot p_{3})\epsilon^{\ast}_{V,\alpha^{\prime}}(l) -(k\cdot \epsilon^{\ast}_{V}(l))p_{3,\alpha^{\prime}}\Big\},\label{Amp3ag}
\end{eqnarray}
where $S^{\alpha\alpha^{\prime}}(k)$ is the standard spin-1 tensor, given by
\begin{eqnarray}
S^{\alpha\alpha^{\prime}}(k)=-g^{\alpha\alpha^{\prime}}+\frac{k^{\alpha}k^{\alpha^{\prime}}}{m_{K_1}^2}=\sum_{r=\pm, 0}\epsilon^{\ast\alpha}_{K_1}(r)\epsilon^{\alpha^{\prime}}_{K_1}(r).\label{S1tensor}
\end{eqnarray}

Assuming $K_1$ decays resonantly, we work in the narrow width approximation, in which we can use the Breit-Wigner distribution
\begin{eqnarray}
BW(k^2)=\frac{1}{k^2-m_{K_1}^2+im_{K_1}\Gamma_{K_1}}\,\,\,\,\Longrightarrow\quad
|BW(k^2)|^2\simeq\frac{\pi}{m_{K_1}\Gamma_{K_1}}\delta\left(k^2-m_{K_1}^2\right).\label{BWigner}
\end{eqnarray}
In this case, the total four-body amplitude for the decay $B\to K_1\,(\to V P)\ell^+\ell^-$ can take the form \footnote{As $r$ is related to $m$ through angular momentum conservation, we ignore the summation over $r$ in Eq. (\ref{Amp4ag}) indicating that we consider only non-zero amplitudes.}

\begin{eqnarray}
\mathcal{M}_{l}^{\lambda_{l^+},\lambda_{l^-}}=
\sum_{m=t, +, -, 0}N_1\,g_{mm}\Big\{L_{V,m}^{{\lambda_{l^+},\lambda_{l^-}}}\,H_m^1
+L_{A,m}^{{\lambda_{l^+},\lambda_{l^-}}}\,H_m^2\Big\}\mathcal{A}_{1}\Big(\epsilon_{K_1}(r),\,\epsilon^{\ast}_{V}(l)\Big)BW(k^2),\label{Amp4ag}
\end{eqnarray}
where $\mathcal{A}_{1}\Big(\epsilon_{K_1}(r),\,\epsilon^{\ast}_{V}(l)\Big)$ is given in Appendix \ref{Casdecay2}. Averaging over the initial spin and summing over the final spins and polarizations, we obtain the squared amplitude for the four-body decay $B\to K_1\,(\to V P)\ell^+\ell^-$ as
\begin{eqnarray}
|\mathcal{M}|^2&=&\sum_{\lambda_{l^+},\lambda_{l^-}}\sum_{l=\pm,0}|\mathcal{M}_{l}^{\lambda_{l^+},\lambda_{l^-}}|^2,\notag\\
&=&\sum_{\lambda_{l^+},\lambda_{l^-}}\sum_{m,n=t, +, -, 0}N_1^2\,g_{mm}g_{nn}\bigg\{L_{V,m}^{{\lambda_{l^+},\lambda_{l^-}}}L_{V,n}^{\ast{\lambda_{l^+},\lambda_{l^-}}}H_m^1 H_n^{1\ast}
+L_{A,m}^{{\lambda_{l^+},\lambda_{l^-}}}L_{A,n}^{\ast{\lambda_{l^+},\lambda_{l^-}}}H_m^2 H_n^{2\ast}\notag\\
&+&L_{V,m}^{{\lambda_{l^+},\lambda_{l^-}}}L_{A,n}^{\ast{\lambda_{l^+},\lambda_{l^-}}}\Big(H_m^1 H_n^{2\ast}
+H_m^2 H_n^{1\ast}\Big)\bigg\}\sum_{l=\pm,0}\mathcal{A}_{1}\Big(\epsilon_{K_1}(r),\,\epsilon^{\ast}_{V}(l)\Big)\mathcal{A}^{\ast}_{1}\Big(\epsilon^{\ast}_{K_1}(s),\,\epsilon_{V}(l)\Big)|BW(k^2)|^2.\label{Amp5ag}
\end{eqnarray}
Here, we can separate the amplitude for longitudinally and transversely polarized final state vector meson as
\begin{eqnarray}
|\mathcal{M}|_{\parallel}^2=\sum_{\lambda_{l^+},\lambda_{l^-}}\sum_{l=0}|\mathcal{M}_{l}^{\lambda_{l^+},\lambda_{l^-}}|^2,\qquad|\mathcal{M}|_{\perp}^2=\sum_{\lambda_{l^+},\lambda_{l^-}}\sum_{l=\pm}|\mathcal{M}_{l}^{\lambda_{l^+},\lambda_{l^-}}|^2,\label{Amp6ag}
\end{eqnarray}
such that
\begin{eqnarray}
|\mathcal{M}|^2=|\mathcal{M}|_{\parallel}^2+|\mathcal{M}|_{\perp}^2.\label{Amp7ag}
\end{eqnarray}

\section{Four-fold angular distributions for \boldmath $B \to K_{1}(1270,1400)(\to V P) \ell^{+}\ell^{-}$ decay}\label{observables}
Integrating over $k^2$ in Eq. (\ref{4diff4}) and combining all the elements in it, the angular decay distribution for the four-fold decay $B\to K_1\,(\to V_{\|(\perp)} P)\ell^+\ell^-$, where the final state vector meson $V$ can be longitudinally or transversely polarized, is obtained as 
\begin{eqnarray}\label{four-folded}
\frac{d^4\Gamma(B\to K_1\,(\to V_{\|(\perp)} P)\ell^+\ell^-)}{dq^2 \ d\cos{\theta_{\ell}} \ d\cos {\theta}_{V} \ d\phi} &=& \frac{9}{32 \pi} \mathcal{B}(K_1\to V_{\|(\perp)} P)
\bigg[I_{1s,\|(\perp)}\sin^2\theta_{V}+I_{1c,\|(\perp)}\cos^2\theta_{V}\notag\\
&+&(I_{2s,\|(\perp)}\sin^2\theta_{V}+I_{2c,\|(\perp)}\cos^2\theta_{V})\cos{2\theta_{\ell}}
\notag\\
&+&I_{3,\|(\perp)}\sin^2\theta_{V}\sin^2\theta_{\ell}\cos{2\phi}+I_{4,\|(\perp)}\sin2\theta_{V}\sin2\theta_{\ell}\cos{\phi}\notag\\
&+&I_{5,\|(\perp)}\sin2\theta_{V}\sin\theta_{\ell}\cos{\phi}\notag\\
&+&(I_{6s,\|(\perp)}\sin^2\theta_{V}+I_{6c,\|(\perp)}\cos^2\theta_{V})\cos{\theta_{\ell}}\notag\\
&+&I_{7,\|(\perp)}\sin2\theta_{V}\sin\theta_{\ell}\sin{\phi}\notag\\
&+&I_{8,\|(\perp)}\sin2\theta_{V}\sin2\theta_{\ell}\sin{\phi}+I_{9,\|(\perp)}\sin^2\theta_{V}\sin^2\theta_{\ell}\sin{2\phi}\bigg].
\end{eqnarray}
In this equation, $I_{1},...,I_{9}$ represent the $q^2$-dependent angular coefficients, with the subscripts $\|$ and $\perp$ specifying the longitudinal and transverse polarizations of the vector meson $V$, respectively. The angles $\theta_V,\; \theta_\ell$ and $\phi$ are illustrated in Fig. \ref{figfourfold}.
The unpolarized angular distribution is recovered by combining
the expressions for the longitudinally and transversely polarized angular distributions. Choosing $VP=\rho K$, and $K^*\pi$, respectively, $\mathcal{B}(K_1\to VP)$ corresponds to the branching ratio of the cascade decay $K_1\to \left(\rho K,\; K^{*}\pi\right)$ and we take its numerical value from the particle data group (PDG) \cite{ParticleDataGroup:2024cfk}. One can adequately say that the angular distribution of $B\to K_{1}(\to \rho_{\|} K,\; K^{*}_{\|}\pi)\ell^{+}\ell^{-}$ is akin to $B\to K^*(\to K\pi)\ell^{+}\ell^{-}$, with the corresponding cascade decays $K_1 \to V_{\|}P$ and $K^{\ast}\to PP$, respectively, affecting the respective angular coefficients $I_{1},...,I_{9}$ in a similar fashion, as discussed in Appendix \ref{transversality-amp}. Similarly, in the angular distribution of $B\to K_1\,(\to V_{\perp} P)\ell^+\ell^-$, cascade decay $K_1\to V_{\perp} P$ can be compared with $a_1\to \rho_{\perp} \pi$ \cite{Colangelo:2019axi} and $D^{\ast}\to \gamma D$ \cite{Colangelo:2018cnj}, showing similar angular distribution with respect to $\theta_{V}$, as shown in $\Gamma_{2,\perp}$ matrix in Appendix \ref{Casdecay2}. Further, by comparing the diagonal elements of the $\Gamma_{2,\|}$, and $\Gamma_{2,\perp}$ matrices, it is clear that by assigning angular coefficients $I_{1c,\|}$, $I_{2c,\|}$, and $I_{6c,\|}$ to $\cos^2{\theta_{V}}$ and $I_{1c,\perp}$, $I_{2c,\perp}$, and $I_{6c,\perp}$ to $3+\cos{2\theta_{V}}$, it is not possible to write the two polarized angular distributions jointly as in Eq. (\ref{four-folded}).\footnote{For the angular distribution, $B\to K_1\,(\to V_{\perp} P)\ell^+\ell^-$, we express the angular structure $9+3\cos{2\theta_{V}}=12\cos^2{\theta_{V}}+6\sin^2{\theta_{V}}$ in the diagonal elements of the $\Gamma_{2,\perp}$ matrix in Appendix \ref{Casdecay2}, so that the transverse and longitudinal polarizations of $K_1$, and consequently that of $j_{\text{eff}}^\mu$ connected by angular momentum conservation, mix in the corresponding angular coefficients $I_{1s,\perp}$, and $I_{2s,\perp}$, and the form of Eq. (\ref{four-folded}) is maintained. In Ref. \cite{Colangelo:2019axi}, helicity amplitudes mix in $I^{a_1}_{1s,\perp}$, $I^{a_1}_{2s,\perp}$, and $I^{a_1}_{6s,\perp}$, however their presentation of angular distribution does not appear to be consistent with ours.} However, with this choice of presentation, it is possible to keep longitudinal and transverse polarizations of $K_1$, and consequently that of $j_{\text{eff}}^\mu$, separate. This allows the longitudinal helicity amplitudes to appear only in $I_{1c,\|}(I_{1s,\perp})$, $I_{2c,\|}(I_{2s,\perp})$, and $I_{6c,\|}(I_{6s,\perp})$, while only the transverse helicity amplitudes can appear in $I_{1s,\|}(I_{1c,\perp})$, $I_{2s,\|}(I_{2c,\perp})$, and $I_{6s,\|}(I_{6c,\perp})$.             



\subsection{Longitudinally polarized vector meson}

In the case of longitudinally polarized vector meson, we find that the angular coefficients $I_{1}$ through $I_{9}$ can be expressed in terms of the helicity amplitudes given in Eq.(\ref{HA8}) as follows:
\begin{eqnarray}\label{30a}
I_{1s,\|} &=& \frac{(2+\beta_\ell^2)}{2}N^2\left(|H_+^1|^2+|H_+^2|^2+|H_-^1|^2+|H_-^2|^2\right)+\frac{4m_\ell^2}{q^2}N^2\left(|H_+^1|^2-|H_+^2|^2+|H_-^1|^2-|H_-^2|^2\right),\\
I_{1c,\|} &=& 2N^2\left(|H_0^1|^2+|H_0^2|^2\right)+\frac{8m_\ell^2}{q^2}N^2\left(|H_0^1|^2-|H_0^2|^2+2|H_t^2|^2\right),\\
I_{2s,\|} &=& \frac{\beta_\ell^2}{2}N^2\left(|H_+^1|^2+|H_+^2|^2+|H_-^1|^2+|H_-^2|^2\right),\\
I_{2c,\|} &=& -2\beta_\ell^2 N^2\left(|H_0^1|^2+|H_0^2|^2\right),\quad I_{3,\|}=-2\beta_\ell^2 N^2\bigg[\mathcal{R}e\left(H_+^{1}H_-^{1\ast}+H_+^{2}H_-^{2\ast}\right)\bigg],\\
I_{4,\|}&=&\beta_\ell^2 N^2\bigg[\mathcal{R}e\left(H_+^{1}H_0^{1\ast}+H_-^{1}H_0^{1\ast}\right)
+\mathcal{R}e\left(H_+^{2}H_0^{2\ast}+H_-^{2}H_0^{2\ast}\right)\bigg],\\
I_{5,\|}&=&-2\beta_\ell N^2\bigg[\mathcal{R}e\left(H_+^{1}H_0^{2\ast}-H_-^{1}H_0^{2\ast}\right)
+\mathcal{R}e\left(H_+^{2}H_0^{1\ast}-H_-^{2}H_0^{1\ast}\right)\bigg],\\
I_{6s,\|}&=&-4\beta_\ell N^2\bigg[\mathcal{R}e\left(H_+^{1}H_+^{2\ast}-H_-^{1}H_-^{2\ast}\right)\bigg],\quad I_{6c,\|}= 0,\\
I_{7,\|}&=&-2\beta_\ell N^2\bigg[\mathcal{I}m\left(H_0^{1}H_+^{2\ast}+H_0^{1}H_-^{2\ast}\right)
+\mathcal{I}m\left(H_0^{2}H_+^{1\ast}+H_0^{2}H_-^{1\ast}\right)\bigg],\\
I_{8,\|}&=&\beta_\ell^2 N^2\bigg[\mathcal{I}m\left(H_0^{1}H_+^{1\ast}-H_0^{1}H_-^{1\ast}\right)
+\mathcal{I}m\left(H_0^{2}H_+^{2\ast}-H_0^{2}H_-^{2\ast}\right)\bigg],\\
I_{9,\|}&=&2\beta_\ell^2 N^2\bigg[\mathcal{I}m\left(H_+^{1}H_-^{1\ast}+H_+^{2}H_-^{2\ast}\right)\bigg],
\end{eqnarray}
where $\beta_{\ell} = \sqrt{1-4m^2_\ell/q^2}$, and $N$, is defined as
\begin{eqnarray}\label{24abc}
N=V_{tb}V^{\ast}_{ts}\Bigg[\frac{G_{F}^2\alpha^2}{3.2^{10} \pi^5 m_{B}^{3}} q^2\sqrt{\lambda}\beta_\ell\Bigg]^{1/2},
\end{eqnarray}
with $\lambda\equiv \lambda(m^2_{B}, m^2_{K_1}, q^2)$.


\subsection{Transversely polarized vector meson}

Similarly, we find that the angular coefficients corresponding to the transversely polarized vector meson take the form
\begin{eqnarray}\label{30df}
I_{1s,\perp} &=&\frac{(2+\beta_\ell^2)}{4}N^2\left(|H_+^1|^2+|H_+^2|^2+|H_-^1|^2+|H_-^2|^2\right)+\left(|H_0^1|^2+|H_0^2|^2\right)\notag\\
&+&\frac{2m_\ell^2}{q^2}N^2\bigg[\left(|H_+^1|^2-|H_+^2|^2+|H_-^1|^2-|H_-^2|^2\right)
+2\left(|H_0^1|^2-|H_0^2|^2+2|H_t^2|^2\right)\bigg],\\
I_{1c,\perp} &=& \frac{(2+\beta_\ell^2)}{2}N^2\left(|H_+^1|^2+|H_+^2|^2+|H_-^1|^2+|H_-^2|^2\right)+\frac{4m_\ell^2}{q^2}N^2\left(|H_+^1|^2-|H_+^2|^2+|H_-^1|^2-|H_-^2|^2\right),
\end{eqnarray}
\begin{eqnarray}
I_{2s,\perp} &=& -\beta_\ell^2 N^2\bigg[\left(|H_0^1|^2+|H_0^2|^2\right)-\frac{1}{4}\left(|H_+^1|^2+|H_+^2|^2+|H_-^1|^2+|H_-^2|^2\right)\bigg],\\
I_{2c,\perp} &=& \frac{\beta_\ell^2}{2} N^2\left(|H_+^1|^2+|H_+^2|^2+|H_-^1|^2+|H_-^2|^2\right),\quad 
I_{3,\perp}=\beta_\ell^2 N^2\bigg[\mathcal{R}e\left(H_+^{1}H_-^{1\ast}+H_+^{2}H_-^{2\ast}\right)\bigg],\\
I_{4,\perp}&=&-\frac{\beta_\ell^2}{2}N^2\bigg[\mathcal{R}e\left(H_+^{1}H_0^{1\ast}+H_-^{1}H_0^{1\ast}\right)
+\mathcal{R}e\left(H_+^{2}H_0^{2\ast}+H_-^{2}H_0^{2\ast}\right)\bigg],\\
I_{5,\perp}&=&\beta_\ell N^2\bigg[\mathcal{R}e\left(H_+^{1}H_0^{2\ast}-H_-^{1}H_0^{2\ast}\right)
+\mathcal{R}e\left(H_+^{2}H_0^{1\ast}-H_-^{2}H_0^{1\ast}\right)\bigg],\\
I_{6s,\perp}&=&-2\beta_\ell N^2\bigg[\mathcal{R}e\left(H_+^{1}H_+^{2\ast}-H_-^{1}H_-^{2\ast}\right)\bigg],\quad 
I_{6c,\perp}=-4\beta_\ell N^2\bigg[\mathcal{R}e\left(H_+^{1}H_+^{2\ast}-H_-^{1}H_-^{2\ast}\right)\bigg],\\
I_{7,\perp}&=&\beta_\ell N^2\bigg[\mathcal{I}m\left(H_0^{1}H_+^{2\ast}+H_0^{1}H_-^{2\ast}\right)
+\mathcal{I}m\left(H_0^{2}H_+^{1\ast}+H_0^{2}H_-^{1\ast}\right)\bigg],\\
I_{8,\perp}&=&-\frac{\beta_\ell^2}{2}N^2\bigg[\mathcal{I}m\left(H_0^{1}H_+^{1\ast}-H_0^{1}H_-^{1\ast}\right)
+\mathcal{I}m\left(H_0^{2}H_+^{2\ast}-H_0^{2}H_-^{2\ast}\right)\bigg],\\
I_{9,\perp}&=&-\beta_\ell^2 N^2\bigg[\mathcal{I}m\left(H_+^{1}H_-^{1\ast}+H_+^{2}H_-^{2\ast}\right)\bigg].
\end{eqnarray}
It is worth mentioning that in addition to expressing the angular coefficients in terms of the helicity amplitudes, they can also be written in terms of the transversality amplitudes \cite{Altmannshofer:2008dz}. We have calculated these relations and provided them in Appendix \ref{transversality-amp}.

\section{Physical observables for \boldmath $B \to K_{1}(1270,1400)(\to V P) \ell^{+}\ell^{-}$ decay}\label{physObs}
This section is devoted to the construction of both polarized and unpolarized physical observables from the four-fold angular decay distribution of $B\to K_1(1270,1400)\left(\to \rho K,\; K^*\pi\right)\ell^{+}\ell^{-}$ decays (c.f. Eq. (\ref{four-folded})). The list of observables includes the differential branching ratios $d\mathcal{B}/{dq^2}$, lepton FB asymmetry $\mathcal{A}_{\text{FB}}^{K_1}$, FB asymmetry for transversely polarized $K_1$ meson $\mathcal{A}_{\text{FB}}^{K_{1T}}$, longitudinal and transverse polarization fractions of $K_1$ meson, $f_{L}^{K_1}$ and $f_{T}^{K_1}$, respectively, and the normalized angular coefficients with longitudinally and transversely polarized final state vector meson $\hat{I}_{\parallel(\perp)}^{K_1}$. The analytical expressions of these observables are discussed separately in the following sections.

\subsection{Differential decay rates and differential branching ratios}
The differential decay rates are important observables from theoretical and experimental perspective, and their expressions can be obtained in terms of the angular coefficients, \textit{i.e.,} the $I$'s. Starting from the four-fold decay distributions given in Eq. (\ref{four-folded}), integrating over $\cos\theta_V = [-1,1]$ and $\phi = [0,2\pi]$ gives the double differential decay rates for the polarized vector meson $\left(\rho, K^*\right)$ in the final state\textit{, i.e.,} 
\begin{eqnarray}
\frac{d^2\Gamma_{\parallel(\perp)}}{dq^2d\cos\theta_{\ell}}&=&\frac{3}{8}\mathcal{B}\left(K_1\to V_{\parallel(\perp)}P\right) \Big(I_{1c,\parallel(\perp)}+2I_{1s,\parallel(\perp)}+\cos2\theta_\ell\left(I_{2c,\parallel(\perp)}+2I_{2s,\parallel(\perp)}\right)\notag
\\
&+&\cos\theta_\ell\left(I_{6c,\parallel(\perp)}+2I_{6s,\parallel(\perp)}\right)\Big).\label{ddrate}
\end{eqnarray}
Further integration over $\cos\theta_\ell = [-1,1]$ in Eq. (\ref{ddrate}) gives us the differential decay rate
\begin{equation}
    \frac{d\Gamma_{\parallel(\perp)}}{dq^2} = \mathcal{B}\left(K_1\to V_{\parallel(\perp)}P\right)\frac{1}{4}\left(3I_{1c,\parallel(\perp)}+6I_{1s,\parallel(\perp)}-I_{2c,\parallel(\perp)}-2I_{2s,\parallel(\perp)}\right),\label{drateag}
\end{equation}
where, in the above equations, $d\Gamma_{\parallel(\perp)}/dq^2\equiv d\Gamma(B\to K_1\,(\to V_{\|(\perp)} P)\ell^+\ell^-)/dq^2$. 
By factoring out the $\mathcal{B}\left(K_1\to V_{\parallel(\perp)}P\right)$, which arises from the subsequent cascade decay $K_1\to V_{\parallel(\perp)} P$ and having the traits of the respective $V$-meson polarization, the differential decay rate of the $B\to K_{1}\ell^{+}\ell^{-}$ decay can be extracted from Eq. (\ref{four-folded}). This gives
\begin{eqnarray}
    \frac{d\Gamma (B\to K_1\ell^+\ell^-)}{dq^2} &=& \frac{1}{4}\left(3I_{1c,\parallel}+6I_{1s,\parallel}-I_{2c,\parallel}-2I_{2s,\parallel}\right)\notag
    \\
    &=&\frac{1}{4}\left(3I_{1c,\perp}+6I_{1s,\perp}-I_{2c,\perp}-2I_{2s,\perp}\right).\label{drateolm}
\end{eqnarray}
Adding the longitudinally and transversely polarized decay rates given in Eq. (\ref{drateag}), and comparing with Eq. (\ref{drateolm}), the total differential decay rate for the four-fold distribution can be expressed as:
\begin{eqnarray}
    \frac{d\Gamma \left(B\to K_1\,(\to V P)\ell^+\ell^-\right)}{dq^2} &=& \frac{d\Gamma_{\parallel}}{dq^2}+\frac{d\Gamma_{\perp}}{dq^2} \notag
    \\
    &=& \mathcal{B}
    \left(K_1\to V P\right)\times\frac{d\Gamma \left(B\to K_1\ell^+\ell^-\right)}{dq^2},
    \label{dratelm}
\end{eqnarray}
where
\begin{equation}
    \mathcal{B}
    \left(K_1\to V P\right) = \mathcal{B}
    \left(K_1\to V_{\|} P\right)+ \mathcal{B}
    \left(K_1\to V_{\perp} P\right).\label{Brparpep}
\end{equation}
The differential branching ratios for $B\to K_1\ell^+\ell^-$ and $B\to K_1\,(\to V_{\|(\perp)} P)\ell^+\ell^-$ decays are given by
\begin{eqnarray}
    \frac{d\mathcal{B} (B\to K_1\ell^+\ell^-)}{dq^2} = \tau_{B} \frac{d\Gamma (B\to K_1\ell^+\ell^-)}{dq^2}, \qquad
   \frac{d\mathcal{B} \left(B\to K_1\,(\to V_{\|(\perp)} P)\ell^+\ell^-\right)}{dq^2} = \tau_{B}  \frac{d\Gamma_{\parallel(\perp)}}{dq^2}.\label{brwithtau}
\end{eqnarray}
The corresponding binned differential branching ratio is defined by
\begin{equation}
\left\langle d\mathcal{B}/dq^2\right\rangle_{\left[q^{2}_{\text{min}},\, q^{2}_{\text{max}}\right]}=\frac{\int^{q^{2}_{\text{max}}}_{q^{2}_{\text{min}}} \left(d\mathcal{B}/dq^2\right)\,dq^2}{q^{2}_{\text{max}}-q^{2}_{\text{min}}}. \label{Binned-BR}
\end{equation}

\subsection{Lepton forward-backward asymmetry}\label{AFB}

The lepton forward-backward asymmetry, $\mathcal{A}_{\text{FB}}^{K_1}$, is related to $\theta_\ell$, which represents the angle between $K_1$ meson and the final state lepton (see Fig. \ref{figfourfold}). In general, $\mathcal{A}_{\text{FB}}^{K_1}$ can be written as $\mathcal{A}_{\text{FB}}^{K_1} = (F-B)/(F+B)$, where $F$ and $B$ correspond to decay rate in the forward and backward hemispheres, respectively. For the decay under discussion, the $q^2$-dependent $\mathcal{A}_{\text{FB}}^{K_1}$ can be obtained from the polarized differential decay rates (\ref{ddrate}), and it reads as follows:
\begin{eqnarray}
\mathcal{A}_{\text{FB}}^{K_1}\left(q^2\right) &\equiv& \left[\int_{0}^1 d\cos\theta_\ell\frac{d^2\Gamma_{\parallel(\perp)}}{dq^2d\cos\theta_{\ell}}-\int_{-1}^0 d\cos\theta_\ell\frac{d^2\Gamma_{\parallel(\perp)}}{dq^2d\cos\theta_{\ell}}\right]\bigg/{\frac{d\Gamma_{\parallel(\perp)}}{dq^2}}.\label{FBexp}
\end{eqnarray} 
In terms of the angular coefficient functions, Eq. (\ref{FBexp}) becomes
\begin{eqnarray}
\mathcal{A}_{\text{FB}}^{K_1}\left(q^2\right) \equiv\frac{6I_{6s,\parallel}}{2(3I_{1c,\parallel}+6I_{1s,\parallel}-I_{2c,\parallel}-2I_{2s,\parallel})}
=\frac{3\left(I_{6c,\perp}+2I_{6s,\perp}\right)}{2(3I_{1c,\perp}+6I_{1s,\perp}-I_{2c,\perp}-2I_{2s,\perp})}.\label{FB}
\end{eqnarray}
Similarly, for the unpolarized vector meson in the final state, $\mathcal{A}_{\text{FB}}^{K_1}$ reads
\begin{eqnarray}
\mathcal{A}_{\text{FB}}^{K_1}\left(q^2\right) &=& \left[\int_{0}^1 d\cos\theta_\ell\left(\frac{d^2\Gamma_{\parallel}}{dq^2d\cos\theta_{\ell}}+\frac{d^2\Gamma_{\perp}}{dq^2d\cos\theta_{\ell}}\right)-\int_{-1}^0 d\cos\theta_\ell\left(\frac{d^2\Gamma_{\parallel}}{dq^2d\cos\theta_{\ell}}+\frac{d^2\Gamma_{\perp}}{dq^2d\cos\theta_{\ell}}\right)\right]\bigg/\left(\frac{d\Gamma_{\parallel}}{dq^2}+\frac{d\Gamma_{\perp}}{dq^2}\right),\notag\\
\label{FBexpunpol}
\end{eqnarray} 
which, in turn, takes the following form
\begin{eqnarray}
\mathcal{A}_{\text{FB}}^{K_1}\left(q^2\right) =\frac{3\Big(\mathcal{B}
    \left(K_1\to V_{\parallel}P\right)2I_{6s,\parallel}+\mathcal{B}
    \left(K_1\to V_{\perp}P\right)\left(I_{6c,\perp}+2I_{6s,\perp}\right)\Big)}{2\Big(\mathcal{B}
    \left(K_1\to V_{\parallel}P\right)(3I_{1c,\parallel}+6I_{1s,\parallel}-I_{2c,\parallel}-2I_{2s,\parallel})+\mathcal{B}
    \left(K_1\to V_{\perp}P\right)
(3I_{1c,\perp}+6I_{1s,\perp}-I_{2c,\perp}-2I_{2s,\perp})\Big)}.\notag\\\label{FBtotun}
\end{eqnarray}
Using the relations, $2I_{6s,\parallel}=I_{6c,\perp}+2I_{6s,\perp}$ in the numerator, and $3I_{1c,\parallel}+6I_{1s,\parallel}-I_{2c,\parallel}-2I_{2s,\parallel}=3I_{1c,\perp}+6I_{1s,\perp}-I_{2c,\perp}-2I_{2s,\perp}$ in the denominator, the result for $\mathcal{A}_{\text{FB}}^{K_1}\left(q^2\right)$ is same for both unpolarized and polarized states of the final vector meson. In other words, this asymmetry is independent of the polarization of vector meson in the cascade decay. 
\subsection{Forward-backward asymmetry for transversely polarized \boldmath $K_1$ meson}\label{FBTPA} 

The expression of transverse forward-backward asymmetry, which arises solely from the transverse polarization of $K_1$ meson, is given by 
\begin{equation}
\mathcal{A}_{\text{FB}}^{K_{1T}}\left(q^2\right) \equiv\frac{3I_{6s,\parallel}}{2(3I_{1s,\parallel}-I_{2s,\parallel})}
=\frac{3\,I_{6c,\perp}}{2(3I_{1c,\perp}-I_{2c,\perp})}.\label{FBTrans}
\end{equation}

\subsection{Longitudinal and Transverse helicity fractions of \boldmath $K_1$ meson}\label{LandTHF}
From the four-fold decay distributions written in Eq. (\ref{four-folded}), integrating over $\cos\theta_\ell = [-1,1]$ and $\phi = [0,2\pi]$ gives the double differential decay rate $\left(\frac{d^{2}\Gamma_{\parallel(\perp)}}{dq^{2}d\cos\theta_{V}}\right)$. Further integrating over $\cos\theta_V = [-1,1]$, with appropriate weight factors, yields the doubly polarized decay rates, expressed in terms of angular coefficients as
\begin{eqnarray}
    \frac{d\Gamma^{K_{1L}}_{\parallel}}{dq^2} &=& \frac{1}{4}\mathcal{B}
    \left(K_1\to V_{\parallel}P\right)\left(3I_{1c,\parallel}-I_{2c,\parallel}\right),\notag
    \\
   \frac{d\Gamma^{K_{1T}}_{\parallel}}{dq^2} &=&\frac{1}{2}\mathcal{B}
    \left(K_1\to V_{\parallel}P\right)\left(3I_{1s,\parallel}-I_{2s,\parallel}\right),\notag\\
   \frac{d\Gamma^{K_{1L}}_{\perp}}{dq^2} &=&\frac{1}{4}\mathcal{B}
    \left(K_1\to V_{\perp}P\right)\Big(\left(6I_{1s,\perp}-3I_{1c,\perp}\right)-\left(2I_{2s,\perp}-I_{2c,\perp}\right)\Big),\notag
    \\
   \frac{d\Gamma^{K_{1T}}_{\perp}}{dq^2} &=&\frac{1}{2}\mathcal{B}
    \left(K_1\to V_{\perp}P\right)\left(3I_{1c,\perp}-I_{2c,\perp}\right).\label{drategamma}
\end{eqnarray}
Here, the superscripts $(K_{1L}, K_{1T})$, represent the polarizations (longitudinal, transverse) of $K_1$ meson, while the subscripts $(\parallel, \perp)$ correspond to the polarizations of the final state vector meson. Additionally, note that
\begin{eqnarray}
    \frac{d\Gamma_{\parallel(\perp)}}{dq^2} = \frac{d\Gamma^{K_{1L}}_{\parallel(\perp)}}{dq^2}+\frac{d\Gamma^{K_{1T}}_{\parallel(\perp)}}{dq^2}.
   \label{relation1}
   \end{eqnarray} 

In $B\to K_1\left(\to V_{\parallel(\perp)}P\right)\ell^{+}\ell^{-}$ decays, the longitudinal helicity fraction corresponds to the longitudinal polarization of $K_1$ meson, \textit{i.e.,}
\begin{eqnarray}
    f^{K_1}_L\left(q^2\right) \equiv \frac{d\Gamma^{K_{1L}}_{\parallel(\perp)}/{dq^2}}{d\Gamma_{\parallel(\perp)}/{dq^2}}.
   \label{flrelation}
   \end{eqnarray}
In terms of angular coefficients, it becomes
\begin{eqnarray}
f^{K_1}_L\left(q^2\right) \equiv\frac{3I_{1c,\parallel}-I_{2c,\parallel}}{3I_{1c,\parallel}+6I_{1s,\parallel}-I_{2c,\parallel}-2I_{2s,\parallel}}
=\frac{\left(6I_{1s,\perp}-3I_{1c,\perp}\right)-\left(2I_{2s,\perp}-I_{2c,\perp}\right)}{3I_{1c,\perp}+6I_{1s,\perp}-I_{2c,\perp}-2I_{2s,\perp}}.\label{flfinal}
\end{eqnarray}
Similarly, for the unpolarized vector meson in the final state, $B\to K_1\left(\to V P\right)\ell^{+}\ell^{-}$, longitudinal helicity fraction is defined as follows
\begin{eqnarray}
f^{K_1}_L\left(q^2\right)= \left(\frac{d\Gamma^{K_{1L}}_{\parallel}}{dq^2}+\frac{d\Gamma^{K_{1L}}_{\perp}}{dq^2}\right)\bigg/\left(\frac{d\Gamma_{\parallel}}{dq^2}+\frac{d\Gamma_{\perp}}{dq^2}\right).
\label{flexpunpol}
\end{eqnarray}
This yields the same result for $f^{K_1}_L$ as obtained in the case of any of the polarized final vector meson state, showing that the longitudinal helicity fraction is independent of final state vector meson polarizations. Similarly, the transverse helicity fraction corresponds to the transverse polarization of $K_1$ meson is
\begin{eqnarray}
    f^{K_1}_T\left(q^2\right) \equiv \frac{d\Gamma^{K_{1T}}_{\parallel(\perp)}/{dq^2}}{d\Gamma_{\parallel(\perp)}/{dq^2}}.
   \label{fTrelation}
   \end{eqnarray}
In terms of angular coefficients, it will take the form
\begin{eqnarray}
f^{K_1}_T\left(q^2\right) \equiv\frac{2\left(3I_{1s,\parallel}-I_{2s,\parallel}\right)}{3I_{1c,\parallel}+6I_{1s,\parallel}-I_{2c,\parallel}-2I_{2s,\parallel}}
=\frac{2\left(3I_{1c,\perp}-I_{2c,\perp}\right)}{3I_{1c,\perp}+6I_{1s,\perp}-I_{2c,\perp}-2I_{2s,\perp}}.\label{fTfinal}
\end{eqnarray}


\subsection{Angular Coefficients} \label{AngCoef}
The various normalized angular coefficients are given as
\begin{eqnarray}
    \hat{I}_{n\kappa,{\parallel\left(\perp\right)}}^{K_1}=  \frac{\mathcal{B}\left(K_1\to V_{\parallel(\perp)}P\right)I_{n\kappa,{\parallel\left(\perp\right)}}}{d\Gamma_{\parallel\left(\perp\right)}/dq^2}.\label{A-Coeffincients}
\end{eqnarray}
where the angular coefficients $I_{n\kappa,{\parallel\left(\perp\right)}}$, with $n=1,...,9$, and $\kappa=s, c$ are functions of $q^2$.
\subsection{Binned Angular Coefficients}\label{binnedAngCoef}
To analyze the angular distributions over specified intervals of $q^2$, the binned angular coefficients are expressed as:
\begin{equation}
\hat{I}_{{n\kappa,\parallel\left(\perp\right)}_{\left[q^{2}_{\text{min}},\, q^{2}_{\text{max}}\right]}}^{K_1}=\frac{\int^{q^{2}_{\text{max}}}_{q^{2}_{\text{min}}}\mathcal{B}\left(K_1\to V_{\parallel(\perp)}P\right)I_{n\kappa,{\parallel\left(\perp\right)}}\,dq^2}{\int^{q^{2}_{\text{max}}}_{q^{2}_{\text{min}}}(d\Gamma_{\parallel\left(\perp\right)}/dq^2)dq^2}. \label{Binned-Acoefficients}
\end{equation}

\section{Numerical Analysis}\label{num-analysis}
In this section, we perform the numerical analysis of $B\to K_{1}\left(1270,\; 1400\right)(\to VP)\ell^+\ell^-$ decay, where $V$ and $P$ in the cascade decay correspond to $\rho,\; K^*$ and $K,\; \pi$, respectively, and the final state lepton is taken as the muon, \textit{i.e.,} $\ell = \mu$. From the PDG \cite{ParticleDataGroup:2024cfk}, the dominant decay mode for $K_1\left(1270\right)$ is $K_1\left(1270\right)\to \rho K$, with the branching ratio of $\left(42\pm 6\right)\%$. Similarly, the probability of $K_1\left(1400\right)$ decaying to $K^*\pi$ is $\left(94\pm6\right)\%$. 

\begin{table*}[tbp] 
\centering
\captionsetup{margin=1.5cm}
\caption{\small The decay form factors for $B\to K_{1A,1B}$ transition \cite{Yang:2008xw}, where
$a$ and $b$ are the form factor parameters in the dipole parametrization.}\label{tabelFFs}
\begin{tabular}{|p{.7in}p{.7in}p{.7in}p{.4in}p{.7in}p{.7in}p{.7in}p{.4in}|}
\hline \hline
$\mathcal{T}^{X}_{i}(q^{2})$&\quad$\mathcal{T}(0)$&\quad$a$&\quad$b$&$\mathcal{T}^{X}_{i}(q^{2})$&\quad$\mathcal{T}(0)$&\quad$a$&\quad$b$\\
\hline
$V_{1}^{K_{1A}}$&$0.34\pm 0.07$&$0.635$&$0.211$&$V_{1}^{K_{1B}}$&$-0.29^{+0.08}_{-0.05}$&$0.729$&$0.074$\\
$V_{2}^{K_{1A}}$&$0.41\pm 0.08$&$1.51$&$1.18$&$V_{2}^{K_{1B}}$&$-0.17^{+0.05}_{-0.03}$&$0.919$&$0.855$\\
$V_{0}^{K_{1A}}$&$0.22\pm 0.04$&$2.40$&$1.78$&$V_{0}^{K_{1B}}$&$-0.45^{+0.12}_{-0.08}$&$1.34$&$0.690$\\
$A^{K_{1A}}$&$0.45\pm 0.09$&$1.60$&$0.974$&$A^{K_{1B}}$&$-0.37^{+0.10}_{-0.06}$&$1.72$&$0.912$\\
$F_{1}^{K_{1A}}$&$0.31^{+0.09}_{-0.05}$&$2.01$&$1.50$&$F_{1}^{K_{1B}}$&$-0.25^{+0.06}_{-0.07}$&$1.59$&$0.790$\\
$F_{2}^{K_{1A}}$&$0.31^{+0.09}_{-0.05}$&$0.629$&$0.387$&$F_{2}^{K_{1B}}$&$-0.25^{+0.06}_{-0.07}$&$0.378$&$-0.755$\\
$F_{3}^{K_{1A}}$&$0.28^{+0.08}_{-0.05}$&$1.36$&$0.720$&$F_{3}^{K_{1B}}$&$-0.11\pm 0.02$&$-1.61$&$10.2$\\
\hline\hline
\end{tabular}
\end{table*}

As an exclusive process, the matrix elements of $B\to K_1$ transition are parameterized in terms of the form factors, which are the non-perturbative (model-dependent) inputs. In our numerical analysis we used the form factors calculated in the light-cone QCD sum rules (LCSR) approach \cite{Yang:2008xw}, and their values at $q^2=0$, along with other relevant parameters are summarized in Table \ref{tabelFFs}. Their extrapolation with respect to $q^2$ is performed through the momentum dependence dipole parameterization, which is given by 
\begin{equation}
\mathcal{T}^{X}_{i}(q^{2})=\frac{\mathcal{T}^{X}_{i}(0)}{1-a_{i}^{X}\left(q^{2}/m^{2}_{B}\right)+b_{i}^{X}\left(q^{2}/m^{2}_{B}\right)^{2}},\label{m11}
\end{equation}
where $\mathcal{T}$ correspond to the form factors $A$, $V$ or $T$, and the
subscript $i$ can take a value $0, 1, 2$, or $3$.  The superscript $X$
denotes the weak eigenstates $K_{1A}$ or $K_{1B}$, and for the form factors of the physical states $K_1\left(1270\right)$ and $K_1\left(1400\right)$, we use the mixing given in Eqs. (\ref{mix1}) and (\ref{mix2}). The vertex renormalization and hard-spectator corrections to these form factors are studied in \cite{Sikandar:2019qyb}, where it has been shown that except for $T_{2,3}\left(q^2\right)$, these corrections are masked by the uncertainties in the input values of the form factors at $q^2=0$. Even in $T_{2,3}\left(q^2\right)$, their contributions are around $10\%$, and hence we will not consider these corrections in our analysis. The numerical values of the SM Wilson coefficients (WCs) are calculated at the renormalization scale $\mu \approx m_b$ and are given in Table \ref{tab:Nvalues}.

\begin{table*}[htp!]
\captionsetup{margin=3.5cm}
\caption{\small The SM values of the WCs with next-to-next-leading logarithmic accuracy, evaluated at the renormalization scale $\mu\sim m_{b}$ \cite{Blake:2016olu}.}
\label{tab:Nvalues}
    \begin{tabular}{|ccccc|}
           \hline\hline
		\, $C_1=-0.294$ ,&  $C_2=1.017$ ,&  $C_3=-0.0059$,&  $C_4=-0.087$,&
    $C_5 =0.0004$ \\ 
             $C_6=0.0011$,&  \, $C_7= -0.324$,&  $C_8=-0.176$,&  $C_9=4.114$,&  $C_{10}=-4.193$\, \\
            \hline\hline
	\end{tabular}
\end{table*}

\begin{table*}[t!]
\renewcommand{\arraystretch}{1}
\begin{center}
\captionsetup{margin=4.5cm}
 \caption{\small Best-fit values of the WCs and the $1\sigma$ ranges of the different NP scenarios with both LFU and LFUV, as presented in \cite{Alguero:2023jeh}.}\label{tab:bestfitWC}
			\begin{tabular}{|clcc|}
				\hline\hline
				Scenario & & Best-fit value & $1\sigma$ \\
				\hline
                S1 & $C_{9\mu}^{\text{V}}$                          &$-1.02$       & $[-1.43, -0.61]$  \\
                 & $C_{10\mu}^{\text{V}}$   &$-0.35$       & $[-0.75, -0.00]$  \\
                 & $C_{9}^{\text{U}}=C_{10}^{\text{U}}$   &$+0.19$       & $[-0.16, +0.58]$  \\
                S2 & $C_{9\mu}^{\text{V}}$   &$-0.21$       & $[-0.39, -0.02]$  \\
                   & $C_{9}^{\text{U}}$                            &$-0.97$       & $[-1.21, -0.72]$  \\
                S3 & $C_{9\mu}^{\text{V}}=-C_{10\mu}^{\text{V}}$     &$-0.08$       & $[-0.14, -0.02]$  \\
                   & $C_{9}^{\text{U}}$                              &$-1.10$       & $[-1.27, -0.91]$  \\
                S4 & $C_{9\mu}^{\text{V}}$                           &$-0.68$       & $[-0.84, -0.52]$  \\
                   & $C_{10^{\prime}}^{\text{U}}$                    &$-0.03$       & $[-0.15, +0.09]$  \\
                \hline\hline
	\end{tabular}
	\end{center}
\end{table*}

Based on the recent global fit analysis to the $b\to s$ data, a comprehensive update incorporating both the experimental inputs and the theoretical framework with several NP scenarios, including those with LFU and lepton-flavor universality violating (LFUV) new physics contributions, are presented in Table 9 of \cite{Alguero:2023jeh}. Out of their various possibilities, we have considered scenarios 5, 7, 8, and 11, which show relatively greater pull and are differentiable in the observables considered and can be easily realized in specific NP models. Their best-fit values with $1\sigma$ intervals are summarized in Table \ref{tab:bestfitWC}. For simplicity, we have renamed them as scenarios 1, 2, 3 and 4 in Table \ref{tab:bestfitWC}.

\begin{figure}[htp!]
\begin{tabular}{cc}
\hspace{0.6cm}($\mathbf{a}$)&\hspace{1.2cm}($\mathbf{b}$)\\
\includegraphics[scale=0.6]{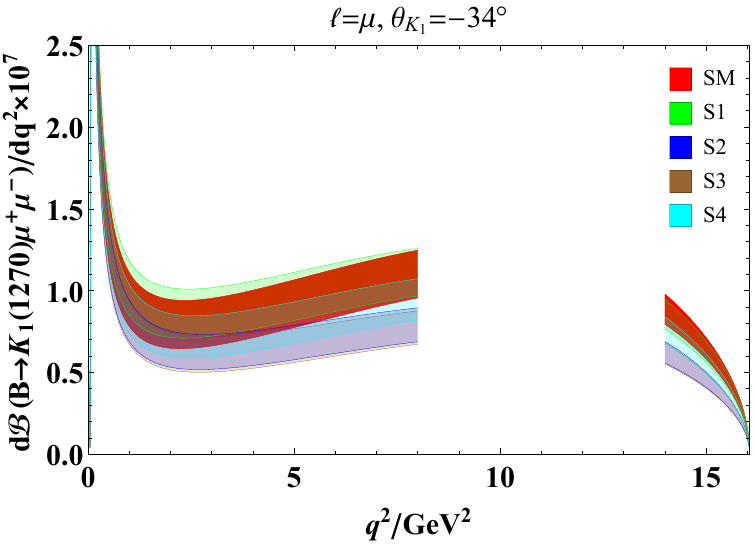} 
&\includegraphics[scale=0.6]{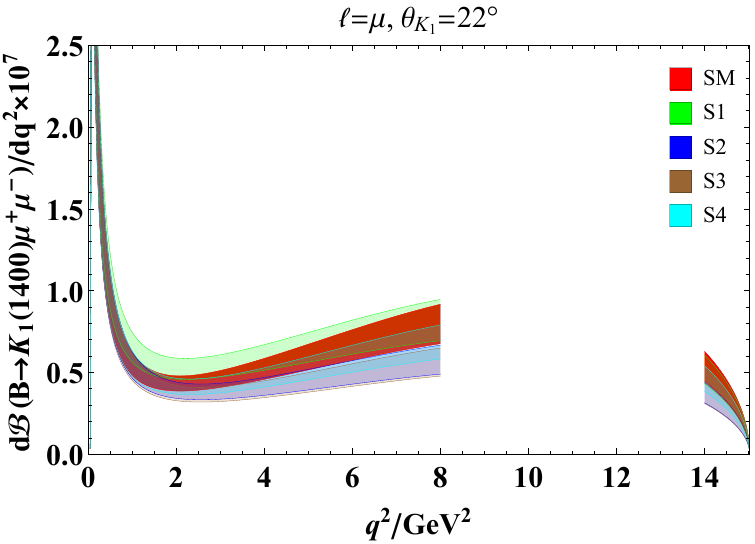}
\end{tabular}
\caption{The profile of differential branching ratios of $\mathcal{B}\to K_1\left(1270,\; 1400\right)\mu^{+}\mu^{-}$ as a function of $q^2$.}
\label{diffbratio}
\end{figure}

With these input parameters of the SM and using the values of WCs in various NP scenarios given in Table \ref{tab:bestfitWC}, we now analyze their impact on different physical observables discussed in Section \ref{physObs}. To illustrate this, we have plotted these observables as a function of the square of the momentum transfer $q^2$ in Figs. \ref{diffbratio} - \ref{I2sK1400}. The bands in these plots correspond to the errors arising from input parameters, where the form factors are the main contributor. To assess the impact of the NP on different observables, we took the lower limit from the $1\sigma$ range. Since the binned analysis of physical observables in these decays is useful for the experimental measurements, we have calculated their numerical values in different bins, both in the SM and in NP scenarios outlined above, and listed them in Tables \ref{Bin1-analysis} - \ref{Bin51400-analysis}. The uncertainties in these results are due to the input values of the form factors.


\begin{itemize}
    \item For the particular values of the mixing angle $\theta_{K_1}$, Figs. \ref{diffbratio}a and \ref{diffbratio}b show the behavior of the differential branching ratio of $B\to K_1\left(1270,\; 1400\right)\mu^{+}\mu^{-}$ decays as a function of $q^2$ in the SM and various NP scenarios. In the SM,  as shown in Fig. \ref{diffbratio}a, the branching ratio with $K_1\left(1270\right)$ meson in the final state is of the order of $10^{-7}$ for $\theta_{K_1}=-34^\circ$, a widely used value from \cite{Hatanaka:2008xj}. It is pertinent to mention that if we take the same value of mixing angle for $B\to K_1\left(1400\right)\mu^{+}\mu^{-}$,  the branching ratio is one to two orders of magnitude smaller than that for $K_1\left(1270\right)$ as final state meson \cite{Hatanaka:2008gu}.
    With $\theta_{K_1}=34^\circ$, the situation reverses, with $B\to K_1\left(1270\right)\mu^{+}\mu^{-}$ being more suppressed \cite{Hatanaka:2008gu}. This suppression in the branching ratio clearly indicates the influence of the mixing angle on the decay form factors. As explained in section \ref{sec1}, there is no unified value of $\theta_{K_1}$, and considering that the highly suppressed decay modes are difficult to measure experimentally, we adopt $\theta_{K_1}=22^\circ$, based on the latest study \cite{Shi:2023kiy}, for $B\to K_1\left(1400\right)\mu^{+}\mu^{-}$ decay. This value enhances the branching ratio of $B\to K_1\left(1400\right)\mu^{+}\mu^{-}$ by two orders of magnitude ($10^{-7}$), bringing it to the same order as that of $B\to K_1\left(1270\right)\mu^{+}\mu^{-}$ decay. Therefore, we have plotted the differential branching ratio of  $B\to K_1\left(1400\right)\mu^{+}\mu^{-}$ by taking $\theta_{K_1} = 22^\circ$ in Fig. \ref{diffbratio}b. The red band in each plot corresponds to the uncertainties, in the SM predictions, arising from the form factors and other input parameters. Except for scenario S1 (green band), the parametric space of other NP scenarios under consideration decreases the value of differential branching ratio in low $q^2$ range, although these effects are masked by the uncertainties in the form factors. Notably, among the NP scenarios, scenario S2 (blue band) shows the most prominent decreasing effects on the value of the differential branching ratio. In addition, for the quantitative analysis, the numerical values of the differential branching ratios of $B\to K_1\left(1270,\; 1400\right)\mu^{+}\mu^{-}$ in the SM and NP scenarios in different bins of $q^2$ are provided in Tables \ref{Bin1-analysis} - \ref{Bin51400-analysis}. As observed in Figs. \ref{diffbratio}a and \ref{diffbratio}b, except for S1, the other NP scenarios tend to decrease the branching ratio, particularly S2 and S3. The SM differential branching ratio is reduced by about $30\%$ in all bins for $K_1\left(1270\right)$ in these two scenarios. However, for $K_1\left(1400\right)$, the maximum reduction is around $(5-10)\% $, and this small deviation from the SM value is almost obscured by the uncertainties in the form factors. 

\begin{figure}[t!]
\begin{tabular}{cc}
\hspace{0.6cm}($\mathbf{a}$)&\hspace{1.2cm}($\mathbf{b}$)\\
\includegraphics[scale=0.60]{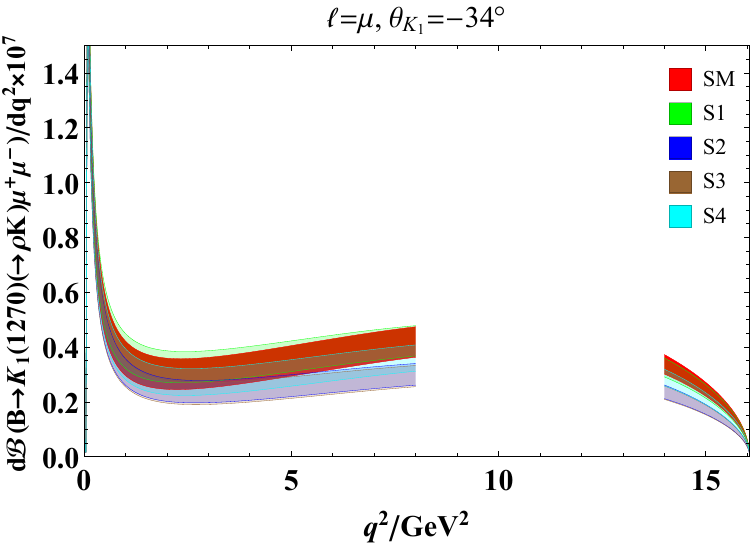} 
&\includegraphics[scale=0.60]{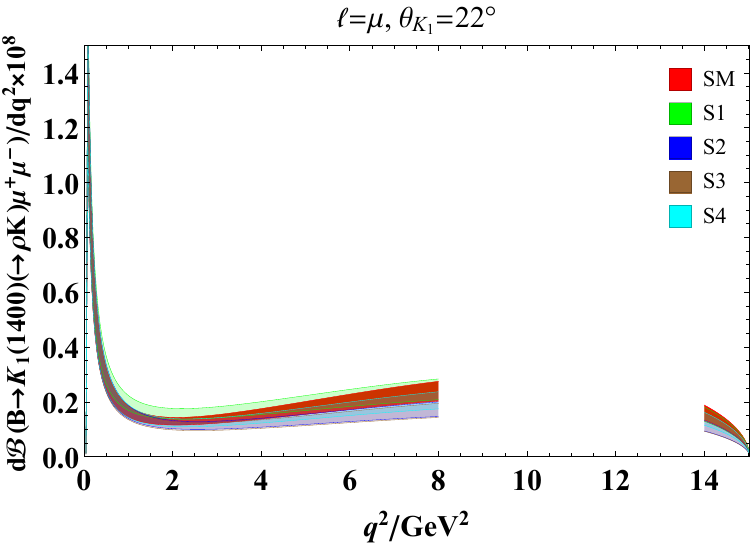}\\
\hspace{0.6cm}($\mathbf{c}$)&\hspace{1.2cm}($\mathbf{d}$)\\
\includegraphics[scale=0.60]{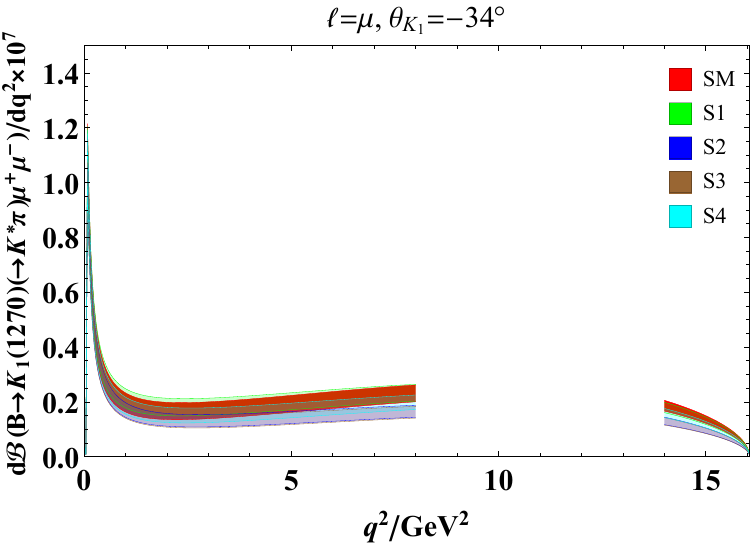} 
&\includegraphics[scale=0.60]{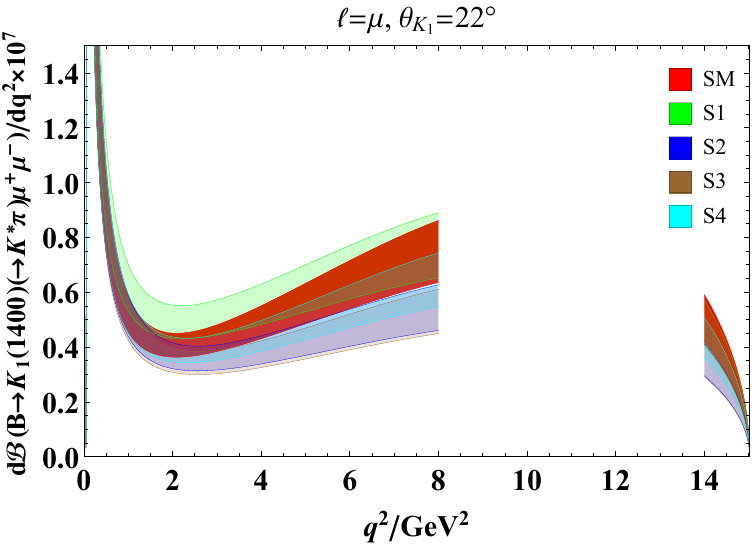}
\end{tabular}
\caption{The differential branching ratios of $B\to K_1\left(1270,\; 1400\right)\mu^{+}\mu^{-}$ as a function of $q^2$, with $K_1\left(1270,\; 1400\right)\to \rho K,\; K^*\pi$ decays.}
\label{BRK1270}
\end{figure}

\item In Figs. \ref{BRK1270}a and \ref{BRK1270}b, we have plotted the differential branching ratio of $B\to K_1\left(1270\right)\mu^{+}\mu^{-}$ as a function of $q^2$, considering the $K_1\left(1270\right)\to \rho K,\; K^*\pi$ decays. With reference to the full decay width of $K_1\left(1270\right)$ meson, branching probabilities to $\rho K$ and $K^*\pi$ are $\left(38\pm 13\right)\%$ and $\left(21\pm 10\right)\%$, respectively. From Eq. (\ref{dratelm}), it is clear that the branching ratio for the cascade decay of the final state meson is factored out, therefore except for an overall suppression factor due to the decay rate corresponding to the relative probability of the cascade decay, the trend of the differential branching ratio remains the same for $ \rho K,$ and $K^*\pi$ as final states. This can be seen in Figs. \ref{BRK1270}(a, b). 
The numerical values of the differential branching ratios, involving $K_1\left(1270\right)$, in different bins of $q^2$, shown in Tables \ref{Bin1-analysis} to \ref{Bin6-analysis}, also highlight that. For the $K_1\left(1400\right)$ meson, the decay probabilities in the $K^*\pi$ and $K\rho$ are $\left(94\pm 6\right)\%$ and $\left(3\pm 3\right)\%$, respectively. In the SM, the dominant $K^*\pi$ parentage is apparent from Tables \ref{Bin11400-analysis} - \ref{Bin51400-analysis}, where the value with $B\to K_1\left(1400\right)\left(\to K^*\pi\right)\mu^{+}\mu^{-}$ is close to the full branching ratio of $B\to K_1\left(1400\right)\mu^{+}\mu^{-}$ across the various $q^2$ bins. Once again, the effects of NP are obscured by the uncertainties in the form factor, and the imprints of the various NP scenarios mirror those seen in the full branching ratio discussed above.
\begin{figure}[ht!]
\begin{tabular}{cc}
\hspace{0.6cm}($\mathbf{a}$)&\hspace{1.2cm}($\mathbf{b}$)\\
\includegraphics[scale=0.60]{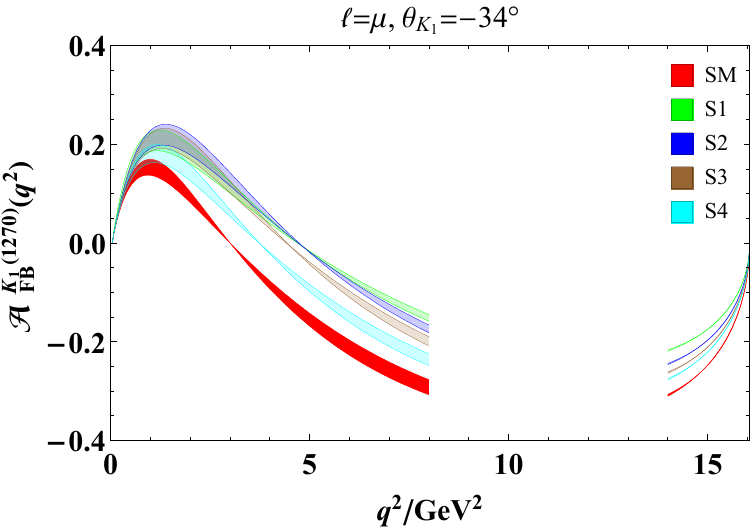} 
&\includegraphics[scale=0.60]{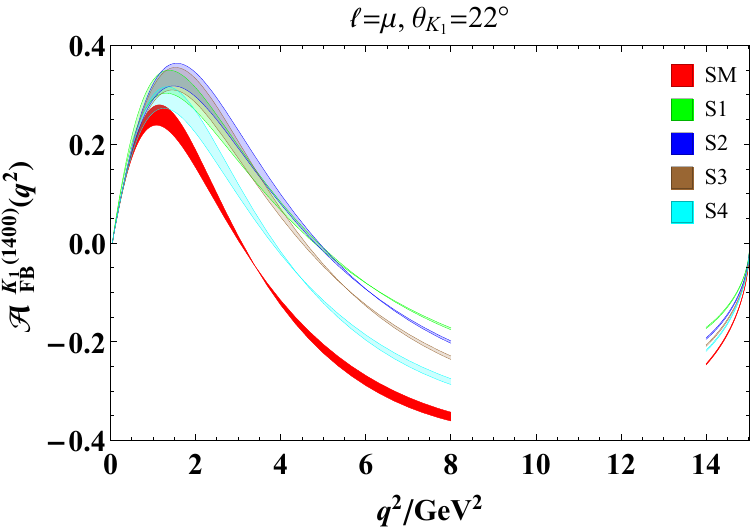}\\
\hspace{0.6cm}($\mathbf{c}$)&\hspace{1.2cm}($\mathbf{d}$)\\
\includegraphics[scale=0.60]{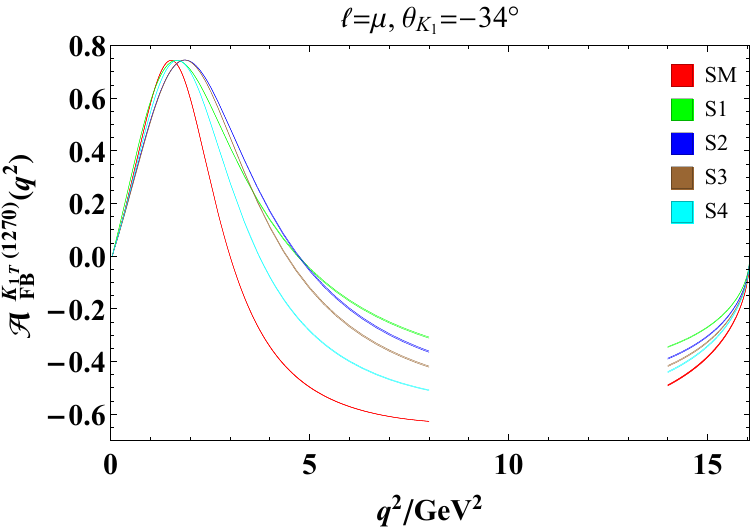} 
&\includegraphics[scale=0.60]{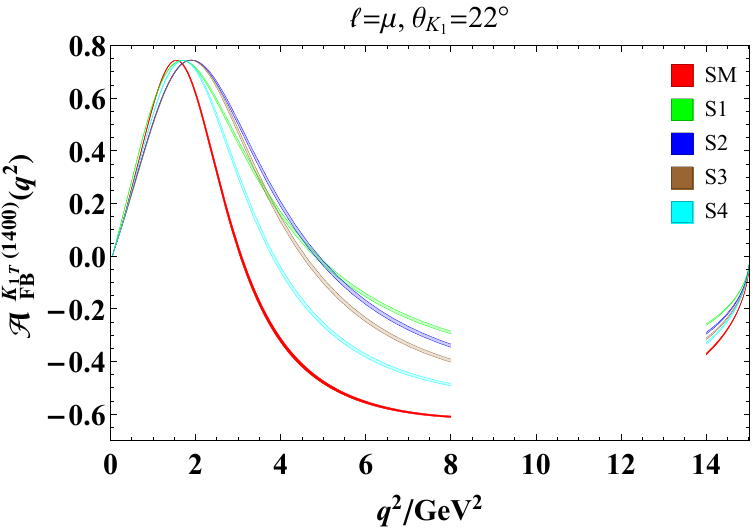}
\end{tabular}
\caption{The $\mathcal{A}_{\text{FB}}^{K_1}$ and $\mathcal{A}_{\text{FB}}^{K_{1T}}$ for unpolarized and transversely polarized $K_1\left(1270,\; 1400\right)$ mesons, respectively. }
\label{FBAK1270}
\end{figure}

\begin{figure}[ht!]
\begin{tabular}{cc}
\hspace{0.6cm}($\mathbf{a}$)&\hspace{1.2cm}($\mathbf{b}$)\\
\includegraphics[scale=0.60]{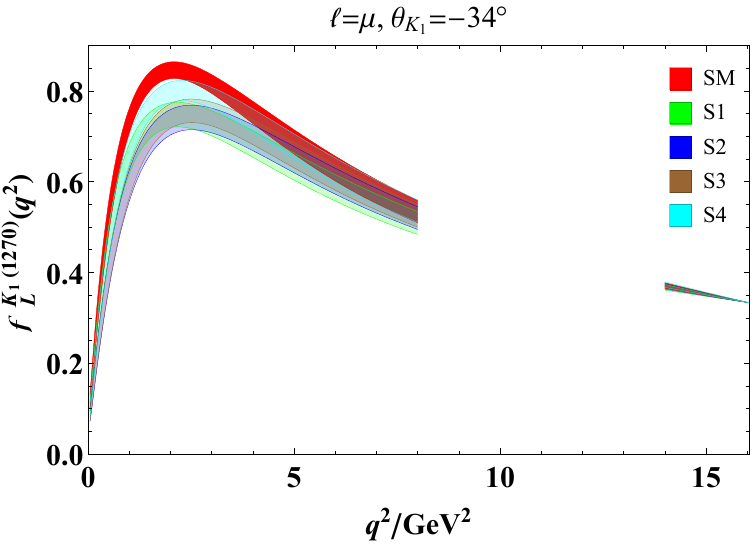} 
&\includegraphics[scale=0.60]{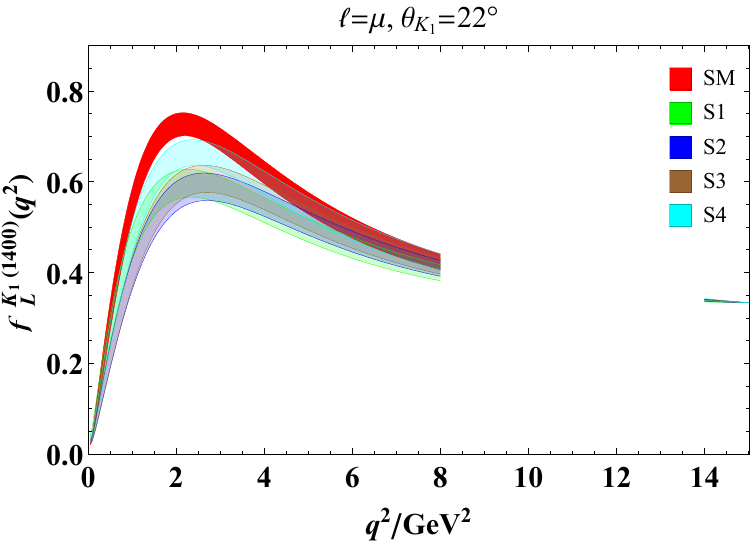}\\
\hspace{0.6cm}($\mathbf{c}$)&\hspace{1.2cm}($\mathbf{d}$)\\
\includegraphics[scale=0.60]{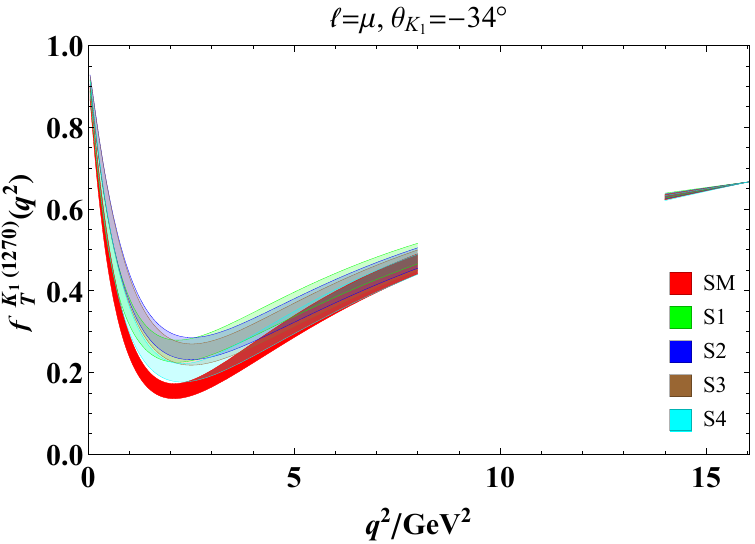} 
&\includegraphics[scale=0.60]{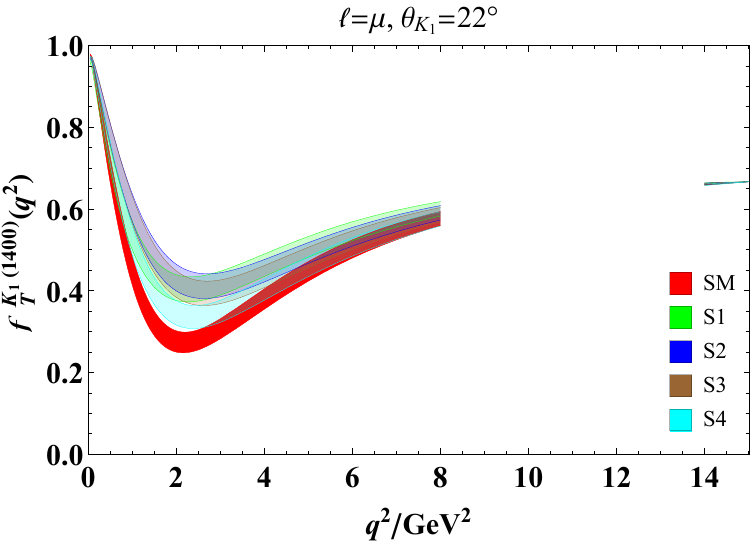}
\end{tabular}
\caption{The longitudinal $f_L^{K_1}(q^2)$ and transverse $f_T^{K_1}(q^2)$ helicity fractions of $K_1\left(1270,\; 1400\right)$ mesons.}
\label{fLK1270}
\end{figure}

\item Fig. \ref{FBAK1270}a shows the profile of the $\mathcal{A}_\text{FB}$ for the $B\to K_1\left(1270\right)\mu^{+}\mu^{-}$ decay, as defined in Eq. (\ref{FBtotun}), as a function of $q^2$. In the SM, the $\mathcal{A}_\text{FB}$ crosses zero at $q^2\approx 2.9 \;\text{GeV}^2$, which is due to the opposite sign of $C_7^{\text{eff}}$ and $C_{9}^{\text{eff}}$, and this crossing point has minimal uncertainty due to the form factors. Therefore, the zero-crossing position is sensitive to the NP contributions in any of these WCs. This can be seen from Fig. \ref{FBAK1270}a, where negative NP contributions to $C_9$ shifts the zero-position of $\mathcal{A}_\text{FB}$ to higher $q^2$ values. The maximum deviation in the zero-position comes from the scenarios S1 and S2, where it shifts to $q^2\approx 5.0\; \text{GeV}^2$, although their effects are intertwined. In S4, the zero-position shifts to $q^2\approx 4.0\; \text{GeV}^2$, and it is distinguishable clearly from both the SM and the other three NP scenarios. Moreover, in all the NP scenarios, the value of $\mathcal{A}_\text{FB}$ is higher than the SM result for the whole $q^2$ range. From the normalized value of $\mathcal{A}_\text{FB}$ in different bins, see Tables \ref{Bin1-analysis}- \ref{Bin6-analysis}, we observe that the errors due to the hadronic uncertainties are almost two orders of magnitude smaller compared to the branching ratios. This implies that even the small deviations from the SM results observed at the experiments would hint for the NP. The trend remains the same when $K_1(1400)$ is the final state meson and this can be seen in Fig. \ref{FBAK1270}b, along with the corresponding values in different bins appended in Tables \ref{Bin11400-analysis} - \ref{Bin51400-analysis}.

The Figs. \ref{FBAK1270}c and \ref{FBAK1270}d, show the $\mathcal{A}_{\text{FB}}^{K_{1T}}$ when $K_1\left(1270\right)$ and $K_1\left(1400\right)$ are transversely polarized 
 (c.f. Eq. (\ref{FBTrans}), respectively. Once again, the position of the zero of the $\mathcal{A}_{\text{FB}}^{K_{1T}}$ is significantly different from the corresponding SM results in both cases. Moreover, here scenarios S3 and S4 are separable from each other and from the other NP scenarios. In addition, the band due to the hadronic uncertainties is quite narrow almost three orders of magnitude smaller than its central values in various bins (see, for instance, Tables \ref{Bin1-analysis} - \ref{Bin51400-analysis}). This makes $\mathcal{A}_{\text{FB}}^{K_{1T}}$ an important observable in the search of NP in the future experiments.

\item The helicity fractions of the final state meson in the semileptonic $B$-meson decays are useful observables to search for NP and to distinguish different NP scenarios. In Figs. \ref{fLK1270}a and \ref{fLK1270}b (\ref{fLK1270}c and \ref{fLK1270}d), we have plotted the $q^2$ dependence of $f_L^{K_1}(q^2)$ and $f_T^{K_1}(q^2)$ for $K_1(1270)$ and $K_1(1400)$ mesons, as calculated in Eqs. (\ref{flexpunpol}) and (\ref{fTrelation}), respectively. One can see that similar to the differential branching ratio, the NP effects interfere destructively with the SM in $f_L^{K_1}(q^2)$, and hence decrease its value in the low $q^2\in [1,5]\;\text{GeV}^2$ range. Compared to this, the NP effects in $f_T^{K_1}(q^2)$ are constructive by the same amount due to the fact that $f_L^{K_1}(q^2)+f_T^{K_1}(q^2) = 1$. This equality can also be ensured from their values in different $q^2$ bins appended in Tables \ref{Bin1-analysis}- \ref{Bin6-analysis} and for $K_1\left(1400\right)$ meson in Tables \ref{Bin11400-analysis} - \ref{Bin51400-analysis}. While the most visible deviations from the SM values occur in scenario S3, with the difference of around $10\%$. In contrast to the branching ratio, this difference in the helicity fraction is significant not only to distinguish the NP effects from the SM but also helpful to segregate different NP scenarios, particularly in the $q^2=[1.0-3.0]$ GeV$^2$ bin, where the uncertainties from the form factors are small, and hence make them the promising candidates for probing NP in these decays.



\begin{figure}[htbp!]
\begin{tabular}{ccc}
\includegraphics[scale=0.405]{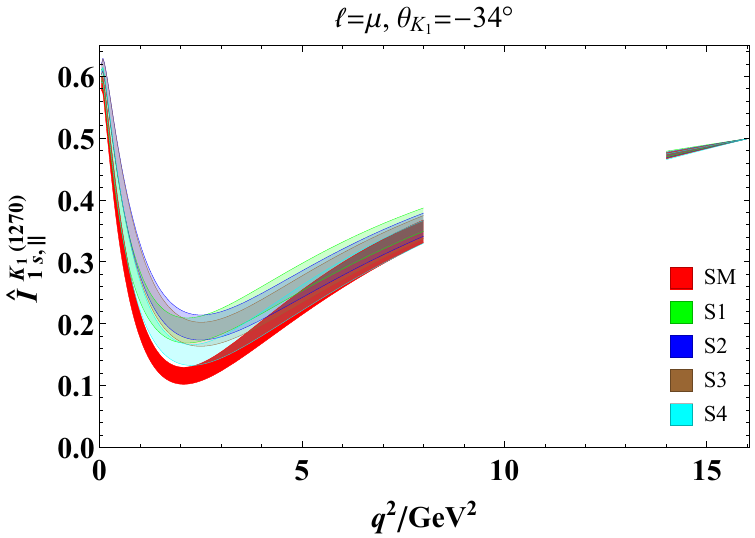} 
&\includegraphics[scale=0.405]{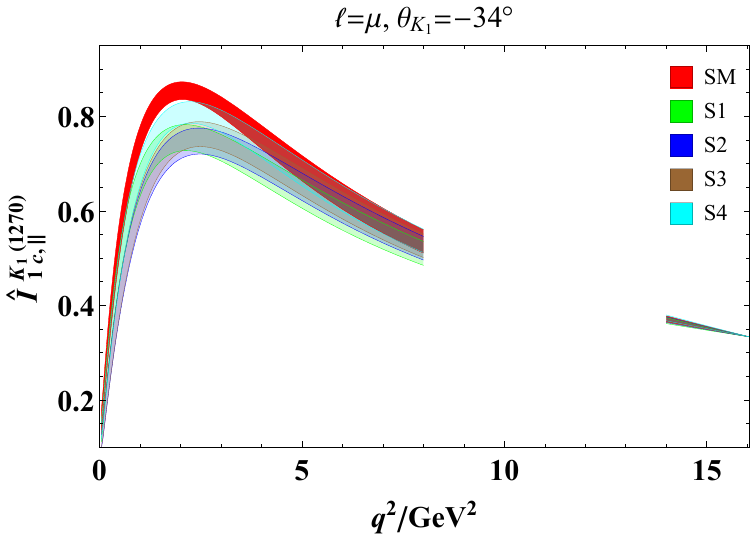}&\includegraphics[scale=0.405]{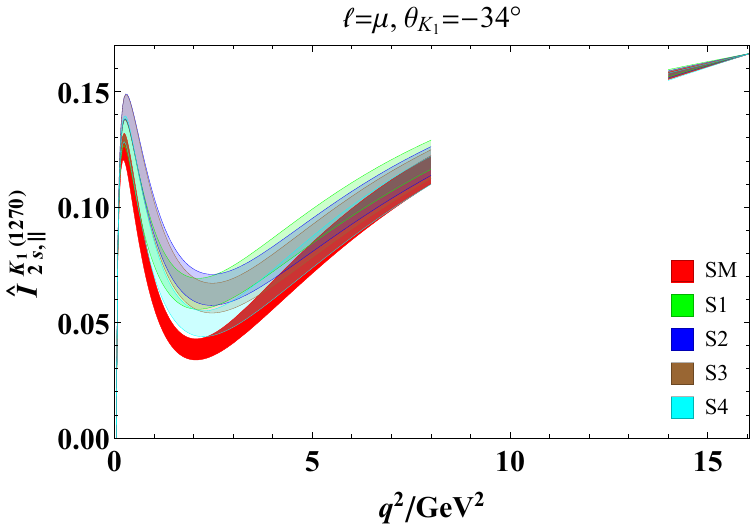}\\
\includegraphics[scale=0.405]{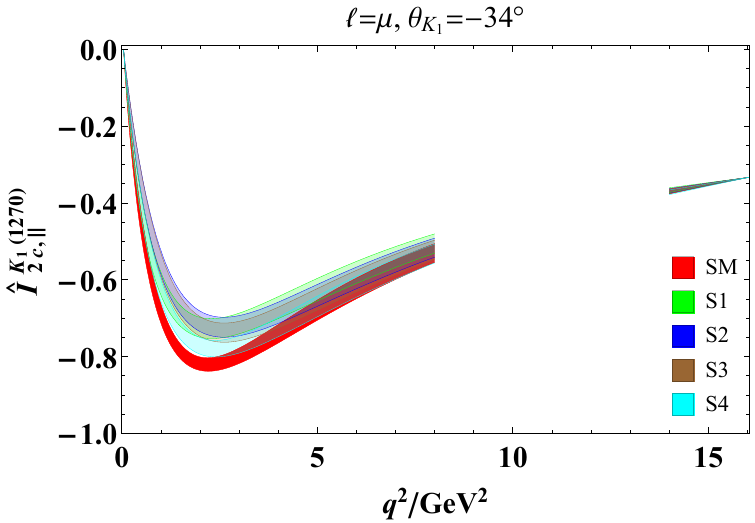}
&\includegraphics[scale=0.405]{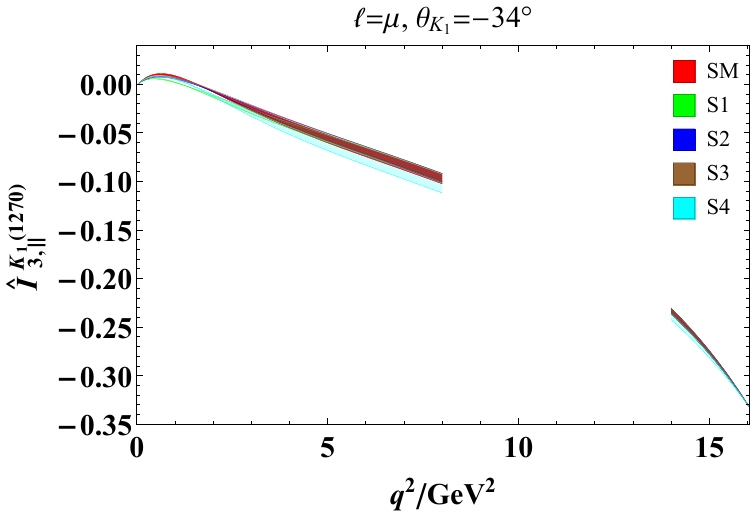}
&\includegraphics[scale=0.405]{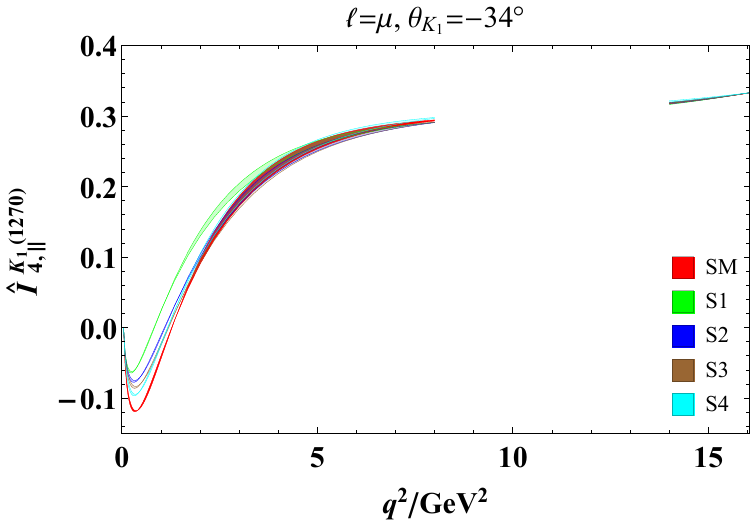}\\
\includegraphics[scale=0.405]{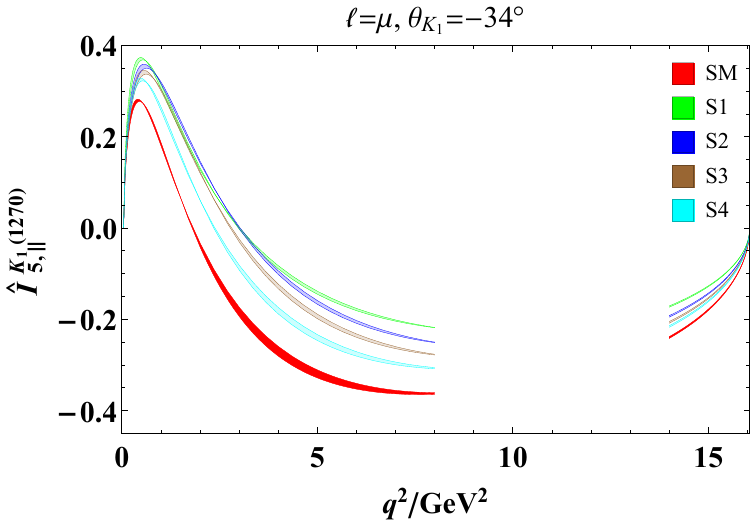}
&\includegraphics[scale=0.405]{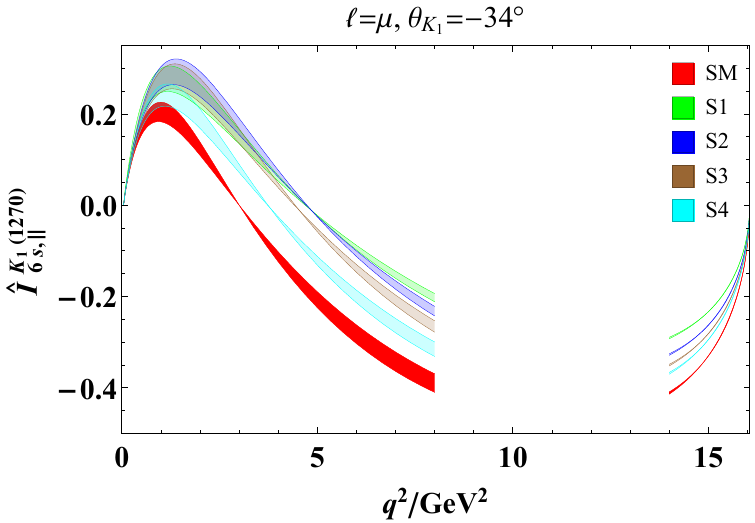}
\end{tabular}
\caption{The normalized angular observables corresponding to the longitudinally polarized vector meson in the decay $B\to K_1\left(1270\right)\left(\to V_{\|}P\right) \mu^+\mu^-$ in the SM and NP scenarios.}
\label{IsK1270}
\end{figure}
\begin{figure}[htbp!]
\begin{tabular}{ccc}
\includegraphics[scale=0.405]{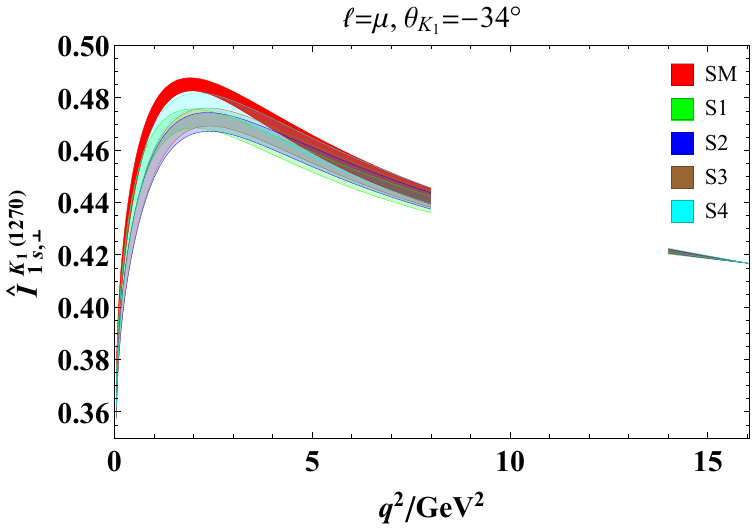}
&\includegraphics[scale=0.405]{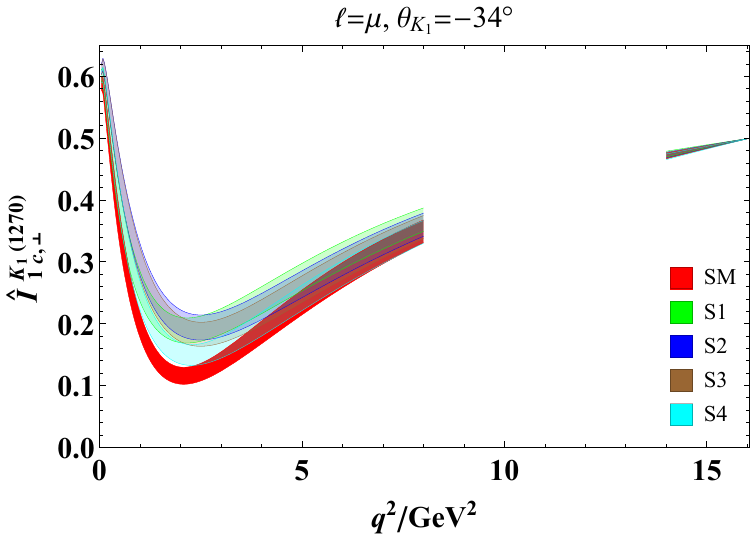}&\includegraphics[scale=0.405]{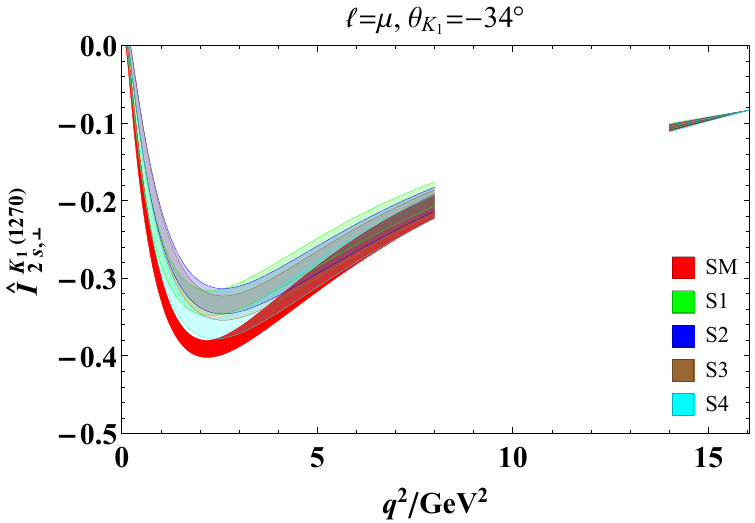}\\
\includegraphics[scale=0.405]{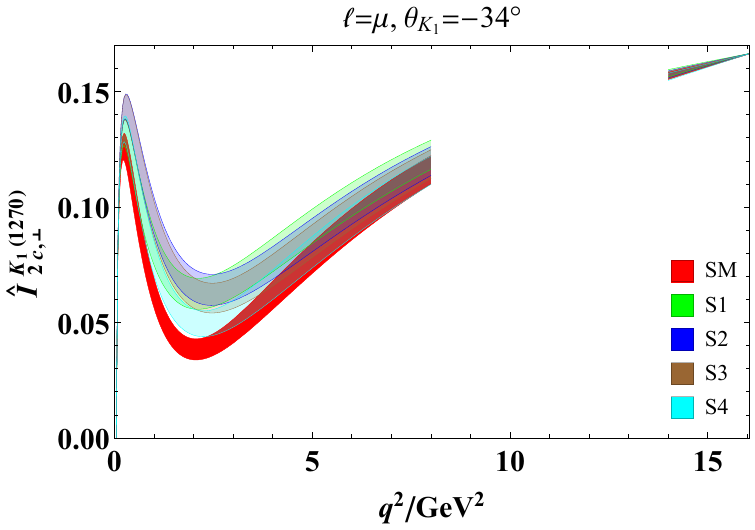}
&\includegraphics[scale=0.405]{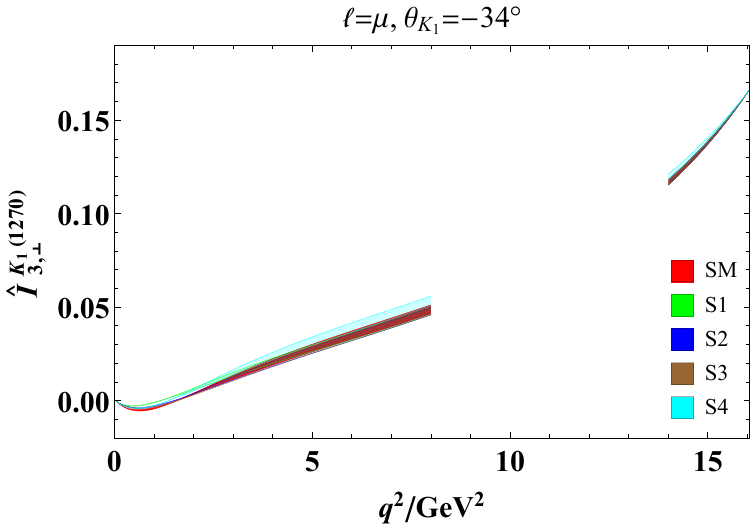}&\includegraphics[scale=0.405]{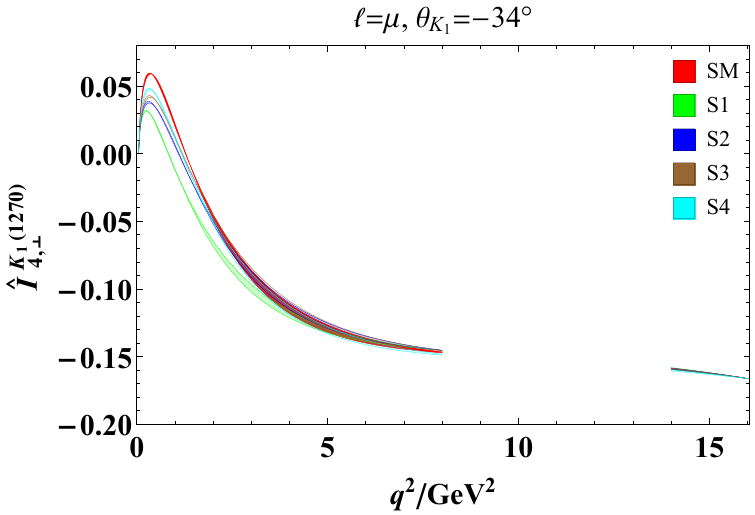}\\
\includegraphics[scale=0.405]{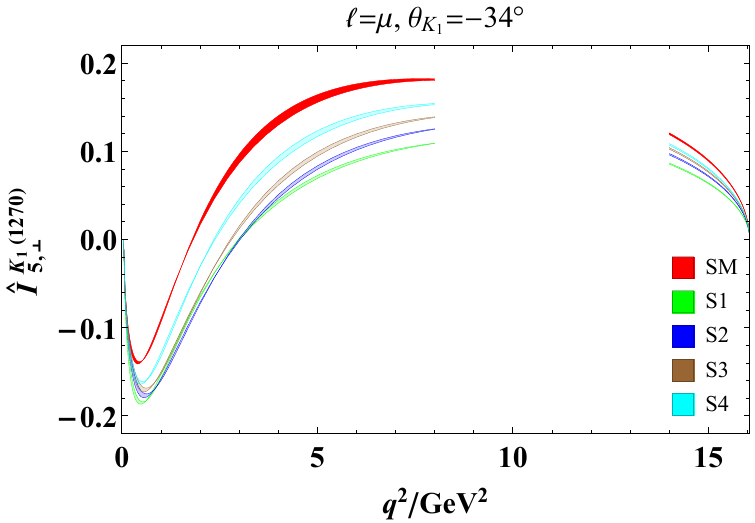}&\includegraphics[scale=0.405]{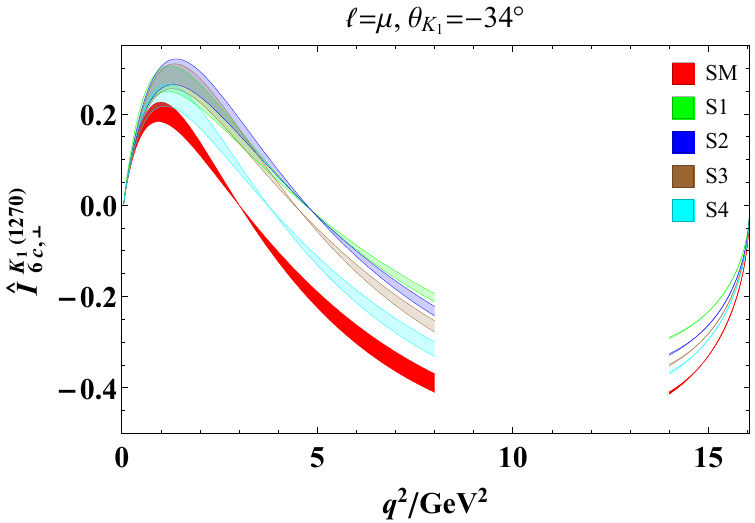}&\includegraphics[scale=0.405]{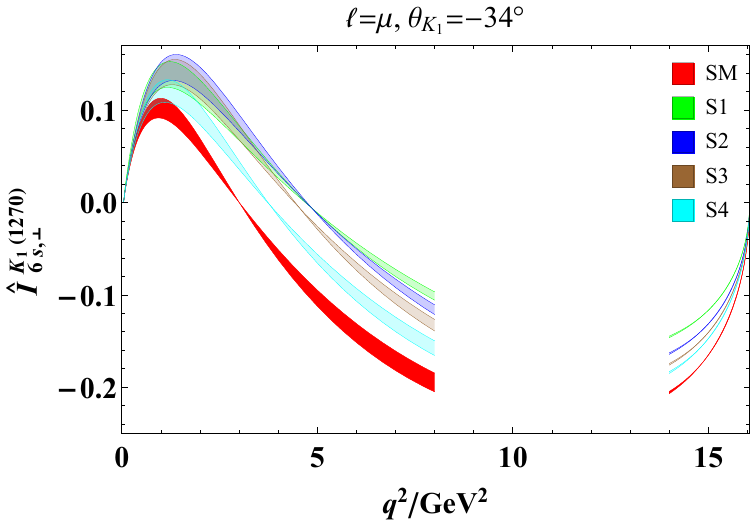}
\end{tabular}
\caption{The normalized angular observables corresponding to the transversely polarized vector meson in the decay $B\to K_1\left(1270\right)\left(\to V_{\perp}P\right) \mu^+\mu^-$ in the SM and NP scenarios.}
\label{I2sK1270}
\end{figure}

\item We now discuss the angular coefficients $\hat{I}_{1s,\parallel},\; \hat{I}_{2s,\parallel},\; \hat{I}_{1c,\parallel},\; \hat{I}_{2c,\parallel},\; \hat{I}_{3,\parallel},...,\hat{I}_{6,\parallel}$, where the subscript $\parallel$ depicts the longitudinal polarization of the vector meson in the cascade decay $K_1\left(1270,\; 1400\right)\to V_{\parallel}P$. The corresponding expressions of these angular coefficients are given in Eq. (\ref{Binned-Acoefficients}), where in the numerator we have factored out the branching ratio of the final state meson, $\mathcal{B}\left(K_1\to V_{\parallel}P\right)$. By looking at Eq. (\ref{drategamma}), we can see that the same factor appears in the denominator, making the values of these $I'$s to be the same whether we have $V=K^*$ or $\rho$ as a final state in the decay of $K_1\left(1270,\; 1400\right)$ meson. Figure \ref{IsK1270}, shows the trend of these angular observables for $K_1\left(1270\right)$ with $q^2$. We observe that, except $\hat{I}_{3,\parallel}$ and $\hat{I}_{4,\parallel}$, the NP effects are significantly different from the SM values, particularly in  $q^2\in [1.0,7.0]\;\text{GeV}^2$ window. Apart from these deviations from the SM predictions of the numerical values and the zero crossing, the angular coefficients $\hat{I}_{5, \parallel}$ and $\hat{I}_{6s, \parallel}$ can be used to discriminate the various NP scenarios, particularly, in $q^2\in [5.0,7.0]\;\text{GeV}^2$ range. We have also calculated the numerical values of $I$'s in different $q^2$ bins and listed in Tables \ref{Bin1-analysis} - \ref{Bin51400-analysis}. Here, it is evident that the uncertainties resulting from hadronic inputs are small in these angular coefficients; as a result, these observables also provide an intriguing means of examining the NP  structure in the $B\to K_1\left(1270,\; 1400\right)$ decays. It is worthwhile to mention here that the angular coefficients $\hat{I}_{7,\parallel},\; \hat{I}_{8, \parallel}$ and $\hat{I}_{9, \parallel}$ are proportional to the imaginary part of the helicity amplitudes; therefore, they remain quite small in both the SM and the selected NP scenarios. For this reason, we have not included their values in the Tables \ref{Bin1-analysis} - \ref{Bin51400-analysis}. 

Based on the numerical results presented in Tables \ref{Bin1-analysis}- \ref{Bin6-analysis}, the quantitative description of the angular observables is as follows:

\begin{itemize}
    \item For $\hat{I}_{1s,\parallel}$ in $[0.1 - 2.0]$ GeV$^2$ bin, the NP contributes constructively with the SM, leading to an increase in its value. The maximum increase of $16\%$ from its SM value comes from the S2 and S3 scenarios, and this difference is increased by $50\%$ in $[2.0 - 4.0]$ GeV$^2$ region. However, in third and fourth bin, \textit{i.e.,} $[4.0 - 6.0]$ $\text{GeV}^2$ and $[6.0 - 8.0]$ GeV$^2$, the SM result is raised by $10\%$ in S1. In the large $q^2$ bins, there is no significant deviation from the SM observed for any NP scenario.
  
    \item  In the case of $\hat{I}_{2c,\parallel}$, the value decreases relative to the SM results in all the scenarios. The maximum decrease of $20\%$ in $[0.1 - 2.0]$ GeV$^2$ is observed for S2 and S3. In the remaining bins when $q^2<8$ GeV$^2$, these deviations are less than $10\%$. Therefore, measuring $\hat{I}_{2c,\parallel}$ in low $q^2$ bin can help to segregate the various NP scenarios. 

    \item The numerical values of $\hat{I}_{2s,\parallel}$ are nearly one-third of those of $\hat{I}_{1s,\parallel}$ in all the $q^2$ bins, and these values are $\mathcal{O}\left(10^{-2}\right)$ in bins with $2.0\leq q^2 \leq 6.0$ GeV$^2$. In other low and intermediate $q^2$ bins, the values are an order of magnitude higher, but the NP effects are not appreciable. 

    \item The angular observable $\hat{I}_{2c,\parallel}$ remains negative in whole $q^2$ range, with a maximum SM value $\left(-0.789\right)$ in $[2.0 - 4.0]$ GeV$^2$ bin. The NP arising from first three scenarios reduce its magnitude by $10\%$ in $0.1\leq q^2\leq 6.0$ GeV$^2$ bins. The scenarios S1, S2, and S3 nearly overlap, while S4 is indistinguishable from the SM.
    
    \item $\hat{I}_{3,\parallel}$ is suppressed both in the SM and the considered NP scenarios for almost all the bins except in $14 \leq q^2 \leq 16$ bin. However, the NP does not enhance enough the SM results in this region, to make it useful for the experimental searches in this region. 
    
    \item In the case of angular observable $\hat{I}_{4,\parallel}$, the SM values in $q^2 \in [4.0 - 6.0]$ GeV$^2$ and $[6.0 - 8.0]$ GeV$^2$ are almost the same and are in the measurable range. The same is the case with $\hat{I}_{5,\parallel}$. For $\hat{I}_{6s,\parallel}$ case, the S1 is distinguishable from the SM and other scenarios, in $[4.0 - 6.0]$ GeV$^2$, where the magnitude of the SM value is reduced by $40\%$. Although the change in the SM results for $\hat{I}_{6s,\parallel}$ is maximum in $[4.0 - 6.0]$ GeV$^2$ bin in S1, S2, and S3, this increment is still small by an order of magnitude in comparison of other angular observables, making it difficult to measure. However, a visible and measurable change in this observable lies in $[6.0 - 8.0]$ GeV$^2$ bin, where similar to the $\hat{I}_{5,\parallel}$, the deviation of $50\%$ is marked for S1.  The other important aspect of these angular coefficients is their zero-crossing that is shifted differently in S4 and other scenarios from their corresponding SM results. Moreover, this observable is almost free from uncertainties, and its value in S4 is distinguishable from SM and other three scenarios.  Therefore, these two angular coefficients can also be used to disentangle the various NP scenarios.
\end{itemize}

\item Fig. \ref{IsK1400} shows the same angular observables (discussed above), but for $B\to K_1\left(1400\right)\left(\to V_\parallel P\right)\mu^{+}\mu^{-}$ against the $q^2$ in the SM and in the NP scenarios S1 to S4. From this figure, we observe that the trend of the NP is similar to that for the decay with $K_1\left(1270\right)$ meson ; however, the numerical values in various low to intermediate $q^2$ bins are different. These numerical values are calculated and presented in Tables \ref{Bin11400-analysis} - \ref{Bin51400-analysis}.

\begin{figure}[b!]
\begin{tabular}{ccc}
\includegraphics[scale=0.425]{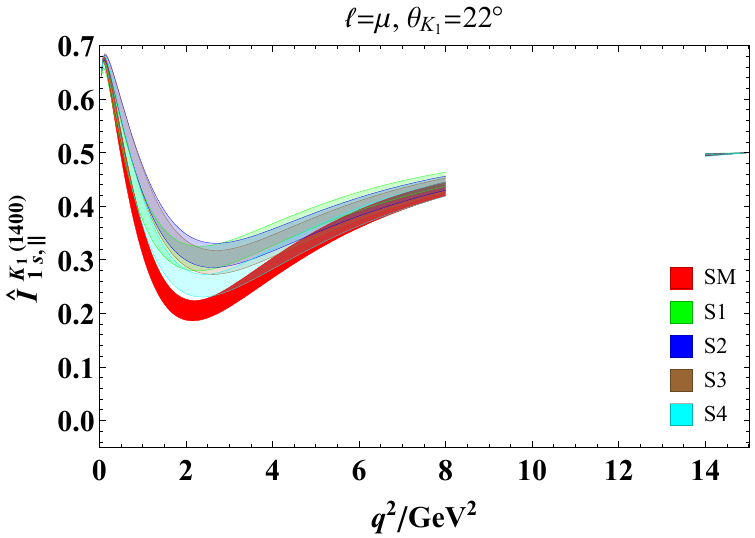} 
&\includegraphics[scale=0.425]{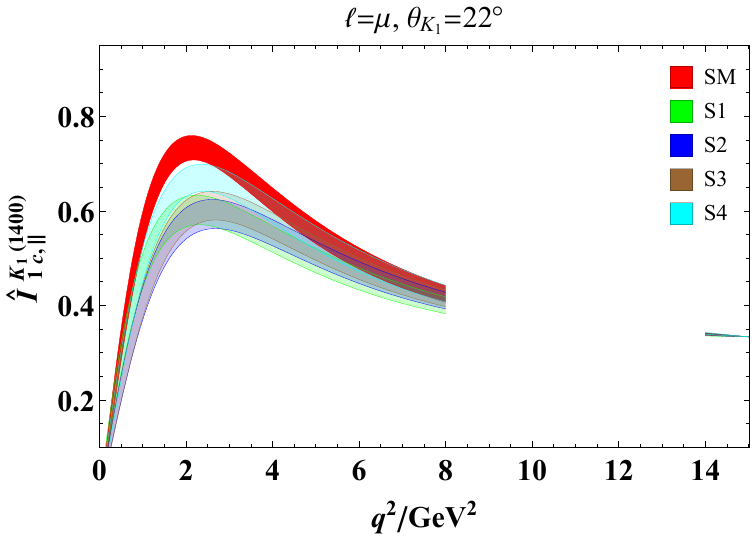}&\includegraphics[scale=0.425]{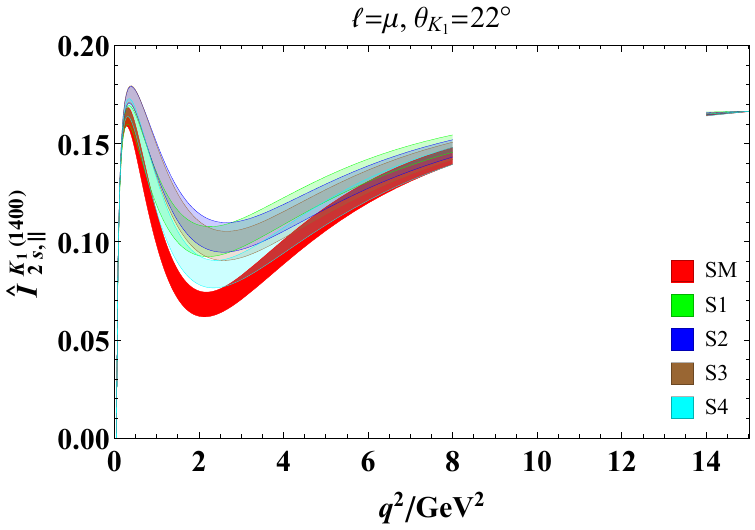}\\
\includegraphics[scale=0.425]{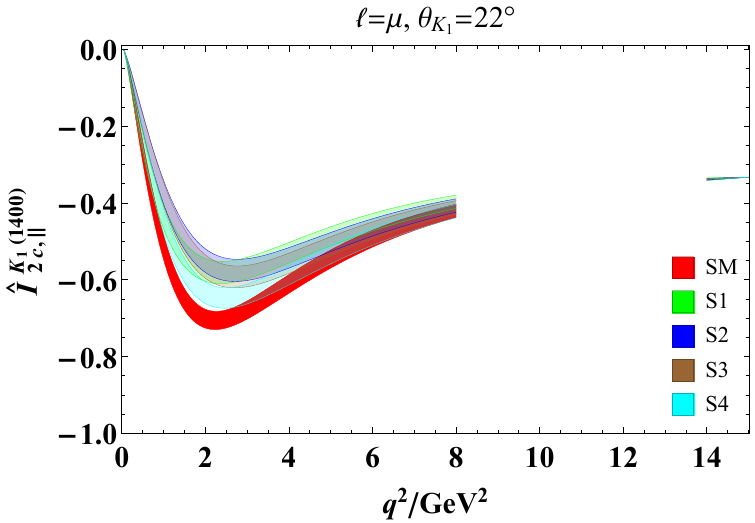}
&\includegraphics[scale=0.425]{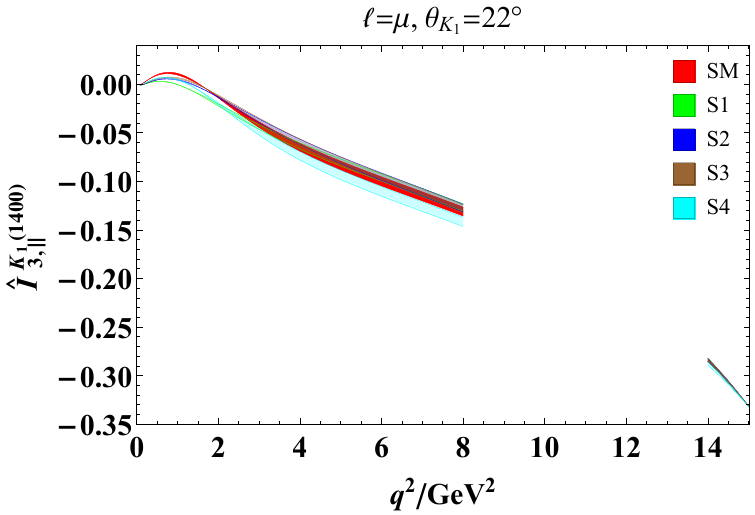}
&\includegraphics[scale=0.425]{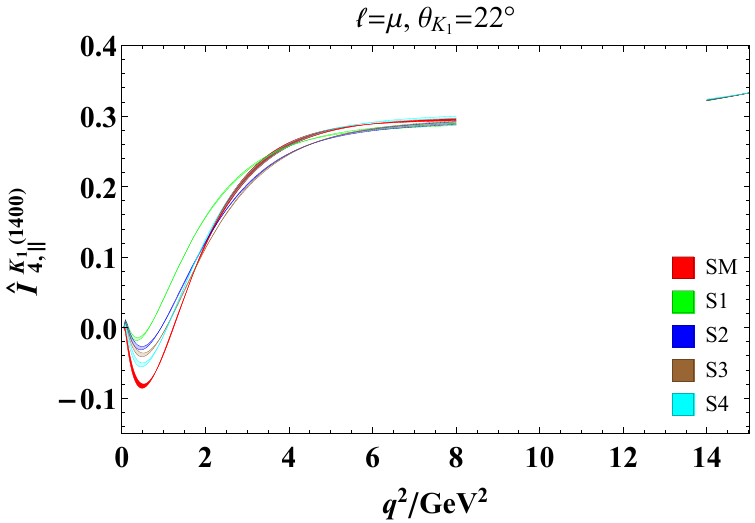}\\
\includegraphics[scale=0.425]{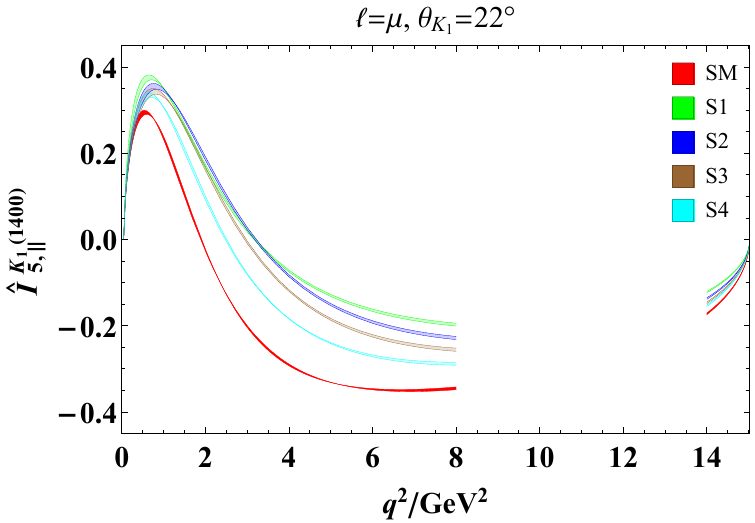}
&\includegraphics[scale=0.425]{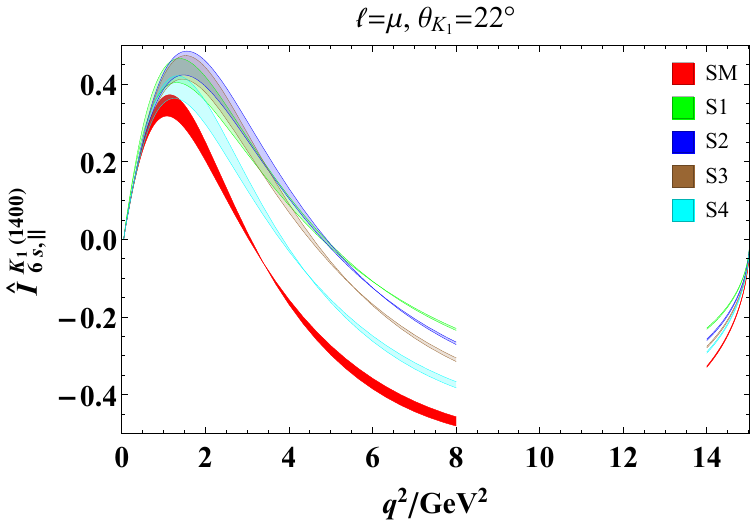}
\end{tabular}
\caption{The normalized angular observables corresponding to the longitudinally polarized vector meson in the decay $B\to K_1\left(1400\right)\left(\to V_{\|}P\right) \mu^+\mu^-$ in the SM and NP scenarios.}
\label{IsK1400}
\end{figure}

\begin{figure}[htbp!]
\begin{tabular}{ccc}
\includegraphics[scale=0.425]{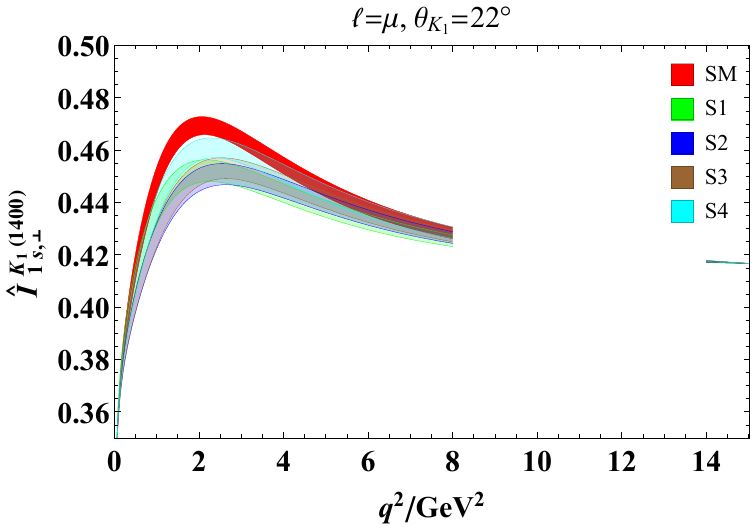}
&\includegraphics[scale=0.425]{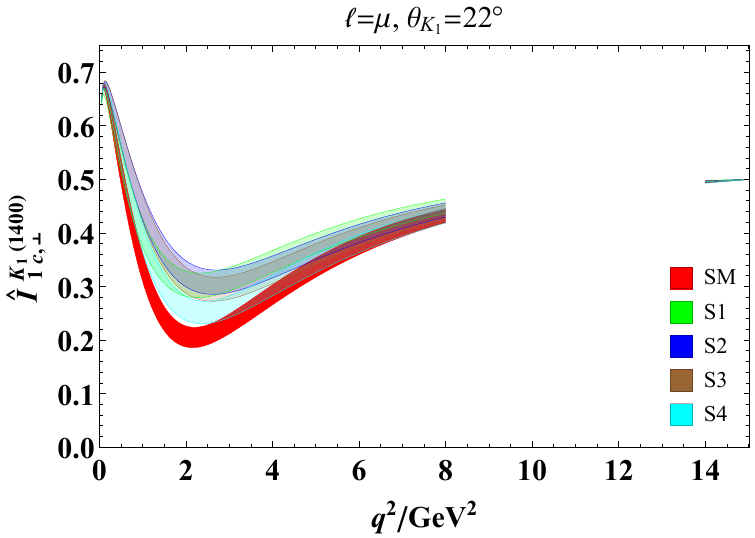}&\includegraphics[scale=0.425]{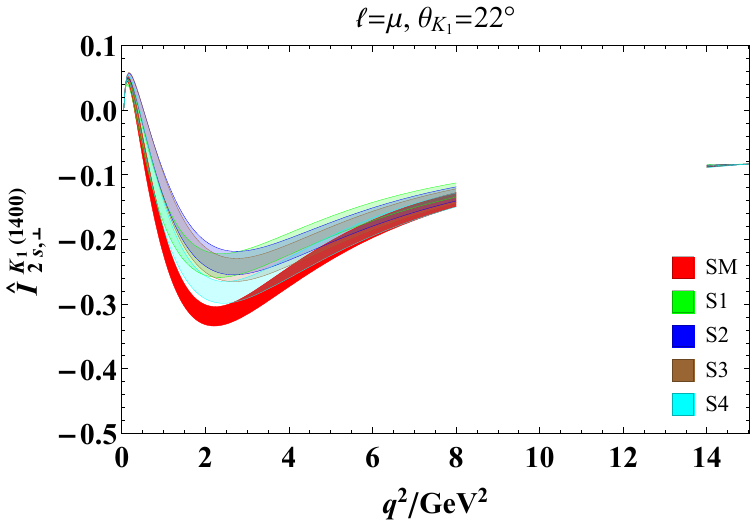}\\
\includegraphics[scale=0.425]{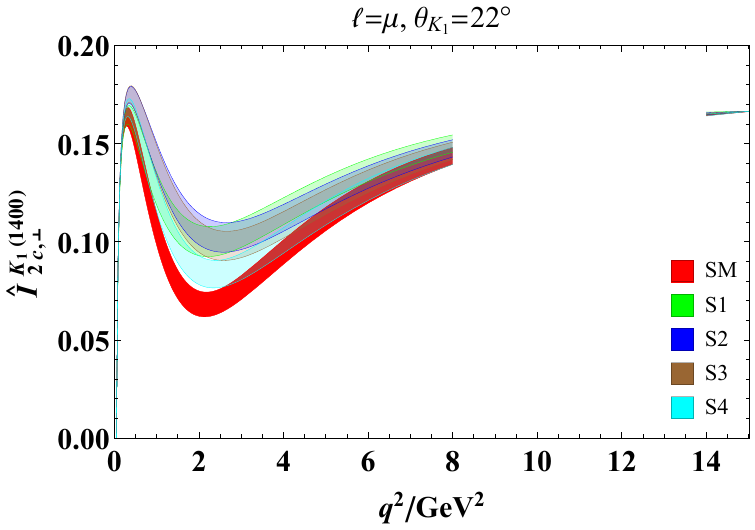}
&\includegraphics[scale=0.425]{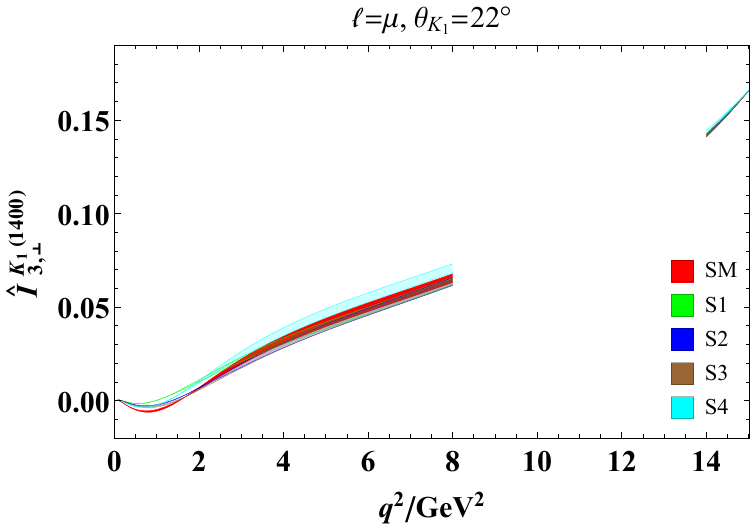}&\includegraphics[scale=0.425]{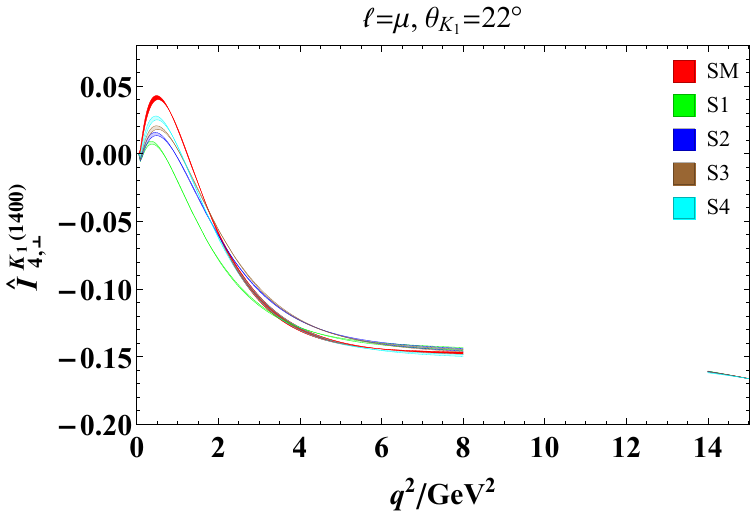}\\
\includegraphics[scale=0.425]{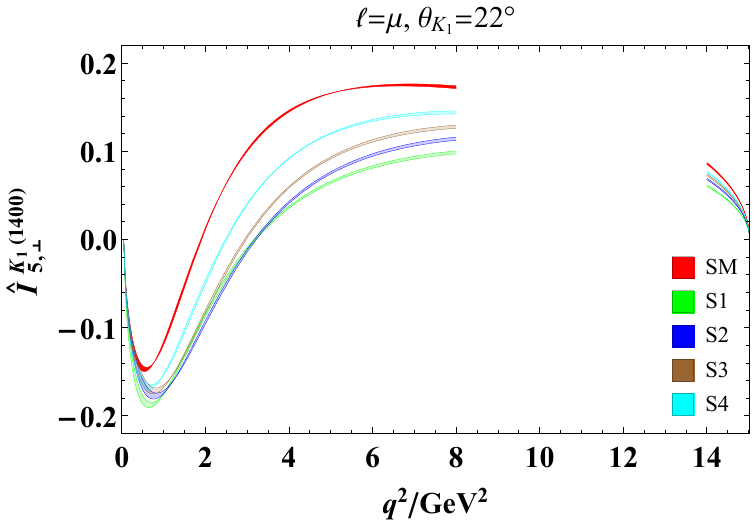}&\includegraphics[scale=0.425]{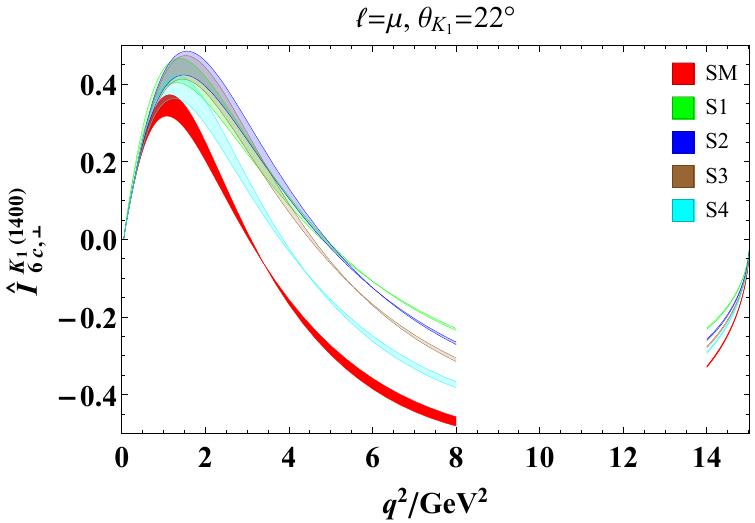}&\includegraphics[scale=0.425]{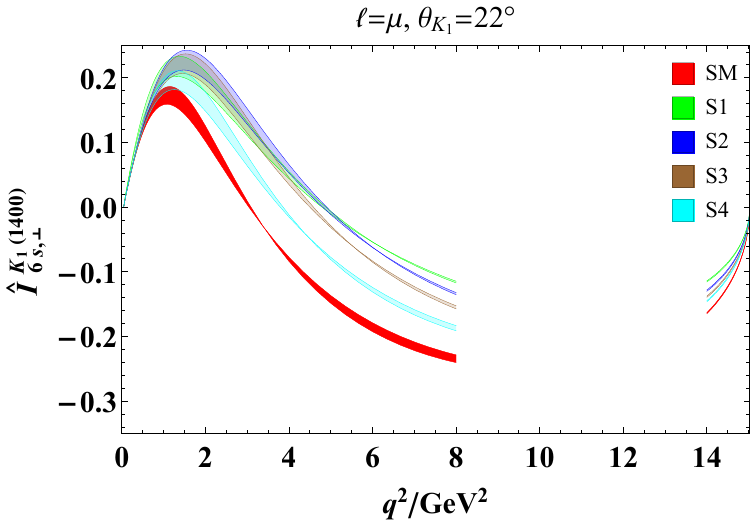}
\end{tabular}
\caption{The normalized angular observables corresponding to the transversely polarized vector meson in the decay $B\to K_1\left(1400\right)\left(\to V_{\perp}P\right) \mu^+\mu^-$ in the SM and NP scenarios.}
\label{I2sK1400}
\end{figure}

\item Similarly, Figs. \ref{I2sK1270} and \ref{I2sK1400}, show the angular coefficients corresponding to the perpendicularly polarized vector meson in $B\to K_1\left(1270,\; 1400\right)\left(\to V_\perp P\right) \mu^+\mu^-$, in the SM and and in the NP scenarios against $q^2$. Except for $\hat{I}_{1s,\perp}$ and $\hat{I}_{5,\perp}$, the values of other angular coefficients increase from their corresponding SM predictions (c.f. Tables \ref{Bin1-analysis} - \ref{Bin51400-analysis}), especially, in the low to intermediate $q^2$ regions. In the presence of NP, the most interesting aspect is the shift of the zero crossing of $\hat{I}_{5,\perp}$, $\hat{I}_{6c,\perp}$ and $I_{6s,\perp}$ from lower to high $q^2$ values. Additionally, these figures indicate that the uncertainties in these angular coefficients  are small, as quantitatively shown in Tables  \ref{Bin1-analysis} - \ref{Bin6-analysis}. In the $q^2\in [0.1-6.0]$ GeV$^2$ range, the presence of NP leads to changes in the values of $I_{\left(1s,\; 1c,\; 2s,\; 2c\right),\perp}$ from their SM values by $(5-15)\%$. Although the values of the angular coefficients $\hat{I}_{5,\perp}$, $\hat{I}_{6c,\perp}$ and $I_{6s,\perp}$ are smaller compared to the other  angular coefficients, their zero crossing has a discriminatory power for the various NP scenarios, particularly for the NP arising due to the WCs of S4. Furthermore, Figs. \ref{I2sK1270} and \ref{I2sK1400} show that the uncertainty in the zero-crossings in these angular coefficients is negligible, making them a good probe for the NP searches. 
 \end{itemize}

\section{Conclusion}\label{conclusions}
We have thoroughly worked out the four-fold decay distribution of the $B \to K_{1}(1270,1400)(\to V P) \ell^{+}\ell^{-}$ decays, where $K_1 $ is an axial-vector meson, $V=\rho, K^{\ast}$ and $P=K, \pi$ mesons, both within the SM and by extending the weak effective Hamiltonian to new vector and axial-vector Wilson Coefficients. We have used the helicity formalism to calculate the four-fold decay distribution of this process for both the longitudinal and transverse polarization of the vector meson $V$, originating from the cascade decay $K_{1}\to VP$. Our formalism retains the lepton mass term, making it valid for all the leptons; however, we have performed the numerical analysis specifically on the muon case. The axial-vector physical states $K_1(1270)$ and $K_1(1400)$ are mixture of the $^{3}P_{1}$ and $^{1}P_{1}$ states, denoted as $K_{1A}$ and $K_{1B}$, respectively; therefore, the form factors given in Table \ref{tabelFFs} depend on the mixing angle $\left(\theta_{K_1}\right)$. Due to lack of consensus on the value of mixing angle, we have adopted the phenomenologically acceptable values of $-34^{\circ}$ and $22^\circ$ for $K_1\left(1270\right)$ and $K_{1}\left(1400\right)$, respectively, aiming to maximize the value of the branching ratio of $B\to K_1\left(1270,1400\right)\mu^{+}\mu^{-}$ decay. Using the form factors computed in the LCSR approach, and the other input parameters, we have presented the SM and NP predictions for physical observables, such as the differential branching ratios $d\mathcal{B}/{dq^2}$, the lepton forward-backward asymmetry $\mathcal{A}_{\text{FB}}^{K_1}$, the $\mathcal{A}_{\text{FB}}^{K_{1T}}$ asymmetry for transversely polarized $K_1$ meson, the longitudinal and transverse polarization fractions of $K_1$ meson, $f_{L}^{K_1}$ and $f_{T}^{K_1}$, respectively, and various normalized angular coefficients $\hat{I}_{\parallel(\perp)}^{K_1}$, for longitudinal and transverse polarization of vector meson, all in the physical region as functions of the momentum transfer squared, $q^2$. Our findings, using the latest global fit analysis data of all $b\to s$ transitions, show that these physical observables are sensitive to the NP and distinguish between different NP scenarios in some kinematical regions. Therefore, the precise measurements of these observables in $B \to K_{1}(1270,1400)(\to V P) \mu^{+}\mu^{-}$ decays at current and future experiments provide an opportunity for complementary searches for physics beyond the SM in $b\to s\ell^+\ell^-$ decays.

\clearpage
\appendix 
\section{The Expressions of Wilson coefficients in the SM}\label{append}
The explicit expressions used for the Wilson coefficients are given as follows \cite{Bobeth:1999mk,Beneke:2001at,Asatrian:2001de,Asatryan:2001zw,Greub:2008cy,Du:2015tda},
\begin{eqnarray}
C_{7}^{\text{eff}}(q^2)&=&C_{7}-\frac{1}{3}\left(C_{3}+\frac{4}{3}C_{4}+20C_{5}+\frac{80}{3}C_{6}\right)
-\frac{\alpha_{s}}{4\pi}\left[(C_{1}-6C_{2})F^{(7)}_{1,c}(q^2)+C_{8}F^{(7)}_{8}(q^2)\right],\notag\\
C_{9}^{\text{eff}}(q^2)&=&C_{9}+\frac{4}{3}\left(C_{3}+\frac{16}{3}C_{5}+\frac{16}{9}C_{6}\right)
-h(0, q^2)\left(\frac{1}{2}C_{3}+\frac{2}{3}C_{4}+8C_{5}+\frac{32}{3}C_{6}\right)\notag\\
&-&h(m_{b}^{\text{pole}}, q^2)\big(\frac{7}{2}C_{3}+\frac{2}{3}C_{4}+38C_{5}+\frac{32}{3}C_{6}\big)+h(m_{c}^{\text{pole}}, q^2)
\big(\frac{4}{3}C_{1}+C_{2}+6C_{3}+60C_{5}\big)\notag\\
&-&\frac{\alpha_{s}}{4\pi}\left[C_{1}F^{(9)}_{1,c}(q^2)+C_{2}F^{(9)}_{2,c}(q^2)+C_{8}F^{(9)}_{8}(q^2)\right],\label{WC3}
\end{eqnarray}where the functions $h(m_{q}^{\text{pole}}, q^2)$ with $q=c, b$ and $F^{(7,9)}_{8}(q^2)$ are
defined in \cite{Beneke:2001at}, while $F^{(7,9)}_{1,c}(q^2)$ and $F^{(7,9)}_{2,c}(q^2)$ are
given for low $q^{2}$ in \cite{Asatryan:2001zw} and for high $q^{2}$ in \cite{Greub:2008cy}. The quark masses in all of these functions are defined in the pole scheme.

\section{Hadronic matrix elements}\label{appendHME}
The matrix elements can be parameterized in terms of transition form factors as follows
\begin{eqnarray}
\left\langle K_1(k,\epsilon_{K_1})\left|\bar{s}\gamma_{\mu}b\right|B(p)\right\rangle&=&-\left(m_{B}+m_{K_1}\right)g_{\mu\nu}\epsilon_{K_1}^{\,\ast\nu}V_{1}^{K_1}(q^{2})
+P_{\mu}(\epsilon_{K_1}^{\,\ast}\cdot q)\frac{V_{2}^{K_1}(q^{2})}{\left(m_{B}+m_{K_1}\right)}\notag\\
&+&\frac{2m_{K_1}}{q^{2}}q_{\mu}(\epsilon_{K_1}^{\,\ast}\cdot q)[V_{3}^{K_1}(q^{2})-V_{0}^{K_1}(q^{2})],\label{F1}\\
\left\langle K_1(k,\epsilon_{K_1})\left|\bar{s}\gamma_{\mu}\gamma_5 b\right|B(p)\right\rangle&=& \frac{2i\epsilon_{\mu\nu\alpha\beta}}{m_{B}+m_{K_1}}
\epsilon_{K_1}^{\,\ast\nu}p^{\alpha}k^{\beta}A^{K_1}(q^{2}),\label{F2}
\end{eqnarray}
where,
\begin{eqnarray}
V_{3}^{K_1}(q^{2})&=&\frac{m_{B}+m_{K_1}}{2m_{K_1}}V_{1}^{K_1}(q^{2})-\frac{m_{B}-m_{K_1}}{2m_{K_1}}V_{2}^{K_1}(q^{2}),\label{V3}\\
V_{3}^{K_1}(0)&=&V_{0}^{K_1}(0).\notag
\end{eqnarray}
In above equations, $m_B$ and $m_{K_1}$ are the masses of $B$ and $K_1$ mesons, respectively, and $q^2=(p_B - p_{K_1})^2 = (p_{\ell^+} + p_{\ell^-})^2$ is the square of momentum transfer, where $p_{\ell^-}$ and $p_{\ell^+}$ denote the momenta of the final state lepton and anti-lepton, respectively. The other contributions from the tensor form factors are
\begin{eqnarray}
\left\langle K_1(k,\epsilon_{K_1})\left|\bar si\sigma_{\mu\nu}q^{\nu}b\right|B(p)\right\rangle&=&\Big[\Big(m_{B}^{2}-m_{K_1}^{2}\Big)g_{\mu\nu}\epsilon_{K_1}^{\,\ast\nu}
-(\epsilon_{K_1}^{\,\ast}\cdot q)P_{\mu}\Big]T_{2}^{K_1}(q^{2})\notag\\
&+&(\epsilon_{K_1}^{\,\ast}\cdot q)\left[q_{\mu}-\frac{q^{2}}{m^2_{B}-m^2_{K_1}}P_{\mu}\right]T_{3}^{K_1}(q^{2}),\label{T1}\\
\left\langle K_1(k,\epsilon_{K_1})\left|\bar si\sigma_{\mu\nu}q^{\nu}\gamma_{5}b\right|B(p)\right\rangle&=&2i\epsilon_{\mu\nu\alpha\beta}\epsilon_{K_1}^{\,\ast\nu}p^{\alpha}k^{\beta}T_{1}^{K_1}(q^{2}),\label{T2}
\end{eqnarray}
where we have adopted $\epsilon_{0123}=1$ convention for the Levi-Civita tensor throughout this study.

\section{Leptonic tensors}\label{leptcal}
The leptonic tensors in Eq. (\ref{Amp5ag}) can be defined as
\begin{eqnarray}
L_{VV,mn}&=&\sum_{\lambda_{l^+},\lambda_{l^-}}L_{V,m}^{{\lambda_{l^+},\lambda_{l^-}}}L_{V,n}^{\ast{\lambda_{l^+},\lambda_{l^-}}}\notag\\
&=&\varepsilon^{\mu^{\prime}}(m)\varepsilon^{\ast\nu^{\prime}}(n) \,\text{tr}\left[\gamma_{\mu^{\prime}}\left(\slashed{p}_{l^{+}}-m_l\right)\gamma_{\nu^{\prime}}\left(\slashed{p}_{l^{-}}+m_l\right)\right]\notag\\
&=& 4\,\varepsilon^{\mu^{\prime}}(m)\varepsilon^{\ast\nu^{\prime}}(n)\left(p_{l^{+}\mu^{\prime}}p_{l^{-}\nu^{\prime}}+p_{l^{+}\nu^{\prime}}p_{l^{-}\mu^{\prime}}-g_{\mu^{\prime}\nu^{\prime}}\left(m_l^2+p_{l^{+}}\cdot p_{l^{-}}\right)\right).\label{Amp8ag}
\end{eqnarray}

\begin{eqnarray}
L_{AA,mn}&=&\sum_{\lambda_{l^+},\lambda_{l^-}}L_{A,m}^{{\lambda_{l^+},\lambda_{l^-}}}L_{A,n}^{\ast{\lambda_{l^+},\lambda_{l^-}}}\notag\\
&=&\varepsilon^{\mu^{\prime}}(m)\varepsilon^{\ast\nu^{\prime}}(n) \,\text{tr}\left[\gamma_{\mu^{\prime}}\gamma_5\left(\slashed{p}_{l^{+}}-m_l\right)\gamma_{\nu^{\prime}}\gamma_5\left(\slashed{p}_{l^{-}}+m_l\right)\right]\notag\\
&=& 4\,\varepsilon^{\mu^{\prime}}(m)\varepsilon^{\ast\nu^{\prime}}(n)\left(p_{l^{+}\mu^{\prime}}p_{l^{-}\nu^{\prime}}+p_{l^{+}\nu^{\prime}}p_{l^{-}\mu^{\prime}}+g_{\mu^{\prime}\nu^{\prime}}\left(m_l^2-p_{l^{+}}\cdot p_{l^{-}}\right)\right).\label{Amp9ag}
\end{eqnarray}

\begin{eqnarray}
L_{VA,mn}&=&\sum_{\lambda_{l^+},\lambda_{l^-}}L_{V,m}^{{\lambda_{l^+},\lambda_{l^-}}}L_{A,n}^{\ast{\lambda_{l^+},\lambda_{l^-}}}=\sum_{\lambda_{l^+},\lambda_{l^-}}L_{A,m}^{{\lambda_{l^+},\lambda_{l^-}}}L_{V,n}^{\ast{\lambda_{l^+},\lambda_{l^-}}}\notag\\
&=&\varepsilon^{\mu^{\prime}}(m)\varepsilon^{\ast\nu^{\prime}}(n) \,\text{tr}\left[\gamma_{\mu^{\prime}}\left(\slashed{p}_{l^{+}}-m_l\right)\gamma_{\nu^{\prime}}\gamma_5\left(\slashed{p}_{l^{-}}+m_l\right)\right]\notag\\
&=& 4 i\,\varepsilon^{\mu^{\prime}}(m)\varepsilon^{\ast\nu^{\prime}}(n)\,\epsilon_{\mu^{\prime}\nu^{\prime}\alpha\beta}\,p_{l^{+}}^{\alpha}p_{l^{-}}^{\beta}.\label{Amp10ag}
\end{eqnarray}
It is convenient to evaluate the lepton tensors in $\ell\bar\ell$ center-of-mass (CM) frame. Following polarization vector conventions corresponding to $j_{\text{eff}}^\mu$ and momenta of leptons as given in \cite{Faessler:2002ut}, 
(c.f. Fig. \ref{figfourfold}), we obtain the expressions of the following leptonic tensors, where the matrix columns and rows are ordered in the sequence $(t, +, -, 0)$:
\begin{eqnarray}
L_{VV,mn}&=&q^2\begin{pmatrix}
0 & 0 & 0 & 0\\
0 & \frac{2+\beta_{\ell}^2}{2}+\frac{4m_{\ell}^2}{q^2}+\frac{\beta_{\ell}^2}{2}\cos{2\theta_{\ell}} & \beta_{\ell}^2\sin^2{\theta_{\ell}}\,e^{2i\phi} & \frac{\beta_{\ell}^2}{\sqrt{2}}\sin{2\theta_{\ell}}\,e^{i\phi}\\
0 & \beta_{\ell}^2\sin^2{\theta_{\ell}}\,e^{-2i\phi} & \frac{2+\beta_{\ell}^2}{2}+\frac{4m_{\ell}^2}{q^2}+\frac{\beta_{\ell}^2}{2}\cos{2\theta_{\ell}} 
& -\frac{\beta_{\ell}^2}{\sqrt{2}}\sin{2\theta_{\ell}}\,e^{-i\phi}\\
0  & \frac{\beta_{\ell}^2}{\sqrt{2}}\sin{2\theta_{\ell}}\,e^{-i\phi}   
& -\frac{\beta_{\ell}^2}{\sqrt{2}}\sin{2\theta_{\ell}}\,e^{i\phi}  & 1+\frac{4m_{\ell}^2}{q^2}-\beta_{\ell}^2\cos{2\theta_{\ell}}
\end{pmatrix},\label{LepTVVcomb}\\
L_{AA,mn}&=&q^2\begin{pmatrix}
\frac{8m_{\ell}^2}{q^2} & 0 & 0 & 0\\
0 & \frac{2+\beta_{\ell}^2}{2}-\frac{4m_{\ell}^2}{q^2}+\frac{\beta_{\ell}^2}{2}\cos{2\theta_{\ell}} & \beta_{\ell}^2\sin^2{\theta_{\ell}}\,e^{2i\phi} & \frac{\beta_{\ell}^2}{\sqrt{2}}\sin{2\theta_{\ell}}\,e^{i\phi}\\
0 & \beta_{\ell}^2\sin^2{\theta_{\ell}}\,e^{-2i\phi} & \frac{2+\beta_{\ell}^2}{2}-\frac{4m_{\ell}^2}{q^2}+\frac{\beta_{\ell}^2}{2}\cos{2\theta_{\ell}} 
& -\frac{\beta_{\ell}^2}{\sqrt{2}}\sin{2\theta_{\ell}}\,e^{-i\phi}\\
0  & \frac{\beta_{\ell}^2}{\sqrt{2}}\sin{2\theta_{\ell}}\,e^{-i\phi}   
& -\frac{\beta_{\ell}^2}{\sqrt{2}}\sin{2\theta_{\ell}}\,e^{i\phi}  
& 1-\frac{4m_{\ell}^2}{q^2}-\beta_{\ell}^2\cos{2\theta_{\ell}}
\end{pmatrix},\label{LepTAAcomb}\\
L_{VA,mn}&=&q^2\begin{pmatrix}
0 & 0 & 0 & 0\\
0 & -2\beta_{\ell}\cos{\theta_{\ell}} & 0 
& -\sqrt{2}\beta_{\ell}\sin{\theta_{\ell}}\,e^{i\phi}\\
0 & 0 & 2\beta_{\ell}\cos{\theta_{\ell}}
& -\sqrt{2}\beta_{\ell}\sin{\theta_{\ell}}\,e^{-i\phi}\\
0  & -\sqrt{2}\beta_{\ell}\sin{\theta_{\ell}}\,e^{-i\phi}   
& -\sqrt{2}\beta_{\ell}\sin{\theta_{\ell}}\,e^{i\phi}  & 0
\end{pmatrix}.\label{LepTVAcomb}
\end{eqnarray}

\section{\boldmath $K_1 \to V P$ decay}\label{Casdecay2}
The matrix element of $K_1(k)\to V (p_3) P (p_4)$ decay can be parameterized as \cite{Roca:2003uk}
\begin{equation}
\mathcal{A}_{1}\left(\epsilon_{K_1}(r),\,\epsilon^{\ast}_{V}(l)\right) = \frac{2\lambda_{K_1 VP}}{m_{K_1}m_V} \left[(k\cdot p_{3})(\epsilon^{\ast}_{V}(l)\cdot \epsilon_{K_1}(r))-(k\cdot \epsilon^{\ast}_{V}(l))(p_3\cdot \epsilon_{K_1}(r))\right],   
\end{equation}
where $\lambda_{K_1 V P}$ is the coupling, $k, \epsilon_{K_1}$, and $p_3, \epsilon_{V}(l)$ are the momenta and polarizations of $K_{1}$ and $V$ mesons, respectively. The corresponding square of amplitude for $K_1(k)\to V (p_3) P (p_4)$ can be written as
\begin{eqnarray}
\left|\mathcal{A}_1\right|^2 & = & \sum_{l,r=\pm,0}\frac{4\lambda^2_{K_1 VP}}{m_{K_1}^2m_V^2}\bigg[
(k\cdot p_{3})(\epsilon^{\ast}_{V}(l)\cdot \epsilon_{K_1}(r))(k\cdot p_{3})(\epsilon_{V}(l)\cdot \epsilon^{\ast}_{K_1}(r))\notag\\
&&-(k\cdot p_{3})(\epsilon^{\ast}_{V}(l)\cdot \epsilon_{K_1}(r))(k\cdot \epsilon_{V}(l))(p_3\cdot \epsilon^{\ast}_{K_1}(r))\notag\\
&&-(k\cdot \epsilon^{\ast}_{V}(l))(p_3\cdot \epsilon_{K_1}(r))(k\cdot p_{3})(\epsilon_{V}(l)\cdot \epsilon^{\ast}_{K_1}(r))\notag\\
&&+(k\cdot \epsilon^{\ast}_{V}(l))(p_3\cdot \epsilon_{K_1}(r))(k\cdot \epsilon_{V}(l))(p_3\cdot \epsilon^{\ast}_{K_1}(r))\bigg].\label{asqure}
\end{eqnarray}
If we open the sum, for each polarization of $V$, \textit{i.e.,} $\pm,\; 0$, there are three possible polarizations of $K_1$ $(\pm,\; 0)$, and there will be 36 possible terms. The 12 terms for $\epsilon_V(+)$ and the $\pm,\; 0$ polarizations of the $K_1$ will give 
\begin{equation}
\left|\mathcal{A}_1\right|_{+}^2 =\lambda^2_{K_1 VP}\left(4+ 4 \frac{|\vec{p}_{3}|^2}{m_{V}^2}\right) . \label{pol1}    
\end{equation}
Similarly, the 12 terms for $\epsilon_{V}(-)$ and the $\pm,\; 0$ polarizations of the $K_1$  are
\begin{equation}
\left|\mathcal{A}_1\right|_{-}^2 =\lambda^2_{K_1 VP}\left(4+ 4 \frac{|\vec{p}_{3}|^2}{m_{V}^2}\right), \label{pol2}    
\end{equation}
and for the longitudinal polarization $\epsilon_V(0)$, one gets
\begin{equation}
\left|\mathcal{A}_1\right|_{0}^2 = 4\lambda^2_{K_1 VP}. \label{pol3}   
\end{equation}
The decay rate is obtained as
\begin{eqnarray}
\Gamma(K_{1}\to V P)=\frac{|\vec{p}_{3}|}{24\pi m^{2}_{K_{1}}}\left(\Tilde{\Gamma}_{\perp}+\Tilde{\Gamma}_{\|}\right),\label{decay231}
\end{eqnarray}
where $\Tilde{\Gamma}_{\perp}=\left|\mathcal{A}_1\right|_{+}^2+\left|\mathcal{A}_1\right|_{-}^2$, $\Tilde{\Gamma}_{\|}=\left|\mathcal{A}_1\right|_{0}^2$, and $|\vec{p}_{3}|=\frac{1}{2m_{K_{1}}}\sqrt{\lambda(m^{2}_{K_{1}},m^{2}_{V},m^{2}_{P})}$, with $\lambda(m_{K_1}^2,m^2_{V},m^2_{P})$ being a K\"all\'en function. Using values in Eq. (\ref{decay231}), one gets
\begin{equation}
\Gamma\left(K_1 \to V P\right) = \frac{|\lambda_{K_1 VP}|^2}{2\pi m_{K_1}^2}|\vec{p}_3|\left(1+\frac{2}{3}\frac{|\vec{p}_{3}|^2}{m_{V}^2}\right),\label{rateK1toKeho}    
\end{equation}
The result obtained in Eq. (\ref{rateK1toKeho}) agree with \cite{Choudhury:2007kx}. Finally, the form of $\mathcal{B}(K_{1}\to V_{\|(\perp)}P)$ given in Eq. (\ref{four-folded}) for the longitudinal and transverse polarization of $V$ meson is given by
\begin{eqnarray}
\mathcal{B}(K_{1}\to V_{\|(\perp)}P)=\frac{1}{\Gamma_{K_{1}}}\frac{|\vec{p}_{3}|}{24\pi m^{2}_{K_{1}}}\Tilde{\Gamma}_{\|(\perp)}.\label{decay23}
\end{eqnarray}
From Eq. (\ref{Amp5ag}), the computation of the decay rate will require the interference terms between matrix elements with different $K_1$ polarizations, which for the longitudinal and transverse polarizations of the final state vector $V$ meson can be defined as
\begin{eqnarray}
\Gamma_{2,\parallel}\Big(\epsilon_{K_1}(r),\,\epsilon^{\ast}_{K_1}(s)\Big)=\frac{\sqrt{\lambda(k^2, m_V^2, m_P^2)}}{16\pi m_{K_1}^3}\sum_{l=0}\mathcal{A}_{1}\Big(\epsilon_{K_1}(r),\,\epsilon^{\ast}_{V}(l)\Big)\mathcal{A}^{\ast}_{1}\Big(\epsilon^{\ast}_{K_1}(s),\,\epsilon_{V}(l)\Big),\label{Amp1Gammma2}\\
\Gamma_{2,\perp}\Big(\epsilon_{K_1}(r),\,\epsilon^{\ast}_{K_1}(s)\Big)=\frac{\sqrt{\lambda(k^2, m_V^2, m_P^2)}}{16\pi m_{K_1}^3}\sum_{l=\pm}\mathcal{A}_{1}\Big(\epsilon_{K_1}(r),\,\epsilon^{\ast}_{V}(l)\Big)\mathcal{A}^{\ast}_{1}\Big(\epsilon^{\ast}_{K_1}(s),\,\epsilon_{V}(l)\Big),\label{Amp2Gammma2}
\end{eqnarray}
where the normalization factor for $\Gamma_{2,\parallel(\perp)}$ in Eqs. (\ref{Amp1Gammma2}) and (\ref{Amp2Gammma2}) comes from the phase space part which is common in $K_{1}\to V_{\|(\perp)}P$ and $B\to K_1\,(\to V_{\|(\perp)} P)\ell^+\ell^-$ decays. Then, the decay rate of $K_{1}\to V_{\|(\perp)}P$ reads as
\begin{eqnarray}
\Gamma(K_{1}\to V_{\|(\perp)}P)=\frac{1}{3}\sum_{r=\pm,0}\Gamma_{2,\parallel(\perp)}\Big(\epsilon_{K_1}(r),\,\epsilon^{\ast}_{K_1}(r)\Big).\label{Gammma2comb}
\end{eqnarray}
The explicit expressions of $\Gamma_{2,\parallel}$ and $\Gamma_{2,\perp}$ are given by
\begin{eqnarray}
\Gamma_{2,\parallel}&=&\frac{\mathcal{B}(K_{1}\to V_{\|}P)\Gamma_{K_1}}{4}\begin{pmatrix}
6\sin^2\theta_{V} & -6\sin^2\theta_{V} & 3\sqrt{2}\sin 2\theta_{V}\\
-6\sin^2\theta_{V} & 6\sin^2\theta_{V} & -3\sqrt{2}\sin 2\theta_{V}\\
3\sqrt{2}\sin 2\theta_{V} & -3\sqrt{2}\sin 2\theta_{V} & 12\cos^2\theta_{V}
\end{pmatrix},\label{Gammma2comba}\\
\Gamma_{2,\perp}&=&\frac{\mathcal{B}(K_{1}\to V_{\perp}P)\Gamma_{K_1}}{8}\begin{pmatrix}
9+3\cos 2\theta_{V} & 6\sin^2\theta_{V} & -3\sqrt{2}\sin 2\theta_{V}\\
6\sin^2\theta_{V} & 9+3\cos 2\theta_{V} &  3\sqrt{2}\sin 2\theta_{V}\\
-3\sqrt{2}\sin 2\theta_{V} & 3\sqrt{2}\sin 2\theta_{V} & 12\sin^2\theta_{V}
\end{pmatrix},\label{Gammma2combb}
\end{eqnarray}
where the rows and columns correspond to the values of $r,s=+, -, 0$.

\section{Four-fold differential decay \boldmath $B \to K_{1}(1270,1400)(\to V_{\|(\perp)} P) \ell^{+}\ell^{-}$ calculations}\label{diffcal}
The polarized four-fold differential decay rate can be written as
\begin{eqnarray}
d^4\Gamma\left(B \to K_{1}(\to V_{\|(\perp)} P) \ell^{+}\ell^{-}\right)=\frac{|\mathcal{M}|_{\parallel(\perp)}^2}{2m_B}\,d\Phi_4\left(p;p_3,p_4,p_{\ell^+},p_{\ell^-}\right),
\label{4diff1}
\end{eqnarray}
where the phase space $d\Phi_4$ can be decomposed iteratively \cite{ParticleDataGroup:2024cfk}, such that
\begin{eqnarray}
\int\,d\Phi_4\frac{|\mathcal{M}|_{\parallel(\perp)}^2}{2m_B}=\int\frac{dq^2}{2\pi}\frac{dk^2}{2\pi}d\Phi_2\left(p;q,k\right)d\Phi_2\left(q;p_{\ell^+},p_{\ell^-}\right)d\Phi_2\left(k;p_3,p_4\right)\frac{|\mathcal{M}|_{\parallel(\perp)}^2}{2m_B},
\label{4diff2}
\end{eqnarray}
with two-body phase space is of the form
\begin{eqnarray}
d\Phi_2\left(k;p_3,p_4\right)=\frac{d\Omega_k}{4\pi}\,\frac{1}{8\pi}\,\frac{\sqrt{\lambda(k^2, m_V^2, m_P^2)}}{k^2},
\label{4diff3}
\end{eqnarray}
 solid angle $d\Omega_k$ for $k$, and $\lambda(k^2, m_V^2, m_P^2)$ as the K\"all\'en function. With the above two-body phase space structure, Eq. (\ref{4diff2}) can be written as
\begin{eqnarray}
\int\,d\Phi_4\frac{|\mathcal{M}|_{\parallel(\perp)}^2}{2m_B}=\frac{1}{(2\pi)^2(8\pi)^4}\int dk^2 \,\beta_{\ell}\frac{\sqrt{\lambda(m^2_{B}, m^2_{K_1}, q^2)}}{m^2_{B}}\,\frac{\sqrt{\lambda(k^2, m_V^2, m_P^2)}}{ m_{K_1}^2}\,dq^2  d\cos{\theta_{\ell}} d\cos {\theta}_{V}  d\phi\,
\frac{|\mathcal{M}|_{\parallel(\perp)}^2}{2m_B}.
\label{4diff4}
\end{eqnarray}


\section{Angular coefficients in terms of transversality amplitude}\label{transversality-amp}
The angular coefficients for the longitudinally polarized $V$ meson in terms of transversality amplitude are
\begin{eqnarray}\label{32}
I_{1s,\|} &=& \frac{(2+\beta_\ell^2)}{4}N^2\bigg[|A_{\perp}^L|^2+|A_{\parallel}^L|^2+(L\to R)\bigg]
+\frac{4m_\ell^2}{q^2}N^2\bigg[\mathcal{R}e\left(A_{\perp}^L A_{\perp}^{R\ast}+A_{\parallel}^L A_{\parallel}^{R\ast}\right)\bigg],\\
I_{1c,\|} &=& N^2\left(|A_0^L|^2+|A_0^R|^2\right)+\frac{4m_\ell^2}{q^2}N^2\left(2\mathcal{R}e(A_0^L A_0^{R\ast})+|A_t|^2\right),\\
I_{2s,\|} &=& \frac{\beta_\ell^2}{4}N^2\left(|A_{\perp}^L|^2+|A_{\parallel}^L|^2+(L\to R)\right)
,\\
I_{2c,\|} &=& -\beta_\ell^2 N^2\left(|A_0^L|^2+(L\to R)\right)
,\quad I_{3,\|}=\frac{\beta_\ell^2}{2}N^2\bigg[|A_{\perp}^L|^2-|A_{\parallel}^L|^2+(L\to R)\bigg],\\
I_{4,\|}&=&\frac{\beta_\ell^2}{\sqrt{2}}N^2\bigg[\mathcal{R}e(A_0^L A_{\parallel}^{L\ast})+(L\to R)\bigg],\quad I_{5,\|}={\sqrt{2}}\beta_\ell N^2\bigg[\mathcal{R}e(A_0^L A_{\perp}^{L\ast})-(L\to R)\bigg],\\
I_{6s,\|}&=&2\beta_\ell N^2\bigg[\mathcal{R}e(A_{\parallel}^L A_{\perp}^{L\ast})-(L\to R)\bigg],\quad\quad
I_{6c,\|}=0,\\
I_{7,\|}&=&{\sqrt{2}}\beta_\ell N^2\bigg[\mathcal{I}m(A_0^L A_{\parallel}^{L\ast})-(L\to R)\bigg],\quad
I_{8,\|}=\frac{\beta_\ell^2}{\sqrt{2}}N^2\bigg[\mathcal{I}m(A_0^L A_{\perp}^{L\ast})+(L\to R)\bigg],\\
I_{9,\|}&=&\beta_\ell^2 N^2\bigg[\mathcal{I}m(A_{\parallel}^{L\ast} A_{\perp}^{L})+(L\to R)\bigg].
\end{eqnarray}
Our results of angular coefficients for the longitudinally polarized vecotr meson $V$ in $B \to K_{1}(\to V_{\|} P) \ell^{+}\ell^{-}$ decay match with the angular coefficients given in \cite{Altmannshofer:2008dz} for the $B \to K^{\ast}(\to K \pi) \mu^{+}\mu^{-}$ decay, except the terms containing contributions of the scalar and pseudo-scalar operators, which we have ignored. Similarly, for the transversely polarized vector meson $V$, we have
\begin{eqnarray}\label{32ac}
I_{1s,\perp} &=& \frac{(2+\beta_\ell^2)}{8}N^2\bigg[|A_{\perp}^L|^2+|A_{\parallel}^L|^2+(L\to R)\bigg]+\frac{N^2}{2}\left(|A_0^L|^2+|A_0^R|^2\right)\notag\\
&+&\frac{2m_\ell^2}{q^2}N^2\bigg[\mathcal{R}e\left(A_{\perp}^L A_{\perp}^{R\ast}+A_{\parallel}^L A_{\parallel}^{R\ast}\right)
+\left(2\mathcal{R}e(A_0^L A_0^{R\ast})+|A_t|^2\right)\bigg],\\
I_{1c,\perp} &=& \frac{(2+\beta_\ell^2)}{4}N^2\bigg[|A_{\perp}^L|^2+|A_{\parallel}^L|^2+(L\to R)\bigg]
+\frac{4m_\ell^2}{q^2}N^2\bigg[\mathcal{R}e\left(A_{\perp}^L A_{\perp}^{R\ast}+A_{\parallel}^L A_{\parallel}^{R\ast}\right)\bigg],\\
I_{2s,\perp} &=& -\frac{\beta_\ell^2}{2}N^2\bigg[\left(|A_0^L|^2+(L\to R)\right)-\frac{1}{4}\left(|A_{\perp}^L|^2+|A_{\parallel}^L|^2+(L\to R)\right)\bigg]
,\\
I_{2c,\perp} &=& \frac{\beta_\ell^2}{4}N^2\bigg[\left(|A_{\perp}^L|^2+|A_{\parallel}^L|^2+(L\to R)\right)\bigg]
,\quad
I_{3,\perp}=-\frac{\beta_\ell^2}{4}N^2\bigg[|A_{\perp}^L|^2-|A_{\parallel}^L|^2+(L\to R)\bigg],\\
I_{4,\perp}&=&-\frac{\beta_\ell^2}{2\sqrt{2}}N^2\bigg[\mathcal{R}e(A_0^L A_{\parallel}^{L\ast})+(L\to R)\bigg],\quad
I_{5,\perp}=-\frac{\beta_\ell}{\sqrt{2}}N^2\bigg[\mathcal{R}e(A_0^L A_{\perp}^{L\ast})-(L\to R)\bigg],\\
I_{6s,\perp}&=&\beta_\ell N^2\bigg[\mathcal{R}e(A_{\parallel}^L A_{\perp}^{L\ast})-(L\to R)\bigg],\quad
I_{6c,\perp}=2\beta_\ell N^2\bigg[\mathcal{R}e(A_{\parallel}^L A_{\perp}^{L\ast})-(L\to R)\bigg],\\
I_{7,\perp}&=&-\frac{\beta_\ell}{\sqrt{2}}N^2\bigg[\mathcal{I}m(A_0^L A_{\parallel}^{L\ast})-(L\to R)\bigg],\quad
I_{8,\perp}=-\frac{\beta_\ell^2}{2\sqrt{2}}N^2\bigg[\mathcal{I}m(A_0^L A_{\perp}^{L\ast})+(L\to R)\bigg],\\
I_{9,\perp}&=&-\frac{\beta_\ell^2}{2}N^2\bigg[\mathcal{I}m(A_{\parallel}^{L\ast} A_{\perp}^{L})+(L\to R)\bigg].
\end{eqnarray}

\begin{table}
\caption{\small Predictions of binned observables for the $B \to K_1(1270)\left(\to VP\right)\mu^+ \mu^-$ decay, including differential branching ratios, $\left\langle d\mathcal{B}/dq^2\right\rangle= \left\langle d\mathcal{B}\left(B\to K_1(1270)\mu^+\mu^-\right)/dq^2\right\rangle$, $\left\langle d\mathcal{B}/dq^2\right\rangle_{\rho K}=\left\langle d\mathcal{B}\left(B\to K_1(1270)\,(\to \rho K)\mu^+\mu^-\right)/dq^2\right\rangle$, and $\left\langle d\mathcal{B}/dq^2\right\rangle_{K^{\ast}\pi}=\left\langle d\mathcal{B}\left(B\to K_1(1270)\,(\to K^{\ast} \pi)\mu^+\mu^-\right)/dq^2\right\rangle$. The other observables include the lepton FB asymmetry $\mathcal{A}_{\text{FB}}^{K_1 (1270)}$, FB asymmetry for transversely polarized $K_1(1270)$, $\mathcal{A}^{K_{1T} (1270)}_{\text{FB}}$, longitudinal helicity fraction $f_{L}^{K_1 (1270)}$, transverse helicity fraction $f_T^{K_1 (1270)}$, and angular coefficients $\hat{I}_{\parallel(\perp)}^{K_1 (1270)}$ with longitudinal and transverse polarization of the vector meson, in the $q^2=[0.1-2.0]$ GeV$^2$ bin, for both the SM and NP scenarios presented in Table \ref{tab:bestfitWC}. The differential branching ratios are given in units of $10^{-7}$ GeV$^{-2}$.}\label{Bin1-analysis}
		\scalebox{0.87}{
    \begin{tabular}{|c|ccccccccc|}
    \hline\hline
        \multicolumn{10}{|c|}{$q^2=[0.1-2.0]$ GeV$^2$}\\
        \hline
				\hline 
				\ \ \  & $\left\langle d\mathcal{B}/dq^2\right\rangle 
    $  & $\left\langle d\mathcal{B}/dq^2\right\rangle_{\rho K} 
    $ & $\ \ \left\langle d\mathcal{B}/dq^2\right\rangle_{K^*\pi} 
    $ & $ \ \ \ \
				 \mathcal{A}_{\text{FB}}^{K_1 (1270)}$  & $ \
				\ \ \ \ \mathcal{A}^{K_{1T} (1270)}_{\text{FB}} $ & $\ \ \ \ \ \  f_{L}^{K_1 (1270)} $  & $\ \ \ \ \ \  f_T^{K_1 (1270)} $ &
    $\hat {I}_{6c,\perp}^{K_1 (1270)} $ & \\ 
    \hline
    SM %
				&$\ \ 1.216\pm 0.187$ 
				& %
				$\ \ 0.462 \pm 0.071$
				& %
				$\ \ 0.255\pm 0.039$ 
				& %
				$\ \ 0.075\pm 0.004$
                &  %
                $\ \ 0.1204 \pm 0.0004$ 
                & %
				$\ \ 0.374 \pm 0.033$ 
                & %
                $\ \ 0.626\pm 0.033$
				& %
				$\ \ 0.100 \pm 0.006$ 
                & \\
    S1 %
				&$\ \ 1.281\pm 0.188$ 
				& %
				$\ \ 0.487 \pm 0.071$
				& %
				$\ \ 0.269\pm 0.039$ 
				& %
				$\ \ 0.099\pm 0.005$
                &  %
                $\ \ 0.1493\pm 0.0006$ 
                & %
				$\ \ 0.338 \pm 0.033$ 
                & %
                $\ \ 0.662\pm 0.033$
				& %
				$\ \ 0.132 \pm 0.007$ 
				&   \\
    S2 %
				&$\ \ 1.079\pm 0.147$
				& %
				$\ \ 0.410 \pm 0.056$
				& %
				$\ \ 0.227 \pm 0.031$ 
                & %
				$\ \ 0.089 \pm 0.004$
                & %
                $\ \ 0.1239 \pm 0.0005$ 
                & %
				$\ \ 0.278 \pm 0.029$ 
                & %
                $\ \ 0.722 \pm 0.029$ 
                & %
				$\ \ 0.119 \pm 0.005$  
                & \\
    S3 %
				&$\ \ 1.061\pm 0.145$
				& %
				$\ \ 0.403 \pm 0.055$
				& %
				$\ \ 0.223 \pm 0.030$
				& %
				$\ \ 0.086 \pm 0.004$
                & %
                $\ \ 0.1196 \pm 0.0005$
                & %
				$\ \ 0.279 \pm 0.029 $
                &  %
				$\ \ 0.721 \pm 0.029$ 
                & %
				$\ \ 0.115 \pm 0.005$
			    & \\
    S4 %
				&$\ \ 1.157\pm 0.169$
				& %
				$\ \ 0.440 \pm 0.064$
				& %
				$\ \ 0.243 \pm 0.035$
				& %
				$\ \ 0.082 \pm 0.004$
                & %
                $\ \ 0.1231 \pm 0.0004$
                & %
				$\ \ 0.330 \pm 0.032 $
                & %
                $\ \ 0.670 \pm 0.032$
				& %
				$\ \ 0.110 \pm 0.006$ 
				& \\     
                 \hline\hline
				& $\hat
                    {I}_{1s,\parallel}^{K_1 (1270)} $ &$\hat
					{I}_{1c,\parallel}^{K_1 (1270)} $ & $ \hat{I}_{2s,\parallel}^{K_1 (1270)} $
					& $\hat{I}_{2c,\parallel}^{K_1 (1270)} $ & $\hat
					{I}_{3,\parallel}^{K_1 (1270)} $ &$\hat
					{I}_{4,\parallel}^{K_1 (1270)} $ &$\hat
					{I}_{5,\parallel}^{K_1 (1270)} $  & $\hat {I}_{6s,\parallel}^{K_1 (1270)} $ &
			  \\ 
				\hline
    SM %
				& $\ \ 0.452 \pm 0.024$
                & %
				$\ \ 0.400 \pm 0.036$
                & %
			$\ \ 0.105 \pm 0.006$
				& %
				$\ \ -0.298\pm 0.025$
                & %
				$\ \ 0.006\pm 0.000 $
				& %
				$\ \ -0.083 \pm 0.001$
                & %
				$\ \ 0.205 \pm 0.005$
                & %
				$\ \ 0.100 \pm 0.006$
                & %
			  \\
    S1 %
				&$\ \ 0.479 \pm 0.024$
                & %
				$\ \ 0.360 \pm 0.035$
                & %
				$\ \ 0.112 \pm 0.006$
				& %
				$\ \ -0.271\pm 0.025$
                & %
				$\ \ 0.003\pm 0.000 $
				& %
				$\ \ -0.037 \pm 0.000$
                & %
				$\ \ 0.278 \pm 0.007$
                & %
				$\ \ 0.132 \pm 0.007$
                & %
			  \\ 
    S2 %
                &$\ \ 0.522 \pm 0.021$ 
                & %
				$\ \ 0.296 \pm 0.031$
                & %
				$\ \ 0.122 \pm 0.005$ 
                & %
				$\ \ -0.222 \pm 0.022$
                & %
				$\ \ 0.004 \pm 0.000$ 
                & %
				$\ \ -0.053 \pm 0.001$ 
                & %
				$\ \ 0.252 \pm 0.009$
                & %
				$\ \ 0.119 \pm 0.005$ 
                & %
                \\
    S3 %
                &$\ \ 0.521 \pm 0.021$
                & %
			$\ \ 0.298 \pm 0.031$
                & %
				$\ \ 0.122 \pm 0.005 $ 
                & %
				$\ \ -0.222 \pm 0.022$ 
                & %
				$\ \ 0.005 \pm 0.000$ 
			    & %
				$\ \ -0.060 \pm 0.001$
                & %
				$\ \ 0.242 \pm 0.009$
                & %
				$\ \ 0.115 \pm 0.005$
                & %
                \\
    S4 %
                &$\ \ 0.484 \pm 0.023$
                & %
				$\ \ 0.352 \pm 0.034$
                & %
				$\ \ 0.113 \pm 0.005 $
                & %
				$\ \ -0.263 \pm 0.024$
                & %
				$\ \ 0.005 \pm 0.000$ 
				& %
				$\ \ -0.066 \pm 0.001$
                & %
				$\ \  0.237 \pm 0.007$
                & %
				$\ \  0.110 \pm 0.006$
                & %
				\\     
                 \hline\hline
				& $\hat
                    {I}_{1s,\perp}^{K_1 (1270)} $ &$\hat
					{I}_{1c,\perp}^{K_1 (1270)} $ & $\hat{I}_{2s,\perp}^{K_1 (1270)} $
					& $\hat{I}_{2c,\perp}^{K_1 (1270)} $ & $\hat
					{I}_{3,\perp}^{K_1 (1270)} $ &$\hat
					{I}_{4,\perp}^{K_1 (1270)} $ &$\hat
					{I}_{5,\perp}^{K_1 (1270)} $  & $\hat {I}_{6s,\perp}^{K_1 (1270)} $ & 
			  \\ 
				\hline
    SM %
				&$\ \ 0.426 \pm 0.006$
                & %
				$\ \ 0.452 \pm 0.024$
                & %
			    $\ \ -0.097 \pm 0.015$
				& %
				$\ \ 0.105\pm 0.006$
                & %
				$\ \ -0.003\pm 0.000$
				& %
				$\ \ 0.041 \pm 0.000$
                & %
				$\ \ -0.102 \pm 0.002$
                & %
				$\ \ 0.050 \pm 0.003$
                & %
               
			  \\
    S1 %
				&$\ \ 0.420 \pm 0.006$
                & %
				$\ \ 0.479 \pm 0.024$
                & %
				$\ \ -0.079 \pm 0.015$
				& %
				$\ \ 0.112\pm 0.006$
                & %
				$\ \ -0.002\pm 0.000 $
				& %
				$\ \ 0.019 \pm 0.000$
                & %
				$\ \ -0.139 \pm 0.004$
                & %
                $\ \ 0.066 \pm 0.003$
                & %
				            
			  \\ 
    S2 %
                &$\ \ 0.409 \pm 0.005$ 
                & %
				$\ \ 0.522 \pm 0.021$
                & %
				$\ \ -0.050 \pm 0.013$ 
                & %
				$\ \ 0.122 \pm 0.005$
                & %
				$\ \ -0.002 \pm 0.000$ 
                & %
				$\ \ 0.027 \pm 0.000$ 
                & %
				$\ \ -0.126 \pm 0.004$
                & %
                $\ \ 0.060 \pm 0.003$
                & %
				
                \\
    S3 %
                &$\ \ 0.409 \pm 0.005$
                & %
			$\ \ 0.521 \pm 0.021$
                & %
				$\ \ -0.050 \pm 0.013 $ 
                & %
				$\ \ 0.122 \pm 0.005$ 
                & %
				$\ \ -0.002 \pm 0.000$ 
			    & %
				$\ \ 0.030 \pm 0.001$
                & %
				$\ \ -0.121 \pm 0.004$
                & %
                $\ \ 0.058 \pm 0.003$
                & %
				
                \\
    S4 %
                &$\ \ 0.418 \pm 0.006$
                & %
				$\ \ 0.484 \pm 0.023$
                & %
				$\ \ -0.075 \pm 0.015 $
                & %
				$\ \ 0.113 \pm 0.005$
                & %
				$\ \ -0.002 \pm 0.000$ 
				& %
				$\ \ 0.033 \pm 0.000$
                & %
				$\ \ -0.119 \pm 0.003$
                & %
                $\ \ 0.055 \pm 0.003$
                & %
				
				\\                    
               \hline \hline
			\end{tabular}
   }%
	\end{table}

\begin{table}
\caption{\small Same as in Table \ref{Bin1-analysis}, but for  the $q^2=[2.0-4.0]$ GeV$^2$ bin.}\label{Bin2-analysis}
		\scalebox{0.87}{
  \begin{tabular}{|c|ccccccccc|}
    \hline\hline
        \multicolumn{10}{|c|}{$q^2=[2.0-4.0]$ GeV$^2$}\\
        \hline
				\hline 
				\ \ \  & $\left\langle d\mathcal{B}/dq^2\right\rangle 
    $  & $\left\langle d\mathcal{B}/dq^2\right\rangle_{\rho K} 
    $ & $\ \ \left\langle d\mathcal{B}/dq^2\right\rangle_{K^*\pi} 
    $ & $ \ \ \ \
				 \mathcal{A}_{\text{FB}}^{K_1 (1270)}$  & $ \
				\ \ \ \ \mathcal{A}^{K_{1T} (1270)}_{\text{FB}} $ & $\ \ \ \ \ \  f_{L}^{K_1 (1270)} $  & $\ \ \ \ \ \  f_T^{K_1 (1270)} $ & $\hat {I}_{6c,\perp}^{K_1 (1270)} $  & \\ 
    \hline
    SM %
				&$\ \ 0.814\pm 0.146$ 
				& %
				$\ \ 0.310 \pm 0.056$
				& %
				$\ \ 0.171\pm 0.031$ 
				& %
				$\ \ 0.009\pm 0.001$
                &  %
                $\ \ 0.0500 \pm 0.0003$ 
                & %
				$\ \ 0.814 \pm 0.021$ 
                & %
                $\ \ 0.186\pm 0.021$
				& %
				$\ \ 0.012 \pm 0.001$ 
                & \\
    S1 %
				&$\ \ 0.876\pm 0.149$ 
				& %
				$\ \ 0.333 \pm 0.056$
				& %
				$\ \ 0.184\pm 0.031$ 
				& %
				$\ \ 0.119\pm 0.012$
                &  %
                $\ \ 0.4346\pm 0.0008$ 
                & %
				$\ \ 0.727 \pm 0.028$ 
                & %
                $\ \ 0.273\pm 0.028$
				& %
				$\ \ 0.158 \pm 0.016$
				&   \\
    S2 %
				&$\ \ 0.633\pm 0.107$
				& %
				$\ \ 0.240 \pm 0.041$
				& %
				$\ \ 0.133 \pm 0.022$ 
                & %
				$\ \ 0.135 \pm 0.014$
                & %
                $\ \ 0.5010 \pm 0.0008$ 
                & %
				$\ \ 0.730 \pm 0.027$ 
                & %
                $\ \ 0.270 \pm 0.027$ 
                & %
				$\ \ 0.181 \pm 0.019$ 
                & \\
    S3 %
				&$\ \ 0.613\pm 0.105$
				& %
				$\ \ 0.233 \pm 0.040$
				& %
				$\ \ 0.129 \pm 0.022$
				& %
				$\ \ 0.120 \pm 0.013$
                & %
                $\ \ 0.4699 \pm 0.0009$
                & %
				$\ \ 0.744 \pm 0.026 $
                &  %
				$\ \ 0.256 \pm 0.026$ 
                & %
				$\ \ 0.161 \pm 0.017$
			    & \\
    S4 %
				&$\ \ 0.730\pm 0.128$
				& %
				$\ \ 0.277 \pm 0.049$
				& %
				$\ \ 0.153 \pm 0.027$
				& %
				$\ \ 0.068 \pm 0.008$
                & %
                $\ \ 0.3124 \pm 0.0008$
                & %
				$\ \ 0.781 \pm 0.024 $
                & %
                $\ \ 0.219 \pm 0.024$
				& %
				$\ \  0.091 \pm 0.010$ 
				& \\     
                 \hline\hline
				& $\hat
                    {I}_{1s,\parallel}^{K_1 (1270)} $ &$\hat
					{I}_{1c,\parallel}^{K_1 (1270)} $ & $ \hat{I}_{2s,\parallel}^{K_1 (1270)} $
					& $\hat{I}_{2c,\parallel}^{K_1 (1270)} $ & $\hat
					{I}_{3,\parallel}^{K_1 (1270)} $ &$\hat
					{I}_{4,\parallel}^{K_1 (1270)} $ &$\hat
					{I}_{5,\parallel}^{K_1 (1270)} $  & $\hat {I}_{6s,\parallel}^{K_1 (1270)} $ &
			  \\ 
				\hline
    SM %
				& $\ \ 0.139 \pm 0.016$
                & %
				$\ \ 0.820 \pm 0.021$
                & %
			$\ \ 0.046 \pm 0.005$
				& %
				$\ \ -0.798\pm 0.020$
                & %
				$\ \ -0.023\pm 0.002 $
				& %
				$\ \ 0.168 \pm 0.006$
                & %
				$\ \ -0.163 \pm 0.009$
                & %
				$\ \ 0.012 \pm 0.001$
                & %
			  \\
    S1 %
				&$\ \ 0.204 \pm 0.021$
                & %
				$\ \ 0.733 \pm 0.028$
                & %
				$\ \ 0.068 \pm 0.007$
				& %
				$\ \ -0.711\pm 0.027$
                & %
				$\ \ -0.027\pm 0.002$
				& %
				$\ \ 0.190 \pm 0.005$
                & %
				$\ \ 0.017 \pm 0.001$
                & %
				$\ \ 0.158 \pm 0.016$
                & %
			  \\ 
    S2 %
                &$\ \ 0.203 \pm 0.021$ 
                & %
				$\ \ 0.735 \pm 0.028$
                & %
				$\ \ 0.067 \pm 0.007$ 
                & %
				$\ \ -0.713 \pm 0.027$
                & %
				$\ \ -0.022 \pm 0.002$ 
                & %
				$\ \ 0.167 \pm 0.004$ 
                & %
				$\ \ 0.021 \pm 0.002$
                & %
				$\ \ 0.181 \pm 0.019$ 
                & %
                \\
    S3 %
                &$\ \ 0.192 \pm 0.020$
                & %
			$\ \ 0.749 \pm 0.027$
                & %
				$\ \ 0.064 \pm 0.007 $ 
                & %
				$\ \ -0.727 \pm 0.026$ 
                & %
				$\ \ -0.021 \pm 0.002$ 
			    & %
				$\ \ 0.162 \pm 0.004$
                & %
				$\ \ -0.007 \pm 0.002$
                & %
				$\ \ 0.161 \pm 0.017$
                & %
                \\
    S4 %
                &$\ \ 0.164 \pm 0.018$
                & %
				$\ \ 0.787 \pm 0.024$
                & %
				$\ \ 0.054 \pm 0.006 $
                & %
				$\ \ -0.764 \pm 0.023$
                & %
				$\ \ -0.028 \pm 0.003$ 
				& %
				$\ \ 0.171 \pm 0.006$
                & %
				$\ \  -0.067 \pm 0.005$
                & %
				$\ \  0.091 \pm 0.010$
                & %
				\\     
                 \hline\hline
				& $\hat
                    {I}_{1s,\perp}^{K_1 (1270)} $ &$\hat
					{I}_{1c,\perp}^{K_1 (1270)} $ & $\hat{I}_{2s,\perp}^{K_1 (1270)} $
					& $\hat{I}_{2c,\perp}^{K_1 (1270)} $ & $\hat
					{I}_{3,\perp}^{K_1 (1270)} $ &$\hat
					{I}_{4,\perp}^{K_1 (1270)} $ &$\hat
					{I}_{5,\perp}^{K_1 (1270)} $  & $\hat {I}_{6s,\perp}^{K_1 (1270)} $ & 
			  \\ 
				\hline
    SM %
				&$\ \ 0.480 \pm 0.003$
                & %
				$\ \ 0.139 \pm 0.016$
                & %
			$\ \ -0.375 \pm 0.013$
				& %
				$\ \ 0.046 \pm 0.005$
                & %
				$\ \ 0.012 \pm 0.001 $
				& %
				$\ \ -0.084 \pm 0.003$
                & %
				$\ \ 0.082 \pm 0.005$
                & %
				$\ \ 0.006 \pm 0.001$
                & %
                
			  \\
    S1 %
				&$\ \ 0.469 \pm 0.004$
                & %
				$\ \ 0.204 \pm 0.021$
                & %
				$\ \ -0.322 \pm 0.017$
				& %
				$\ \ 0.068 \pm 0.007$
                & %
				$\ \ 0.013\pm 0.001 $
				& %
				$\ \ -0.095 \pm 0.003$
                & %
				$\ \ -0.008 \pm 0.007$
                & %
				$\ \ 0.079 \pm 0.008$
                & %
                
			  \\ 
    S2 %
                &$\ \ 0.469 \pm 0.002$ 
                & %
				$\ \ 0.203 \pm 0.021$
                & %
				$\ \ -0.323 \pm 0.017$ 
                & %
				$\ \ 0.067 \pm 0.007$ 
                & %
				$\ \ 0.011 \pm 0.001$ 
                & %
				$\ \ -0.084 \pm 0.002$ 
                & %
				$\ \ -0.010 \pm 0.001$
                & %
				$\ \ 0.090 \pm 0.009$ 
                & %
               
                \\
    S3 %
                &$\ \ 0.471 \pm 0.003$
                & %
			    $\ \ 0.192 \pm 0.020$
                & %
				$\ \ -0.332 \pm 0.016 $ 
                & %
				$\ \ 0.064 \pm 0.007 $  
                & %
				$\ \ 0.011 \pm 0.001$ 
			    & %
				$\ \ -0.081 \pm 0.002$
                & %
				$\ \ 0.003 \pm 0.001$
                & %
				$\ \ 0.080 \pm 0.008$
                & %
                
                \\
    S4 %
                &$\ \ 0.476 \pm 0.003$
                & %
				$\ \ 0.164 \pm 0.018$
                & %
				$\ \ -0.355 \pm 0.014$
                & %
				$\ \ 0.054 \pm 0.006$
                & %
				$\ \ 0.014 \pm 0.001$ 
				& %
				$\ \ -0.086 \pm 0.003$
                & %
				$\ \ 0.033 \pm 0.002$
                & %
				$\ \ 0.046 \pm 0.005$
                & %
                
				\\                    
               \hline \hline
			\end{tabular}
   }%
	\end{table}

\begin{table}
		\caption{\small Same as in Table \ref{Bin1-analysis}, but for the $q^2=[4.0-6.0]$ GeV$^2$ bin.}\label{Bin3-analysis}
		\scalebox{0.87}{
  \begin{tabular}{|c|ccccccccc|}
    \hline\hline
        \multicolumn{10}{|c|}{$q^2=[4.0-6.0]$ GeV$^2$}\\
        \hline
				\hline 
				\ \ \  & $\left\langle d\mathcal{B}/dq^2\right\rangle 
    $  & $\left\langle d\mathcal{B}/dq^2\right\rangle_{\rho K} 
    $ & $\ \ \left\langle d\mathcal{B}/dq^2\right\rangle_{K^*\pi} 
    $ & $ \ \ \ \
				 \mathcal{A}_{\text{FB}}^{K_1 (1270)}$  & $ \
				\ \ \ \ \mathcal{A}^{K_{1T} (1270)}_{\text{FB}} $ & $\ \ \ \ \ \  f_{L}^{K_1 (1270)} $  & $\ \ \ \ \ \  f_T^{K_1 (1270)} $ & $\hat {I}_{6c,\perp}^{K_1 (1270)} $  & \\ 
    \hline
    SM %
				&$\ \ 0.921\pm 0.146$ 
				& %
				$\ \ 0.350 \pm 0.056$
				& %
				$\ \ 0.193\pm 0.031$ 
				& %
				$\ \ -0.151\pm 0.013$
                &  %
                $\ \ -0.489 \pm 0.001$ 
                & %
				$\ \ 0.690 \pm 0.027$ 
                & %
                $\ \ 0.310\pm 0.027$
				& %
				$\ \ -0.202 \pm 0.017$ 
                & \\
    S1 %
				&$\ \ 0.966\pm 0.147$ 
				& %
				$\ \ 0.367 \pm 0.056$
				& %
				$\ \ 0.203\pm 0.031$ 
				& %
				$\ \ -0.014\pm 0.000$
                &  %
                $\ \ -0.039\pm 0.002$ 
                & %
				$\ \ 0.635 \pm 0.030$ 
                & %
                $\ \ 0.365\pm 0.030$
				& %
				$\ \ -0.019 \pm 0.000$
				&   \\
    S2 %
				&$\ \ 0.684\pm 0.105$
				& %
				$\ \ 0.260 \pm 0.040$
				& %
				$\ \ 0.144 \pm 0.022$ 
                & %
				$\ \ -0.015 \pm 0.000$
                & %
                $\ \ -0.044 \pm 0.003$ 
                & %
				$\ \ 0.650 \pm 0.029$ 
                & %
                $\ \ 0.350 \pm 0.029$ 
                & %
				$\ \ -0.020 \pm 0.000$
                & \\
    S3 %
				&$\ \ 0.665\pm 0.103$
				& %
				$\ \ 0.253 \pm 0.039$
				& %
				$\ \ 0.140 \pm 0.022$
				& %
				$\ \ -0.037 \pm 0.002$
                & %
                $\ \ -0.110 \pm 0.003$
                & %
				$\ \ 0.660 \pm 0.028 $
                &  %
				$\ \ 0.340 \pm 0.028$ 
                & %
				$\ \ -0.050 \pm 0.003$ 
			    & \\
    S4 %
				&$\ \ 0.806\pm 0.127$
				& %
				$\ \ 0.306 \pm 0.048$
				& %
				$\ \ 0.169 \pm 0.027$
				& %
				$\ \ -0.087 \pm 0.007$
                & %
                $\ \ -0.273 \pm 0.002$
                & %
				$\ \ 0.680 \pm 0.027 $
                & %
                $\ \ 0.320 \pm 0.027$
				& %
				$\ \  -0.117 \pm 0.009$ 
				& \\     
                 \hline\hline
				& $\hat
                    {I}_{1s,\parallel}^{K_1 (1270)} $ &$\hat
					{I}_{1c,\parallel}^{K_1 (1270)} $ & $ \hat{I}_{2s,\parallel}^{K_1 (1270)} $
					& $\hat{I}_{2c,\parallel}^{K_1 (1270)} $ & $\hat
					{I}_{3,\parallel}^{K_1 (1270)} $ &$\hat
					{I}_{4,\parallel}^{K_1 (1270)} $ &$\hat
					{I}_{5,\parallel}^{K_1 (1270)} $  & $\hat {I}_{6s,\parallel}^{K_1 (1270)} $ &
			  \\ 
				\hline
    SM %
				& $\ \ 0.232 \pm 0.020$
                & %
				$\ \ 0.693 \pm 0.027$
                & %
			$\ \ 0.077 \pm 0.007$
				& %
				$\ \ -0.681\pm 0.026$
                & %
				$\ \ -0.056\pm 0.004$
				& %
				$\ \ 0.256 \pm 0.005$
                & %
				$\ \ -0.315 \pm 0.009$
                & %
				$\ \ -0.202 \pm 0.017$
                & %
			  \\
    S1 %
				&$\ \ 0.274 \pm 0.022$
                & %
				$\ \ 0.638 \pm 0.030$
                & %
				$\ \ 0.091 \pm 0.007$
				& %
				$\ \ -0.627\pm 0.029$
                & %
				$\ \ -0.056\pm 0.004$
				& %
				$\ \ 0.259 \pm 0.004$
                & %
				$\ \ -0.137 \pm 0.003$
                & %
				$\ \ -0.019 \pm 0.000$
                & %
			  \\ 
    S2 %
                &$\ \ 0.262 \pm 0.022$ 
                & %
				$\ \ 0.653 \pm 0.029$
                & %
				$\ \ 0.087 \pm 0.007$ 
                & %
				$\ \ -0.642 \pm 0.028$
                & %
				$\ \ -0.053 \pm 0.003$ 
                & %
				$\ \ 0.252 \pm 0.004$ 
                & %
				$\ \ -0.158 \pm 0.004$
                & %
				$\ \ -0.020 \pm 0.000$ 
                & %
                \\
    S3 %
                &$\ \ 0.255 \pm 0.021$
                & %
			$\ \ 0.663 \pm 0.028$
                & %
				$\ \ 0.085 \pm 0.007$ 
                & %
				$\ \ -0.652 \pm 0.028$ 
                & %
				$\ \ -0.053 \pm 0.004$ 
			    & %
				$\ \ 0.251 \pm 0.004$
                & %
				$\ \ -0.188 \pm 0.005$
                & %
				$\ \ -0.050 \pm 0.003$
                & %
                \\
    S4 %
                &$\ \ 0.240 \pm 0.020$
                & %
				$\ \ 0.683 \pm 0.027$
                & %
				$\ \ 0.080 \pm 0.007$
                & %
				$\ \ -0.671 \pm 0.027$
                & %
				$\ \ -0.063 \pm 0.005$ 
				& %
				$\ \ 0.259 \pm 0.005$
                & %
				$\ \  -0.235 \pm 0.006$
                & %
				$\ \  -0.117 \pm 0.009$
                & %
				\\     
                 \hline\hline
				& $\hat
                    {I}_{1s,\perp}^{K_1 (1270)} $ &$\hat
					{I}_{1c,\perp}^{K_1 (1270)} $ & $\hat{I}_{2s,\perp}^{K_1 (1270)} $
					& $\hat{I}_{2c,\perp}^{K_1 (1270)} $ & $\hat
					{I}_{3,\perp}^{K_1 (1270)} $ &$\hat
					{I}_{4,\perp}^{K_1 (1270)} $ &$\hat
					{I}_{5,\perp}^{K_1 (1270)} $  & $\hat {I}_{6s,\perp}^{K_1 (1270)} $ & 
			  \\ 
				\hline
    SM %
				&$\ \ 0.463 \pm 0.003$
                & %
				$\ \ 0.232 \pm 0.020$
                & %
			$\ \ -0.302 \pm 0.016$
				& %
				$\ \ 0.077 \pm 0.007$
                & %
				$\ \ 0.028 \pm 0.002 $
				& %
				$\ \ -0.128 \pm 0.003$
                & %
				$\ \ 0.157 \pm 0.004$
                & %
				$\ \ -0.101 \pm 0.009$
                & %
                
			  \\
    S1 %
				&$\ \ 0.456 \pm 0.004$
                & %
				$\ \ 0.274 \pm 0.022$
                & %
				$\ \ -0.268 \pm 0.018$
				& %
				$\ \ 0.091 \pm 0.007$
                & %
				$\ \ 0.028\pm 0.002$
				& %
				$\ \ -0.130 \pm 0.002$
                & %
				$\ \ 0.069 \pm 0.002$
                & %
				$\ \ -0.009 \pm 0.000$
                & %
                
			  \\ 
    S2 %
                &$\ \ 0.458 \pm 0.004$ 
                & %
				$\ \ 0.262 \pm 0.022$
                & %
				$\ \ -0.277 \pm 0.018$ 
                & %
				$\ \ 0.087 \pm 0.007$
                & %
				$\ \ 0.027 \pm 0.002$ 
                & %
				$\ \ -0.126 \pm 0.002$ 
                & %
				$\ \ 0.079 \pm 0.002$
                & %
				$\ \ -0.010 \pm 0.000$ 
                & %
                
                \\
    S3 %
                &$\ \ 0.459 \pm 0.004$
                & %
			$\ \ 0.255 \pm 0.021$
                & %
				$\ \ -0.284 \pm 0.017 $ 
                & %
				$\ \ 0.085 \pm 0.007 $ 
                & %
				$\ \ 0.027 \pm 0.002$ 
			    & %
				$\ \ -0.125 \pm 0.002$
                & %
				$\ \ 0.094 \pm 0.002$
                & %
				$\ \ -0.025 \pm 0.001$
                & %
                
                \\
    S4 %
                &$\ \ 0.461 \pm 0.003$
                & %
				$\ \ 0.240 \pm 0.020$
                & %
				$\ \ -0.296 \pm 0.017$
                & %
				$\ \ 0.080 \pm 0.007$
                & %
				$\ \ 0.031 \pm 0.002$ 
				& %
				$\ \ -0.129 \pm 0.003$
                & %
				$\ \ 0.117 \pm 0.003$
                & %
				$\ \ -0.058 \pm 0.005$
                & %
                
				\\                    
               \hline \hline
			\end{tabular}
   }%
	\end{table}

\begin{table}
		\caption{\small Same as in Table \ref{Bin1-analysis}, but for the $q^2=[6.0-8.0]$ GeV$^2$ bin.}\label{Bin4-analysis}
		\scalebox{0.87}{
  \begin{tabular}{|c|ccccccccc|}
    \hline\hline
        \multicolumn{10}{|c|}{$q^2=[6.0-8.0]$ GeV$^2$}\\
        \hline
				\hline 
				\ \ \  & $\left\langle d\mathcal{B}/dq^2\right\rangle 
    $  & $\left\langle d\mathcal{B}/dq^2\right\rangle_{\rho K} 
    $ & $\ \ \left\langle d\mathcal{B}/dq^2\right\rangle_{K^*\pi} 
    $ & $ \ \ \ \
				 \mathcal{A}_{\text{FB}}^{K_1 (1270)}$  & $ \
				\ \ \ \ \mathcal{A}^{K_{1T} (1270)}_{\text{FB}} $ & $\ \ \ \ \ \  f_{L}^{K_1 (1270)} $  & $\ \ \ \ \ \  f_T^{K_1 (1270)} $ & $\hat {I}_{6c,\perp}^{K_1 (1270)} $  & \\ 
    \hline
    SM %
				&$\ \ 1.047\pm 0.147$ 
				& %
				$\ \ 0.398 \pm 0.056$
				& %
				$\ \ 0.220\pm 0.031$ 
				& %
				$\ \ -0.255\pm 0.015$
                &  %
                $\ \ -0.607 \pm 0.000$ 
                & %
				$\ \ 0.579 \pm 0.026$ 
                & %
                $\ \ 0.421\pm 0.026$
				& %
				$\ \ -0.340 \pm 0.021$ 
                & \\
    S1 %
				&$\ \ 1.070\pm 0.146$ 
				& %
				$\ \ 0.407 \pm 0.056$
				& %
				$\ \ 0.225\pm 0.031$ 
				& %
				$\ \ -0.113\pm 0.006$
                &  %
                $\ \ -0.249\pm 0.002$ 
                & %
				$\ \ 0.546 \pm 0.028$ 
                & %
                $\ \ 0.454\pm 0.028$
				& %
				$\ \ -0.150 \pm 0.008$
				&   \\
    S2 %
				&$\ \ 0.759\pm 0.105$
				& %
				$\ \ 0.288 \pm 0.040$
				& %
				$\ \ 0.159 \pm 0.022$ 
                & %
				$\ \ -0.129 \pm 0.007$
                & %
                $\ \ -0.293 \pm 0.003$ 
                & %
				$\ \ 0.559 \pm 0.027$ 
                & %
                $\ \ 0.441 \pm 0.027$ 
                & %
				$\ \ -0.172 \pm 0.009$ 
                & \\
    S3 %
				&$\ \ 0.741\pm 0.103$
				& %
				$\ \ 0.281 \pm 0.039$
				& %
				$\ \ 0.156 \pm 0.022$
				& %
				$\ \ -0.154 \pm 0.008$
                & %
                $\ \ -0.355 \pm 0.003$
                & %
				$\ \ 0.566 \pm 0.027$
                &  %
				$\ \ 0.434 \pm 0.027$ 
                & %
				$\ \ -0.206 \pm 0.011$ 
			    & \\
    S4 %
				&$\ \ 0.904\pm 0.127$
				& %
				$\ \ 0.344 \pm 0.048$
				& %
				$\ \ 0.190 \pm 0.027$
				& %
				$\ \ -0.196 \pm 0.012$
                & %
                $\ \ -0.465 \pm 0.002$
                & %
				$\ \ 0.578 \pm 0.026$
                & %
                $\ \ 0.422 \pm 0.026$
				& %
				$\ \  -0.261 \pm 0.016$ 
				& \\     
                 \hline\hline
				& $\hat
                    {I}_{1s,\parallel}^{K_1 (1270)} $ &$\hat
					{I}_{1c,\parallel}^{K_1 (1270)} $ & $ \hat{I}_{2s,\parallel}^{K_1 (1270)} $
					& $\hat{I}_{2c,\parallel}^{K_1 (1270)} $ & $\hat
					{I}_{3,\parallel}^{K_1 (1270)} $ &$\hat
					{I}_{4,\parallel}^{K_1 (1270)} $ &$\hat
					{I}_{5,\parallel}^{K_1 (1270)} $  & $\hat {I}_{6s,\parallel}^{K_1 (1270)} $ &
			  \\ 
				\hline
    SM %
				& $\ \ 0.315 \pm 0.020$
                & %
				$\ \ 0.581 \pm 0.026$
                & %
			$\ \ 0.105 \pm 0.007$
				& %
				$\ \ -0.574\pm 0.026$
                & %
				$\ \ -0.084\pm 0.005$
				& %
				$\ \ 0.286 \pm 0.002$
                & %
				$\ \ -0.358 \pm 0.004$
                & %
				$\ \ -0.340 \pm 0.021$
                & %
			  \\
    S1 %
				&$\ \ 0.340 \pm 0.021$
                & %
				$\ \ 0.548 \pm 0.028$
                & %
				$\ \ 0.113 \pm 0.007$
				& %
				$\ \ -0.542\pm 0.027$
                & %
				$\ \ -0.082\pm 0.004$
				& %
				$\ \ 0.284 \pm 0.001$
                & %
				$\ \ -0.201 \pm 0.001$
                & %
				$\ \ -0.150 \pm 0.008$
                & %
			  \\ 
    S2 %
                &$\ \ 0.331 \pm 0.020$ 
                & %
				$\ \ 0.561 \pm 0.027$
                & %
				$\ \ 0.110 \pm 0.007$ 
                & %
				$\ \ -0.555 \pm 0.027$
                & %
				$\ \ -0.082 \pm 0.004$ 
                & %
				$\ \ 0.283 \pm 0.001$ 
                & %
				$\ \ -0.231 \pm 0.002$
                & %
				$\ \ -0.172 \pm 0.009$ 
                & %
                \\
    S3 %
                &$\ \ 0.326 \pm 0.020$
                & %
			$\ \ 0.567 \pm 0.027$
                & %
				$\ \ 0.108 \pm 0.007$ 
                & %
				$\ \ -0.561 \pm 0.027$ 
                & %
				$\ \ -0.082 \pm 0.004$ 
			    & %
				$\ \ 0.283 \pm 0.001$
                & %
				$\ \ -0.260 \pm 0.002$
                & %
				$\ \ -0.206 \pm 0.011$
                & %
                \\
    S4 %
                &$\ \ 0.316 \pm 0.020$
                & %
				$\ \ 0.580 \pm 0.027$
                & %
				$\ \ 0.105 \pm 0.007$
                & %
				$\ \ -0.573 \pm 0.026$
                & %
				$\ \ -0.092 \pm 0.005$ 
				& %
				$\ \ 0.289 \pm 0.002$
                & %
				$\ \  -0.295 \pm 0.003$
                & %
				$\ \  -0.261 \pm 0.016$
                & %
				\\     
                 \hline\hline
				& $\hat
                    {I}_{1s,\perp}^{K_1 (1270)} $ &$\hat
					{I}_{1c,\perp}^{K_1 (1270)} $ & $\hat{I}_{2s,\perp}^{K_1 (1270)} $
					& $\hat{I}_{2c,\perp}^{K_1 (1270)} $ & $\hat
					{I}_{3,\perp}^{K_1 (1270)} $ &$\hat
					{I}_{4,\perp}^{K_1 (1270)} $ &$\hat
					{I}_{5,\perp}^{K_1 (1270)} $  & $\hat {I}_{6s,\perp}^{K_1 (1270)} $ & 
			  \\ 
				\hline
    SM %
				&$\ \ 0.448 \pm 0.003$
                & %
				$\ \ 0.315 \pm 0.020$
                & %
			$\ \ -0.235 \pm 0.016$
				& %
				$\ \ 0.105 \pm 0.007$
                & %
				$\ \ 0.042 \pm 0.002 $
				& %
				$\ \ -0.143 \pm 0.001$
                & %
				$\ \ 0.179 \pm 0.002$
                & %
				$\ \ -0.170 \pm 0.011$
                & %
                
			  \\
    S1 %
				&$\ \ 0.444 \pm 0.003$
                & %
				$\ \ 0.340 \pm 0.021$
                & %
				$\ \ -0.214 \pm 0.017$
				& %
				$\ \ 0.113 \pm 0.007$
                & %
				$\ \ 0.041\pm 0.002 $
				& %
				$\ \ -0.142 \pm 0.000$
                & %
				$\ \ 0.100 \pm 0.001$
                & %
				$\ \ -0.075 \pm 0.004$
                & %
                
			  \\ 
    S2 %
                &$\ \ 0.446 \pm 0.003$ 
                & %
				$\ \ 0.331 \pm 0.020$
                & %
				$\ \ -0.222 \pm 0.017$ 
                & %
				$\ \ 0.110 \pm 0.007$
                & %
				$\ \ 0.041 \pm 0.002$ 
                & %
				$\ \ -0.141 \pm 0.001$ 
                & %
				$\ \ 0.116 \pm 0.001$
                & %
				$\ \ -0.086 \pm 0.004$ 
                & %
               
                \\
    S3 %
                &$\ \ 0.446 \pm 0.003$
                & %
			$\ \ 0.326 \pm 0.020$
                & %
				$\ \ -0.226 \pm 0.017$ 
                & %
				$\ \ 0.108 \pm 0.007$ 
                & %
				$\ \ 0.041 \pm 0.002$ 
			    & %
				$\ \ -0.141 \pm 0.001$
                & %
				$\ \ 0.130 \pm 0.001$
                & %
				$\ \ -0.103 \pm 0.006$
                & %
                
                \\
    S4 %
                &$\ \ 0.448 \pm 0.003$
                & %
				$\ \ 0.316 \pm 0.020$
                & %
				$\ \ -0.234 \pm 0.016 $
                & %
				$\ \ 0.105 \pm 0.007$
                & %
				$\ \ 0.046 \pm 0.003$ 
				& %
				$\ \ -0.145 \pm 0.001$
                & %
				$\ \ 0.147 \pm 0.002$
                & %
				$\ \ -0.131 \pm 0.008$
                & %
                
				\\                    
               \hline \hline
			\end{tabular}
   }%
	\end{table}

\begin{table}
		\caption{\small Same as in Table \ref{Bin1-analysis}, but for the $q^2=[14.0-15.0]$ GeV$^2$ bin.}\label{Bin5-analysis}
		\scalebox{0.87}{
  \begin{tabular}{|c|ccccccccc|}
    \hline\hline
        \multicolumn{10}{|c|}{$q^2=[14.0-15.0]$ GeV$^2$}\\
        \hline
				\hline 
				\ \ \  & $\left\langle d\mathcal{B}/dq^2\right\rangle 
    $  & $\left\langle d\mathcal{B}/dq^2\right\rangle_{\rho K} 
    $ & $\ \ \left\langle d\mathcal{B}/dq^2\right\rangle_{K^*\pi} 
    $ & $ \ \ \ \
				 \mathcal{A}_{\text{FB}}^{K_1 (1270)}$  & $ \
				\ \ \ \ \mathcal{A}^{K_{1T} (1270)}_{\text{FB}} $ & $\ \ \ \ \ \  f_{L}^{K_1 (1270)} $  & $\ \ \ \ \ \  f_T^{K_1 (1270)} $ & $\hat {I}_{6c,\perp}^{K_1 (1270)} $ & \\ 
    \hline
    SM %
				&$\ \ 0.776\pm 0.080$ 
				& %
				$\ \ 0.295 \pm 0.031$
				& %
				$\ \ 0.163\pm 0.017$ 
				& %
				$\ \ -0.282\pm 0.001$
                &  %
                $\ \ -0.442 \pm 0.002$ 
                & %
				$\ \ 0.362 \pm 0.005$ 
                & %
                $\ \ 0.638\pm 0.005$
				& %
				$\ \ -0.376 \pm 0.001$
                & \\
    S1 %
				&$\ \ 0.750\pm 0.079$ 
				& %
				$\ \ 0.285 \pm 0.030$
				& %
				$\ \ 0.158\pm 0.016$ 
				& %
				$\ \ -0.199\pm 0.001$
                &  %
                $\ \ -0.311\pm 0.001$ 
                & %
				$\ \ 0.359 \pm 0.005$ 
                & %
                $\ \ 0.641\pm 0.005$
				& %
				$\ \ -0.266 \pm 0.001$
				&   \\
    S2 %
				&$\ \ 0.545\pm 0.057$
				& %
				$\ \ 0.207 \pm 0.022$
				& %
				$\ \ 0.114 \pm 0.012$ 
                & %
				$\ \ -0.224 \pm 0.000$
                & %
                $\ \ -0.351 \pm 0.002$ 
                & %
				$\ \ 0.361 \pm 0.005$ 
                & %
                $\ \ 0.639 \pm 0.005$ 
                & %
				$\ \ -0.299 \pm 0.001$ 
                & \\
    S3 %
				&$\ \ 0.538\pm 0.056$
				& %
				$\ \ 0.205 \pm 0.021$
				& %
				$\ \ 0.113 \pm 0.012$
				& %
				$\ \ -0.376 \pm 0.002$
                & %
                $\ \ -0.355 \pm 0.003$
                & %
				$\ \ 0.361 \pm 0.005$
                &  %
				$\ \ 0.639 \pm 0.005$ 
                & %
				$\ \ -0.320 \pm 0.001$
			    & \\
    S4 %
				&$\ \ 0.666\pm 0.069$
				& %
				$\ \ 0.253 \pm 0.026$
				& %
				$\ \ 0.140 \pm 0.015$
				& %
				$\ \ -0.252 \pm 0.001$
                & %
                $\ \ -0.396 \pm 0.002$
                & %
				$\ \ 0.364 \pm 0.005$
                & %
                $\ \ 0.636 \pm 0.005$
				& %
				$\ \  -0.336 \pm 0.001$
				& \\     
                 \hline\hline
				& $\hat
                    {I}_{1s,\parallel}^{K_1 (1270)} $ &$\hat
					{I}_{1c,\parallel}^{K_1 (1270)} $ & $ \hat{I}_{2s,\parallel}^{K_1 (1270)} $
					& $\hat{I}_{2c,\parallel}^{K_1 (1270)} $ & $\hat
					{I}_{3,\parallel}^{K_1 (1270)} $ &$\hat
					{I}_{4,\parallel}^{K_1 (1270)} $ &$\hat
					{I}_{5,\parallel}^{K_1 (1270)} $  & $\hat {I}_{6s,\parallel}^{K_1 (1270)} $ &
			  \\ 
				\hline
    SM %
				& $\ \ 0.478 \pm 0.003$
                & %
				$\ \ 0.362 \pm 0.005$
                & %
			$\ \ 0.159 \pm 0.001$
				& %
				$\ \ -0.361\pm 0.005$
                & %
				$\ \ -0.254\pm 0.002$
				& %
				$\ \ 0.321 \pm 0.001$
                & %
				$\ \ -0.212 \pm 0.002$
                & %
				$\ \ -0.376 \pm 0.001$
                & %
			  \\
    S1 %
				&$\ \ 0.481 \pm 0.004$
                & %
				$\ \ 0.359 \pm 0.005$
                & %
				$\ \ 0.160 \pm 0.001$
				& %
				$\ \ -0.358\pm 0.005$
                & %
				$\ \ -0.254\pm 0.002$
				& %
				$\ \ 0.321 \pm 0.001$
                & %
				$\ \ -0.151 \pm 0.001$
                & %
				$\ \ -0.266 \pm 0.001$
                & %
			  \\ 
    S2 %
                &$\ \ 0.479 \pm 0.003$ 
                & %
				$\ \ 0.361 \pm 0.005$
                & %
				$\ \ 0.160 \pm 0.001$ 
                & %
				$\ \ -0.360 \pm 0.005$
                & %
				$\ \ -0.254 \pm 0.002$ 
                & %
				$\ \ 0.321 \pm 0.001$ 
                & %
				$\ \ -0.170 \pm 0.001$
                & %
				$\ \ -0.299 \pm 0.001$ 
                & %
                \\
    S3 %
                &$\ \ 0.479 \pm 0.003$
                & %
			$\ \ 0.362 \pm 0.005$
                & %
				$\ \ 0.159 \pm 0.001$ 
                & %
				$\ \ -0.360 \pm 0.005$ 
                & %
				$\ \ -0.254 \pm 0.002$ 
			    & %
				$\ \ 0.321 \pm 0.001$
                & %
				$\ \ -0.182 \pm 0.002$
                & %
				$\ \ -0.320 \pm 0.001$
                & %
                \\
    S4 %
                &$\ \ 0.477 \pm 0.003$
                & %
				$\ \ 0.364 \pm 0.005$
                & %
				$\ \ 0.159 \pm 0.001$
                & %
				$\ \ -0.363 \pm 0.005$
                & %
				$\ \ -0.259 \pm 0.002$ 
				& %
				$\ \ 0.323 \pm 0.001$
                & %
				$\ \  -0.190 \pm 0.002$
                & %
				$\ \  -0.336 \pm 0.001$
                & %
				\\     
                 \hline\hline
				& $\hat
                    {I}_{1s,\perp}^{K_1 (1270)} $ &$\hat
					{I}_{1c,\perp}^{K_1 (1270)} $ & $\hat{I}_{2s,\perp}^{K_1 (1270)} $
					& $\hat{I}_{2c,\perp}^{K_1 (1270)} $ & $\hat
					{I}_{3,\perp}^{K_1 (1270)} $ &$\hat
					{I}_{4,\perp}^{K_1 (1270)} $ &$\hat
					{I}_{5,\perp}^{K_1 (1270)} $  & $\hat {I}_{6s,\perp}^{K_1 (1270)} $ & 
			  \\ 
				\hline
    SM %
				&$\ \ 0.420 \pm 0.001$
                & %
				$\ \ 0.478 \pm 0.003$
                & %
			$\ \ -0.101 \pm 0.003$
				& %
				$\ \ 0.159 \pm 0.001$
                & %
				$\ \ 0.127 \pm 0.001 $
				& %
				$\ \ -0.161 \pm 0.000$
                & %
				$\ \ 0.106 \pm 0.001$
                & %
				$\ \ -0.188 \pm 0.000$
                & %
                
			  \\
    S1 %
				&$\ \ 0.420 \pm 0.001$
                & %
				$\ \ 0.481 \pm 0.004$
                & %
				$\ \ -0.099 \pm 0.003$
				& %
				$\ \ 0.160 \pm 0.001$
                & %
				$\ \ 0.127\pm 0.001$
				& %
				$\ \ -0.160 \pm 0.000$
                & %
				$\ \ 0.076 \pm 0.001$
                & %
				$\ \ -0.133 \pm 0.000$
                & %
               
			  \\ 
    S2 %
                &$\ \ 0.420 \pm 0.001$ 
                & %
				$\ \ 0.479 \pm 0.003$
                & %
				$\ \ -0.100 \pm 0.003$ 
                & %
				$\ \ 0.160 \pm 0.001$
                & %
				$\ \ 0.127 \pm 0.001$ 
                & %
				$\ \ -0.161 \pm 0.000$ 
                & %
				$\ \ 0.085 \pm 0.001$
                & %
				$\ \ -0.150 \pm 0.000$ 
                & %
               
                \\
    S3 %
                &$\ \ 0.420 \pm 0.001$
                & %
			$\ \ 0.479 \pm 0.003$
                & %
				$\ \ -0.101 \pm 0.003$ 
                & %
				$\ \ 0.159 \pm 0.001$ 
                & %
				$\ \ 0.127 \pm 0.001$ 
			    & %
				$\ \ -0.161 \pm 0.000$
                & %
				$\ \ 0.091 \pm 0.001$
                & %
				$\ \ -0.160 \pm 0.000$
                & %
                
                \\
    S4 %
                &$\ \ 0.421 \pm 0.001$
                & %
				$\ \ 0.477 \pm 0.003$
                & %
				$\ \ -0.102 \pm 0.003$
                & %
				$\ \ 0.159 \pm 0.001$
                & %
				$\ \ 0.129 \pm 0.001$ 
				& %
				$\ \ -0.162 \pm 0.000$
                & %
				$\ \ 0.095 \pm 0.001$
                & %
				$\ \ -0.168 \pm 0.000$
                & %
               
				\\                    
               \hline \hline
			\end{tabular}
   }%
	\end{table}

\begin{table}
		\caption{\small Same as in Table \ref{Bin1-analysis}, but for the $q^2=[15.0-16.0]$ GeV$^2$ bin.}\label{Bin6-analysis}
		\scalebox{0.87}{
  \begin{tabular}{|c|ccccccccc|}
    \hline\hline
        \multicolumn{10}{|c|}{$q^2=[15.0-16.0]$ GeV$^2$}\\
        \hline
				\hline 
				\ \ \  & $\left\langle d\mathcal{B}/dq^2\right\rangle 
    $  & $\left\langle d\mathcal{B}/dq^2\right\rangle_{\rho K} 
    $ & $\ \ \left\langle d\mathcal{B}/dq^2\right\rangle_{K^*\pi} 
    $ & $ \ \ \ \
				 \mathcal{A}_{\text{FB}}^{K_1 (1270)}$  & $ \
				\ \ \ \ \mathcal{A}^{K_{1T} (1270)}_{\text{FB}} $ & $\ \ \ \ \ \  f_{L}^{K_1 (1270)} $  & $\ \ \ \ \ \  f_T^{K_1 (1270)} $ & $\hat {I}_{6c,\perp}^{K_1 (1270)} $ & \\ 
    \hline
    SM %
				&$\ \ 0.452\pm 0.046$ 
				& %
				$\ \ 0.172 \pm 0.017$
				& %
				$\ \ 0.095\pm 0.010$ 
				& %
				$\ \ -0.182\pm 0.001$
                &  %
                $\ \ -0.277 \pm 0.002$ 
                & %
				$\ \ 0.343 \pm 0.002$ 
                & %
                $\ \ 0.657\pm 0.002$
				& %
				$\ \ -0.242 \pm 0.001$ 
                & \\
    S1 %
				&$\ \ 0.438\pm 0.045$ 
				& %
				$\ \ 0.166 \pm 0.017$
				& %
				$\ \ 0.092\pm 0.009$ 
				& %
				$\ \ -0.128\pm 0.000$
                &  %
                $\ \ -0.195\pm 0.001$ 
                & %
				$\ \ 0.342 \pm 0.002$ 
                & %
                $\ \ 0.658\pm 0.002$
				& %
				$\ \ -0.171 \pm 0.000$
				&   \\
    S2 %
				&$\ \ 0.318\pm 0.032$
				& %
				$\ \ 0.121 \pm 0.012$
				& %
				$\ \ 0.067 \pm 0.007$ 
                & %
				$\ \ -0.145 \pm 0.001$
                & %
                $\ \ -0.220 \pm 0.001$ 
                & %
				$\ \ 0.343 \pm 0.002$ 
                & %
                $\ \ 0.657 \pm 0.002$ 
                & %
				$\ \ -0.193 \pm 0.001$ 
                & \\
    S3 %
				&$\ \ 0.314\pm 0.032$
				& %
				$\ \ 0.119 \pm 0.012$
				& %
				$\ \ 0.066 \pm 0.007$
				& %
				$\ \ -0.155 \pm 0.001$
                & %
                $\ \ -0.236 \pm 0.002$
                & %
				$\ \ 0.343 \pm 0.002$
                &  %
				$\ \ 0.657 \pm 0.002$ 
                & %
				$\ \ -0.206 \pm 0.001$ 
			    & \\
    S4 %
				&$\ \ 0.390\pm 0.039$
				& %
				$\ \ 0.148 \pm 0.015$
				& %
				$\ \ 0.082 \pm 0.008$
				& %
				$\ \ -0.162 \pm 0.001$
                & %
                $\ \ -0.247 \pm 0.002$
                & %
				$\ \ 0.344 \pm 0.002$
                & %
                $\ \ 0.656 \pm 0.002$
				& %
				$\ \  -0.216 \pm 0.001$
				& \\     
                 \hline\hline
				& $\hat
                    {I}_{1s,\parallel}^{K_1 (1270)} $ &$\hat
					{I}_{1c,\parallel}^{K_1 (1270)} $ & $ \hat{I}_{2s,\parallel}^{K_1 (1270)} $
					& $\hat{I}_{2c,\parallel}^{K_1 (1270)} $ & $\hat
					{I}_{3,\parallel}^{K_1 (1270)} $ &$\hat
					{I}_{4,\parallel}^{K_1 (1270)} $ &$\hat
					{I}_{5,\parallel}^{K_1 (1270)} $  & $\hat {I}_{6s,\parallel}^{K_1 (1270)} $ &
			  \\ 
				\hline
    SM %
				& $\ \ 0.492 \pm 0.001$
                & %
				$\ \ 0.344 \pm 0.002$
                & %
			$\ \ 0.164 \pm 0.000$
				& %
				$\ \ -0.343\pm 0.002$
                & %
				$\ \ -0.301\pm 0.001$
				& %
				$\ \ 0.328 \pm 0.000$
                & %
				$\ \ -0.127 \pm 0.001$
                & %
				$\ \ -0.242 \pm 0.001$
                & %
			  \\
    S1 %
				&$\ \ 0.493 \pm 0.001$
                & %
				$\ \ 0.343 \pm 0.002$
                & %
				$\ \ 0.164 \pm 0.000$
				& %
				$\ \ -0.342\pm 0.002$
                & %
				$\ \ -0.302\pm 0.001$
				& %
				$\ \ 0.328 \pm 0.000$
                & %
				$\ \ -0.090 \pm 0.001$
                & %
				$\ \ -0.171 \pm 0.000$
                & %
			  \\ 
    S2 %
                &$\ \ 0.493 \pm 0.001$ 
                & %
				$\ \ 0.343 \pm 0.002$
                & %
				$\ \ 0.164 \pm 0.000$ 
                & %
				$\ \ -0.343 \pm 0.002$
                & %
				$\ \ -0.302 \pm 0.001$ 
                & %
				$\ \ 0.328 \pm 0.000$ 
                & %
				$\ \ -0.102 \pm 0.001$
                & %
				$\ \ -0.193 \pm 0.001$ 
                & %
                \\
    S3 %
                &$\ \ 0.493 \pm 0.001$
                & %
			$\ \ 0.343 \pm 0.002$
                & %
				$\ \ 0.164 \pm 0.000$ 
                & %
				$\ \ -0.343 \pm 0.002$ 
                & %
				$\ \ -0.302 \pm 0.001$ 
			    & %
				$\ \ 0.328 \pm 0.000$
                & %
				$\ \ -0.109 \pm 0.001$
                & %
				$\ \ -0.206 \pm 0.001$
                & %
                \\
    S4 %
                &$\ \ 0.492 \pm 0.001$
                & %
				$\ \ 0.344 \pm 0.002$
                & %
				$\ \ 0.164 \pm 0.000$
                & %
				$\ \ -0.344 \pm 0.002$
                & %
				$\ \ -0.304 \pm 0.001$ 
				& %
				$\ \ 0.329 \pm 0.000$
                & %
				$\ \  -0.114 \pm 0.001$
                & %
				$\ \  -0.216 \pm 0.001$
                & %
				\\     
                 \hline\hline
				& $\hat
                    {I}_{1s,\perp}^{K_1 (1270)} $ &$\hat
					{I}_{1c,\perp}^{K_1 (1270)} $ & $\hat{I}_{2s,\perp}^{K_1 (1270)} $
					& $\hat{I}_{2c,\perp}^{K_1 (1270)} $ & $\hat
					{I}_{3,\perp}^{K_1 (1270)} $ &$\hat
					{I}_{4,\perp}^{K_1 (1270)} $ &$\hat
					{I}_{5,\perp}^{K_1 (1270)} $  & $\hat {I}_{6s,\perp}^{K_1 (1270)} $ & 
			  \\ 
				\hline
    SM %
				&$\ \ 0.418 \pm 0.000$
                & %
				$\ \ 0.492 \pm 0.001$
                & %
			$\ \ -0.089 \pm 0.001$
				& %
				$\ \ 0.164 \pm 0.000$
                & %
				$\ \ 0.151 \pm 0.001$
				& %
				$\ \ -0.164 \pm 0.000$
                & %
				$\ \ 0.064 \pm 0.001$
                & %
				$\ \ -0.121 \pm 0.001$
                & %
                
			  \\
    S1 %
				&$\ \ 0.418 \pm 0.000$
                & %
				$\ \ 0.493 \pm 0.001$
                & %
				$\ \ -0.089 \pm 0.001$
				& %
				$\ \ 0.164 \pm 0.000$
                & %
				$\ \ 0.151\pm 0.000$
				& %
				$\ \ -0.164 \pm 0.000$
                & %
				$\ \ 0.045 \pm 0.000$
                & %
				$\ \ -0.086 \pm 0.000$
                & %
                
			  \\ 
    S2 %
                &$\ \ 0.418 \pm 0.000$ 
                & %
				$\ \ 0.493 \pm 0.001$
                & %
				$\ \ -0.089 \pm 0.001$ 
                & %
				$\ \ 0.164 \pm 0.000$
                & %
				$\ \ 0.151 \pm 0.001$ 
                & %
				$\ \ -0.164 \pm 0.000$ 
                & %
				$\ \ 0.051 \pm 0.000$
                & %
				$\ \ -0.096 \pm 0.000$ 
                & %
                
                \\
    S3 %
                &$\ \ 0.418 \pm 0.000$
                & %
			$\ \ 0.493 \pm 0.001$
                & %
				$\ \ -0.089 \pm 0.001 $ 
                & %
				$\ \ 0.164 \pm 0.000$ 
                & %
				$\ \ 0.151 \pm 0.001$ 
			    & %
				$\ \ -0.164 \pm 0.000$
                & %
				$\ \ 0.054 \pm 0.000$
                & %
				$\ \ -0.103 \pm 0.000$
                & %
                
                \\
    S4 %
                &$\ \ 0.418 \pm 0.000$
                & %
				$\ \ 0.492 \pm 0.001$
                & %
				$\ \ -0.090 \pm 0.001 $
                & %
				$\ \ 0.164 \pm 0.000$
                & %
				$\ \ 0.152 \pm 0.001$ 
				& %
				$\ \ -0.165 \pm 0.000$
                & %
				$\ \ 0.057 \pm 0.000$
                & %
				$\ \ -0.108 \pm 0.000$
                & %
                
				\\                    
               \hline \hline
			\end{tabular}
   }%
	\end{table}

\begin{table}
\caption{\small Predictions of binned observables for the $B \to K_1(1400)\left(\to VP\right)\mu^+ \mu^-$ decay, including differential branching ratios, $\left\langle d\mathcal{B}/dq^2\right\rangle= \left\langle d\mathcal{B}\left(B\to K_1(1400)\mu^+\mu^-\right)/dq^2\right\rangle$, $\left\langle d\mathcal{B}/dq^2\right\rangle_{\rho K}=\left\langle d\mathcal{B}\left(B\to K_1(1400)\,(\to \rho K)\mu^+\mu^-\right)/dq^2\right\rangle$, and $\left\langle d\mathcal{B}/dq^2\right\rangle_{K^{\ast}\pi}=\left\langle d\mathcal{B}\left(B\to K_1(1400)\,(\to K^{\ast} \pi)\mu^+\mu^-\right)/dq^2\right\rangle$. The other observables include the lepton FB asymmetry $\mathcal{A}_{\text{FB}}^{K_1 (1400)}$, FB asymmetry for transversely polarized $K_1(1400)$, $\mathcal{A}^{K_{1T} (1400)}_{\text{FB}}$, longitudinal helicity fraction $f_{L}^{K_1 (1400)}$, transverse helicity fraction $f_T^{K_1 (1400)}$, and angular coefficients $\hat{I}_{\parallel(\perp)}^{K_1 (1400)}$ with longitudinal and transverse polarization of the vector meson, in the $q^2=[0.1-2.0]$ GeV$^2$ bin, for both the SM and NP scenarios presented in Table \ref{tab:bestfitWC}. The differential branching ratios $\left\langle d\mathcal{B}/dq^2\right\rangle$ and $\left\langle d\mathcal{B}/dq^2\right\rangle_{K^{\ast}\pi}$ are given in units of $10^{-7}$ GeV$^{-2}$, whereas $\left\langle d\mathcal{B}/dq^2\right\rangle_{\rho K}$ is given in units of $10^{-8}$ GeV$^{-2}$.}\label{Bin11400-analysis}
		\scalebox{0.87}{
    \begin{tabular}{|c|ccccccccc|}
    \hline\hline
        \multicolumn{10}{|c|}{$q^2=[0.1-2.0]$ GeV$^2$}\\
        \hline
				\hline 
				\ \ \  & $\left\langle d\mathcal{B}/dq^2\right\rangle 
    $  & $\left\langle d\mathcal{B}/dq^2\right\rangle_{\rho K} 
    $ & $\ \ \left\langle d\mathcal{B}/dq^2\right\rangle_{K^*\pi} 
    $ & $ \ \ \ \
				 \mathcal{A}_{\text{FB}}^{K_1 (1400)}$  & $ \
				\ \ \ \ \mathcal{A}^{K_{1T} (1400)}_{\text{FB}} $ & $\ \ \ \ \ \  f_{L}^{K_1 (1400)} $  & $\ \ \ \ \ \  f_T^{K_1 (1400)} $ & $\hat {I}_{6c,\perp}^{K_1 (1400)} $  & \\ 
    \hline
    SM %
				&$\ \ 0.805\pm 0.122$ 
				& %
				$\ \ 0.242\pm 0.036$
				& %
				$\ \ 0.757\pm 0.114$ 
				& %
				$\ \ 0.098\pm 0.002$
                &  %
                $\ \ 0.1195 \pm 0.0004$ 
                & %
				$\ \ 0.178 \pm 0.023$ 
                & %
                $\ \ 0.822\pm 0.023$
				& %
				$\ \ 0.131 \pm 0.003$ 
                & \\
    S1 %
				&$\ \ 0.912\pm 0.142$ 
				& %
				$\ \ 0.274 \pm 0.043$
				& %
				$\ \ 0.857\pm 0.134$ 
				& %
				$\ \ 0.122\pm 0.003$
                &  %
                $\ \ 0.1480\pm 0.0005$ 
                & %
				$\ \ 0.176 \pm 0.022$ 
                & %
                $\ \ 0.824\pm 0.022$
				& %
				$\ \ 0.163 \pm 0.004$
				&   \\
    S2 %
				&$\ \ 0.804\pm 0.133$
				& %
				$\ \ 0.241 \pm 0.040$
				& %
				$\ \ 0.755 \pm 0.125$ 
                & %
				$\ \ 0.106 \pm 0.002$
                & %
                $\ \ 0.1228 \pm 0.0005$ 
                & %
				$\ \ 0.131 \pm 0.017$ 
                & %
                $\ \ 0.869 \pm 0.017$ 
                & %
				$\ \ 0.142 \pm 0.002$ 
                & \\
    S3 %
				&$\ \ 0.786\pm 0.130$
				& %
				$\ \ 0.236 \pm 0.039$
				& %
				$\ \ 0.738 \pm 0.122$
				& %
				$\ \ 0.103 \pm 0.002$
                & %
                $\ \ 0.1186 \pm 0.0004$
                & %
				$\ \ 0.128 \pm 0.017 $
                &  %
				$\ \ 0.872 \pm 0.017$ 
                & %
				$\ \ 0.138 \pm 0.002$ 
			    & \\
    S4 %
				&$\ \ 0.810\pm 0.128$
				& %
				$\ \ 0.243 \pm 0.038$
				& %
				$\ \ 0.761 \pm 0.120$
				& %
				$\ \ 0.103 \pm 0.002$
                & %
                $\ \ 0.1220 \pm 0.0004$
                & %
				$\ \ 0.157 \pm 0.020 $
                & %
                $\ \ 0.843 \pm 0.020$
				& %
				$\ \  0.137 \pm 0.003$ 
				& \\     
                 \hline\hline
				& $\hat
                    {I}_{1s,\parallel}^{K_1 (1400)} $ &$\hat
					{I}_{1c,\parallel}^{K_1 (1400)} $ & $ \hat{I}_{2s,\parallel}^{K_1 (1400)} $
					& $\hat{I}_{2c,\parallel}^{K_1 (1400)} $ & $\hat
					{I}_{3,\parallel}^{K_1 (1400)} $ &$\hat
					{I}_{4,\parallel}^{K_1 (1400)} $ &$\hat
					{I}_{5,\parallel}^{K_1 (1400)} $  & $\hat {I}_{6s,\parallel}^{K_1 (1400)} $ &
			  \\ 
				\hline
    SM %
				& $\ \ 0.594 \pm 0.016$
                & %
				$\ \ 0.188 \pm 0.025$
                & %
			    $\ \ 0.137 \pm 0.004$
				& %
				$\ \ -0.148\pm 0.017$
                & %
				$\ \ 0.003\pm 0.000 $
				& %
				$\ \ -0.038 \pm 0.003$
                & %
				$\ \ 0.195 \pm 0.008$
                & %
				$\ \ 0.131 \pm 0.003$
                & %
			  \\
    S1 %
				&$\ \ 0.596 \pm 0.016$
                & %
				$\ \ 0.186 \pm 0.024$
                & %
				$\ \ 0.140 \pm 0.004$
				& %
				$\ \ -0.144\pm 0.016$
                & %
				$\ \ 0.001\pm 0.000 $
				& %
				$\ \ \ \ 0.005 \pm 0.002$
                & %
				$\ \ 0.250 \pm 0.010$
                & %
				$\ \ 0.163 \pm 0.004$
                & %
			  \\ 
    S2 %
                &$\ \ 0.628 \pm 0.012$ 
                & %
				$\ \ 0.138 \pm 0.018$
                & %
				$\ \ 0.147 \pm 0.003$ 
                & %
				$\ \ -0.108 \pm 0.013$
                & %
				$\ \ 0.001 \pm 0.000$ 
                & %
				$\ \ -0.004 \pm 0.002$ 
                & %
				$\ \ 0.218 \pm 0.010$
                & %
				$\ \ 0.142 \pm 0.002$ 
                & %
                \\
    S3 %
                &$\ \ 0.630 \pm 0.012$
                & %
			    $\ \ 0.136 \pm 0.018$
                & %
				$\ \ 0.147 \pm 0.003$ 
                & %
				$\ \ -0.106 \pm 0.013$ 
                & %
				$\ \ 0.001 \pm 0.000$ 
			    & %
				$\ \ -0.010 \pm 0.003$
                & %
				$\ \ 0.211 \pm 0.010$
                & %
				$\ \ 0.138 \pm 0.002$
                & %
                \\
    S4 %
                &$\ \ 0.610 \pm 0.015$
                & %
				$\ \ 0.166 \pm 0.022$
                & %
				$\ \ 0.142 \pm 0.003$
                & %
				$\ \ -0.130 \pm 0.015$
                & %
				$\ \ 0.001 \pm 0.000$ 
				& %
				$\ \ -0.019 \pm 0.003$
                & %
				$\ \  0.215 \pm 0.009$
                & %
				$\ \  0.137 \pm 0.003$
                & %
				\\     
                 \hline\hline
				& $\hat
                    {I}_{1s,\perp}^{K_1 (1400)} $ &$\hat
					{I}_{1c,\perp}^{K_1 (1400)} $ & $\hat{I}_{2s,\perp}^{K_1 (1400)} $
					& $\hat{I}_{2c,\perp}^{K_1 (1400)} $ & $\hat
					{I}_{3,\perp}^{K_1 (1400)} $ &$\hat
					{I}_{4,\perp}^{K_1 (1400)} $ &$\hat
					{I}_{5,\perp}^{K_1 (1400)} $  & $\hat {I}_{6s,\perp}^{K_1 (1400)} $ & 
			  \\ 
				\hline
    SM %
				&$\ \ 0.391 \pm 0.004$
                & %
				$\ \ 0.594 \pm 0.016$
                & %
			    $\ \ -0.018 \pm 0.010$
				& %
				$\ \ 0.137 \pm 0.004$
                & %
				$\ \ -0.002\pm 0.000$
				& %
				$\ \ \ \ 0.019 \pm 0.002$
                & %
				$\ \ -0.098 \pm 0.004$
                & %
				$\ \ 0.065 \pm 0.002$
                & %
                
			  \\
    S1 %
				&$\ \ 0.391 \pm 0.004$
                & %
				$\ \ 0.596 \pm 0.016$
                & %
				$\ \ -0.013 \pm 0.010$
				& %
				$\ \ 0.140 \pm 0.004$
                & %
				$\ \ \ \ 0.000\pm 0.000 $
				& %
				$\ \ -0.003 \pm 0.001$
                & %
				$\ \ -0.125 \pm 0.005$
                & %
                $\ \ 0.081 \pm 0.002$
                & %
				             
			  \\ 
    S2 %
                &$\ \ 0.383 \pm 0.003$ 
                & %
				$\ \ 0.628 \pm 0.012$
                & %
				$\ \ \ \ 0.020 \pm 0.008$ 
                & %
				$\ \ 0.147 \pm 0.003$
                & %
				$\ \ \ \ 0.000 \pm 0.000$ 
                & %
				$\ \ \ \ 0.002 \pm 0.001$ 
                & %
				$\ \ -0.109 \pm 0.005$
                & %
                $\ \ 0.071 \pm 0.001$
                & %
				  
                \\
    S3 %
                &$\ \ 0.383 \pm 0.003$
                & %
			$\ \ 0.630 \pm 0.012$
                & %
				$\ \ \ \ 0.021 \pm 0.008 $ 
                & %
				$\ \ 0.147 \pm 0.003$ 
                & %
				$\ \ -0.001 \pm 0.000$ 
			    & %
				$\ \ \ \ 0.005 \pm 0.001$
                & %
				$\ \ -0.105 \pm 0.005$
                & %
                $\ \ 0.069 \pm 0.001$
                & %
				
                \\
    S4 %
                &$\ \ 0.388 \pm 0.004$
                & %
				$\ \ 0.610 \pm 0.015$
                & %
				$\ \ \ \ 0.015 \pm 0.009 $
                & %
				$\ \ 0.142 \pm 0.003$
                & %
				$\ \ -0.001 \pm 0.000$ 
				& %
				$\ \ \ \ 0.010 \pm 0.001$
                & %
				$\ \ -0.108 \pm 0.005$
                & %
                $\ \ 0.069 \pm 0.001$
                & %
				
				\\                    
               \hline \hline
			\end{tabular}
   }%
	\end{table}

\begin{table}
		\caption{\small Same as in Table \ref{Bin11400-analysis}, but for the $q^2=[2.0-4.0]$ GeV$^2$ bin.}\label{Bin21400-analysis}
		\scalebox{0.87}{
    \begin{tabular}{|c|ccccccccc|}
    \hline\hline
        \multicolumn{10}{|c|}{$q^2=[2.0-4.0]$ GeV$^2$}\\
        \hline
				\hline 
				\ \ \  & $\left\langle d\mathcal{B}/dq^2\right\rangle 
    $  & $\left\langle d\mathcal{B}/dq^2\right\rangle_{\rho K} 
    $ & $\ \ \left\langle d\mathcal{B}/dq^2\right\rangle_{K^*\pi} 
    $ & $ \ \ \ \
				 \mathcal{A}_{\text{FB}}^{K_1 (1400)}$  & $ \
				\ \ \ \ \mathcal{A}^{K_{1T} (1400)}_{\text{FB}} $ & $\ \ \ \ \ \  f_{L}^{K_1 (1400)} $  & $\ \ \ \ \ \  f_T^{K_1 (1400)} $ & $\hat {I}_{6c,\perp}^{K_1 (1400)} $  & \\ 
    \hline
    SM %
				&$\ \ 0.466\pm 0.053$ 
				& %
				$\ \ 0.140\pm 0.016$
				& %
				$\ \ 0.438\pm 0.050$ 
				& %
				$\ \ 0.025\pm 0.006$
                &  %
                $\ \ 0.080 \pm 0.011$ 
                & %
				$\ \ 0.690 \pm 0.025$ 
                & %
                $\ \ 0.310\pm 0.025$
				& %
				$\ \ 0.033 \pm 0.007$ 
                & \\
    S1 %
				&$\ \ 0.546\pm 0.069$ 
				& %
				$\ \ 0.164 \pm 0.021$
				& %
				$\ \ 0.513\pm 0.065$ 
				& %
				$\ \ 0.191\pm 0.016$
                &  %
                $\ \ 0.452\pm 0.007$ 
                & %
				$\ \ 0.579 \pm 0.029$ 
                & %
                $\ \ 0.421\pm 0.029$
				& %
				$\ \ 0.254 \pm 0.022$
				&   \\
    S2 %
				&$\ \ 0.391\pm 0.050$
				& %
				$\ \ 0.117 \pm 0.015$
				& %
				$\ \ 0.368 \pm 0.047$ 
                & %
				$\ \ 0.219 \pm 0.018$
                & %
                $\ \ 0.519 \pm 0.007$ 
                & %
				$\ \ 0.578 \pm 0.030$ 
                & %
                $\ \ 0.422 \pm 0.030$ 
                & %
				$\ \ 0.292 \pm 0.025$  
                & \\
    S3 %
				&$\ \ 0.374\pm 0.047$
				& %
				$\ \ 0.112 \pm 0.014$
				& %
				$\ \ 0.352 \pm 0.044$
				& %
				$\ \ 0.199 \pm 0.018$
                & %
                $\ \ 0.490 \pm 0.008$
                & %
				$\ \ 0.594 \pm 0.029 $
                &  %
				$\ \ 0.406 \pm 0.029$ 
                & %
				$\ \ 0.265 \pm 0.023$
			    & \\
    S4 %
				&$\ \ 0.431\pm 0.051$
				& %
				$\ \ 0.129 \pm 0.015$
				& %
				$\ \ 0.405 \pm 0.048$
				& %
				$\ \ 0.120 \pm 0.013$
                & %
                $\ \ 0.337 \pm 0.010$
                & %
				$\ \ 0.645 \pm 0.027 $
                & %
                $\ \ 0.355 \pm 0.027$
				& %
				$\ \  0.160 \pm 0.017$ 
				& \\     
                 \hline\hline
				& $\hat
                    {I}_{1s,\parallel}^{K_1 (1400)} $ &$\hat
					{I}_{1c,\parallel}^{K_1 (1400)} $ & $ \hat{I}_{2s,\parallel}^{K_1 (1400)} $
					& $\hat{I}_{2c,\parallel}^{K_1 (1400)} $ & $\hat
					{I}_{3,\parallel}^{K_1 (1400)} $ &$\hat
					{I}_{4,\parallel}^{K_1 (1400)} $ &$\hat
					{I}_{5,\parallel}^{K_1 (1400)} $  & $\hat {I}_{6s,\parallel}^{K_1 (1400)} $ &
			  \\ 
				\hline
    SM %
				& $\ \ 0.232 \pm 0.019$
                & %
				$\ \ 0.696 \pm 0.025$
                & %
			    $\ \ 0.077 \pm 0.006$
				& %
				$\ \ -0.675\pm 0.024$
                & %
				$\ \ -0.039\pm 0.003$
				& %
				$\ \ 0.201 \pm 0.003$
                & %
				$\ \ -0.178 \pm 0.002$
                & %
				$\ \ 0.033 \pm 0.007$
                & %
			  \\
    S1 %
				&$\ \ 0.316 \pm 0.022$
                & %
				$\ \ 0.583 \pm 0.030$
                & %
				$\ \ 0.105 \pm 0.007$
				& %
				$\ \ -0.566\pm 0.028$
                & %
				$\ \ -0.042\pm 0.002$
				& %
				$\ \ 0.214 \pm 0.001$
                & %
				$\ \ \ \ 0.039 \pm 0.003$
                & %
				$\ \ 0.254 \pm 0.022$
                & %
			  \\ 
    S2 %
                &$\ \ 0.316 \pm 0.022$ 
                & %
				$\ \ 0.582 \pm 0.030$
                & %
				$\ \ 0.105 \pm 0.007$ 
                & %
				$\ \ -0.565 \pm 0.029$
                & %
				$\ \ -0.035 \pm 0.002$ 
                & %
				$\ \ 0.191 \pm 0.000$ 
                & %
				$\ \ \ \ 0.047 \pm 0.003$
                & %
				$\ \ 0.292 \pm 0.025$ 
                & %
                \\
    S3 %
                &$\ \ 0.304 \pm 0.022$
                & %
			    $\ \ 0.599 \pm 0.030$
                & %
				$\ \ 0.101 \pm 0.007$ 
                & %
				$\ \ -0.581 \pm 0.028$ 
                & %
				$\ \ -0.034 \pm 0.002$ 
			    & %
				$\ \ 0.186 \pm 0.001$
                & %
				$\ \ \ \ 0.016 \pm 0.003$
                & %
				$\ \ 0.265 \pm 0.023$
                & %
                \\
    S4 %
                &$\ \ 0.266 \pm 0.021$
                & %
				$\ \ 0.650 \pm 0.028$
                & %
				$\ \ 0.088 \pm 0.007$
                & %
				$\ \ -0.631 \pm 0.026$
                & %
				$\ \ -0.046 \pm 0.003$ 
				& %
				$\ \ 0.201 \pm 0.002$
                & %
				$\ \  -0.057 \pm 0.002$
                & %
				$\ \  0.160 \pm 0.017$
                & %
				\\     
                 \hline\hline
				& $\hat
                    {I}_{1s,\perp}^{K_1 (1400)} $ &$\hat
					{I}_{1c,\perp}^{K_1 (1400)} $ & $\hat{I}_{2s,\perp}^{K_1 (1400)} $
					& $\hat{I}_{2c,\perp}^{K_1 (1400)} $ & $\hat
					{I}_{3,\perp}^{K_1 (1400)} $ &$\hat
					{I}_{4,\perp}^{K_1 (1400)} $ &$\hat
					{I}_{5,\perp}^{K_1 (1400)} $  & $\hat {I}_{6s,\perp}^{K_1 (1400)} $ & 
			  \\ 
				\hline
    SM %
				&$\ \ 0.464 \pm 0.003$
                & %
				$\ \ 0.232 \pm 0.019$
                & %
			    $\ \ -0.299 \pm 0.015$
				& %
				$\ \ 0.077 \pm 0.006$
                & %
				$\ \ 0.020\pm 0.001$
				& %
				$\ \ -0.100 \pm 0.002$
                & %
				$\ \ \ \ 0.089\pm 0.001$
                & %
				$\ \ 0.017 \pm 0.004$
                & %
                
			  \\
    S1 %
				&$\ \ 0.449 \pm 0.004$
                & %
				$\ \ 0.316 \pm 0.022$
                & %
				$\ \ -0.230 \pm 0.018$
				& %
				$\ \ 0.105 \pm 0.007$
                & %
				$\ \ 0.021\pm 0.001 $
				& %
				$\ \ -0.107 \pm 0.000$
                & %
				$\ \ -0.020 \pm 0.001$
                & %
                $\ \ 0.127 \pm 0.011$
                & %
				             
			  \\ 
    S2 %
                &$\ \ 0.449 \pm 0.004$ 
                & %
				$\ \ 0.316 \pm 0.022$
                & %
				$\ \ -0.230 \pm 0.018$ 
                & %
				$\ \ 0.105 \pm 0.007$
                & %
				$\ \ 0.018 \pm 0.001$ 
                & %
				$\ \ -0.095 \pm 0.000$ 
                & %
				$\ \ -0.023 \pm 0.002$
                & %
                $\ \ 0.146 \pm 0.012$
                & %
				  
                \\
    S3 %
                &$\ \ 0.451 \pm 0.004$
                & %
			$\ \ 0.304 \pm 0.022$
                & %
				$\ \ -0.240 \pm 0.018 $ 
                & %
				$\ \ 0.101 \pm 0.007$ 
                & %
				$\ \ 0.017 \pm 0.001$ 
			    & %
				$\ \ -0.093 \pm 0.000$
                & %
				$\ \ -0.008 \pm 0.002$
                & %
                $\ \ 0.133 \pm 0.012$
                & %
				
                \\
    S4 %
                &$\ \ 0.458 \pm 0.004$
                & %
				$\ \ 0.266 \pm 0.021$
                & %
				$\ \ -0.271 \pm 0.017 $
                & %
				$\ \ 0.088 \pm 0.007$
                & %
				$\ \ 0.023 \pm 0.001$ 
				& %
				$\ \ -0.101 \pm 0.001$
                & %
				$\ \ \ \ 0.028 \pm 0.001$
                & %
                $\ \ 0.080 \pm 0.008$
                & %
				
				\\                    
               \hline \hline
			\end{tabular}
   }%
	\end{table}

\begin{table}
		\caption{\small Same as in Table \ref{Bin11400-analysis}, but for the $q^2=[4.0-6.0]$ GeV$^2$ bin.}\label{Bin31400-analysis}
		\scalebox{0.87}{
    \begin{tabular}{|c|ccccccccc|}
    \hline\hline
        \multicolumn{10}{|c|}{$q^2=[4.0-6.0]$ GeV$^2$}\\
        \hline
				\hline 
				\ \ \  & $\left\langle d\mathcal{B}/dq^2\right\rangle 
    $  & $\left\langle d\mathcal{B}/dq^2\right\rangle_{\rho K} 
    $ & $\ \ \left\langle d\mathcal{B}/dq^2\right\rangle_{K^*\pi} 
    $ & $ \ \ \ \
				 \mathcal{A}_{\text{FB}}^{K_1 (1400)}$  & $ \
				\ \ \ \ \mathcal{A}^{K_{1T} (1400)}_{\text{FB}} $ & $\ \ \ \ \ \  f_{L}^{K_1 (1400)} $  & $\ \ \ \ \ \  f_T^{K_1 (1400)} $ & $\hat {I}_{6c,\perp}^{K_1 (1400)} $  & \\ 
    \hline
    SM %
				&$\ \ 0.595\pm 0.079$ 
				& %
				$\ \ 0.178\pm 0.024$
				& %
				$\ \ 0.559\pm 0.074$ 
				& %
				$\ \ -0.209\pm 0.009$
                &  %
                $\ \ -0.470 \pm 0.006$ 
                & %
				$\ \ 0.555 \pm 0.024$ 
                & %
                $\ \ 0.445\pm 0.024$
				& %
				$\ \ -0.278 \pm 0.012$ 
                & \\
    S1 %
				&$\ \ 0.654\pm 0.091$ 
				& %
				$\ \ 0.196 \pm 0.027$
				& %
				$\ \ 0.615\pm 0.086$ 
				& %
				$\ \ -0.008\pm 0.003$
                &  %
                $\ \ -0.017 \pm 0.008$ 
                & %
				$\ \ 0.494 \pm 0.025$ 
                & %
                $\ \ 0.506\pm 0.025$
				& %
				$\ \ -0.011 \pm 0.004$
				&   \\
    S2 %
				&$\ \ 0.457\pm 0.063$
				& %
				$\ \ 0.137 \pm 0.019$
				& %
				$\ \ 0.429 \pm 0.059$ 
                & %
				$\ \ -0.008 \pm 0.004$
                & %
                $\ \ -0.017 \pm 0.009$ 
                & %
				$\ \ 0.509 \pm 0.025$ 
                & %
                $\ \ 0.491 \pm 0.025$ 
                & %
				$\ \ -0.011 \pm 0.005$
                & \\
    S3 %
				&$\ \ 0.440\pm 0.060$
				& %
				$\ \ 0.132 \pm 0.018$
				& %
				$\ \ 0.414 \pm 0.057$
				& %
				$\ \ -0.040 \pm 0.002$
                & %
                $\ \ -0.083 \pm 0.009$
                & %
				$\ \ 0.520 \pm 0.025 $
                &  %
				$\ \ 0.480\pm 0.025$ 
                & %
				$\ \ -0.053 \pm 0.003$ 
			    & \\
    S4 %
				&$\ \ 0.526\pm 0.070$
				& %
				$\ \ 0.158 \pm 0.021$
				& %
				$\ \ 0.495 \pm 0.066$
				& %
				$\ \ -0.114 \pm 0.002$
                & %
                $\ \ -0.249 \pm 0.008$
                & %
				$\ \ 0.543 \pm 0.025 $
                & %
                $\ \ 0.457\pm 0.025$
				& %
				$\ \  -0.152 \pm 0.003$ 
				& \\     
                 \hline\hline
				& $\hat
                    {I}_{1s,\parallel}^{K_1 (1400)} $ &$\hat
					{I}_{1c,\parallel}^{K_1 (1400)} $ & $ \hat{I}_{2s,\parallel}^{K_1 (1400)} $
					& $\hat{I}_{2c,\parallel}^{K_1 (1400)} $ & $\hat
					{I}_{3,\parallel}^{K_1 (1400)} $ &$\hat
					{I}_{4,\parallel}^{K_1 (1400)} $ &$\hat
					{I}_{5,\parallel}^{K_1 (1400)} $  & $\hat {I}_{6s,\parallel}^{K_1 (1400)} $ &
			  \\ 
				\hline
    SM %
				& $\ \ 0.334 \pm 0.018$
                & %
				$\ \ 0.557 \pm 0.024$
                & %
			    $\ \ 0.111 \pm 0.006$
				& %
				$\ \ -0.548\pm 0.024$
                & %
				$\ \ -0.083\pm 0.004$
				& %
				$\ \ 0.278 \pm 0.001$
                & %
				$\ \ -0.327 \pm 0.000$
                & %
				$\ \ -0.278 \pm 0.012$
                & %
			  \\
    S1 %
				&$\ \ 0.379 \pm 0.019$
                & %
				$\ \ 0.496 \pm 0.026$
                & %
				$\ \ 0.126 \pm 0.006$
				& %
				$\ \ -0.489\pm 0.025$
                & %
				$\ \ -0.080\pm 0.004 $
				& %
				$\ \ 0.272 \pm 0.001$
                & %
				$\ \ -0.125 \pm 0.002$
                & %
				$\ \ -0.011 \pm 0.004$
                & %
			  \\ 
    S2 %
                &$\ \ 0.368 \pm 0.019$ 
                & %
				$\ \ 0.511 \pm 0.026$
                & %
				$\ \ 0.123 \pm 0.006$ 
                & %
				$\ \ -0.503 \pm 0.025$
                & %
				$\ \ -0.078 \pm 0.004$ 
                & %
				$\ \ 0.267 \pm 0.000$ 
                & %
				$\ \ -0.146 \pm 0.002$
                & %
				$\ \ -0.011 \pm 0.005$ 
                & %
                \\
    S3 %
                &$\ \ 0.360 \pm 0.019$
                & %
			    $\ \ 0.522 \pm 0.025$
                & %
				$\ \ 0.120 \pm 0.006$ 
                & %
				$\ \ -0.514 \pm 0.025$ 
                & %
				$\ \ -0.078 \pm 0.004$ 
			    & %
				$\ \ 0.267\pm 0.000$
                & %
				$\ \ -0.180 \pm 0.002$
                & %
				$\ \ -0.053 \pm 0.003$
                & %
                \\
    S4 %
                &$\ \ 0.343 \pm 0.018$
                & %
				$\ \ 0.545 \pm 0.025$
                & %
				$\ \ 0.114 \pm 0.006$
                & %
				$\ \ -0.536 \pm 0.024$
                & %
				$\ \ -0.093 \pm 0.005$ 
				& %
				$\ \ 0.279 \pm 0.001$
                & %
				$\ \  -0.235 \pm 0.001$
                & %
				$\ \  -0.152 \pm 0.003$
                & %
				\\     
                 \hline\hline
				& $\hat
                    {I}_{1s,\perp}^{K_1 (1400)} $ &$\hat
					{I}_{1c,\perp}^{K_1 (1400)} $ & $\hat{I}_{2s,\perp}^{K_1 (1400)} $
					& $\hat{I}_{2c,\perp}^{K_1 (1400)} $ & $\hat
					{I}_{3,\perp}^{K_1 (1400)} $ &$\hat
					{I}_{4,\perp}^{K_1 (1400)} $ &$\hat
					{I}_{5,\perp}^{K_1 (1400)} $  & $\hat {I}_{6s,\perp}^{K_1 (1400)} $ & 
			  \\ 
				\hline
    SM %
				&$\ \ 0.446 \pm 0.003$
                & %
				$\ \ 0.334 \pm 0.018$
                & %
			    $\ \ -0.219 \pm 0.015$
				& %
				$\ \ 0.111 \pm 0.006$
                & %
				$\ \ 0.042\pm 0.002$
				& %
				$\ \ -0.139 \pm 0.001$
                & %
				$\ \ 0.164 \pm 0.000$
                & %
				$\ \ -0.139 \pm 0.006$
                & %
                
			  \\
    S1 %
				&$\ \ 0.438 \pm 0.003$
                & %
				$\ \ 0.379 \pm 0.019$
                & %
				$\ \ -0.181 \pm 0.016$
				& %
				$\ \ 0.126 \pm 0.006$
                & %
				$\ \ 0.040\pm 0.002$
				& %
				$\ \ -0.136 \pm 0.000$
                & %
				$\ \ 0.063 \pm 0.001$
                & %
                $\ \ -0.006\pm 0.002$
                & %
				             
			  \\ 
    S2 %
                &$\ \ 0.440 \pm 0.003$ 
                & %
				$\ \ 0.368 \pm 0.019$
                & %
				$\ \ -0.190 \pm 0.016$ 
                & %
				$\ \ 0.123 \pm 0.006$
                & %
				$\ \ 0.039 \pm 0.002$ 
                & %
				$\ \ -0.133 \pm 0.001$ 
                & %
				$\ \ 0.073 \pm 0.001$
                & %
                $\ \ -0.006 \pm 0.003$
                & %
				
                \\
    S3 %
                &$\ \ 0.441 \pm 0.003$
                & %
			$\ \ 0.360 \pm 0.019$
                & %
				$\ \ -0.197 \pm 0.015$ 
                & %
				$\ \ 0.120 \pm 0.006$ 
                & %
				$\ \ 0.039 \pm 0.002$ 
			    & %
				$\ \ -0.134 \pm 0.000$
                & %
				$\ \ 0.090 \pm 0.001$
                & %
                $\ \ -0.026 \pm 0.002$
                & %
				
                \\
    S4 %
                &$\ \ 0.444 \pm 0.003$
                & %
				$\ \ 0.343 \pm 0.018$
                & %
				$\ \ -0.211 \pm 0.015$
                & %
				$\ \ 0.114 \pm 0.006$
                & %
				$\ \ 0.046 \pm 0.002$ 
				& %
				$\ \ -0.139 \pm 0.000$
                & %
				$\ \ 0.118 \pm 0.001$
                & %
                $\ \ -0.076 \pm 0.002$
                & %
				
				\\                    
               \hline \hline
			\end{tabular}
   }%
	\end{table}

\begin{table}
		\caption{\small Same as in Table \ref{Bin11400-analysis}, but for the $q^2=[6.0-8.0]$ GeV$^2$ bin.}\label{Bin41400-analysis}
		\scalebox{0.87}{
    \begin{tabular}{|c|ccccccccc|}
    \hline\hline
        \multicolumn{10}{|c|}{$q^2=[6.0-8.0]$ GeV$^2$}\\
        \hline
				\hline 
				\ \ \  & $\left\langle d\mathcal{B}/dq^2\right\rangle 
    $  & $\left\langle d\mathcal{B}/dq^2\right\rangle_{\rho K} 
    $ & $\ \ \left\langle d\mathcal{B}/dq^2\right\rangle_{K^*\pi} 
    $ & $ \ \ \ \
				 \mathcal{A}_{\text{FB}}^{K_1 (1400)}$  & $ \
				\ \ \ \ \mathcal{A}^{K_{1T} (1400)}_{\text{FB}} $ & $\ \ \ \ \ \  f_{L}^{K_1 (1400)} $  & $\ \ \ \ \ \  f_T^{K_1 (1400)} $ & $\hat {I}_{6c,\perp}^{K_1 (1400)} $  & \\ 
    \hline
    SM %
				&$\ \ 0.736\pm 0.107$ 
				& %
				$\ \ 0.221\pm 0.032$
				& %
				$\ \ 0.692\pm 0.101$ 
				& %
				$\ \ -0.320\pm 0.010$
                &  %
                $\ \ -0.590 \pm 0.003$ 
                & %
				$\ \ 0.457\pm 0.020$ 
                & %
                $\ \ 0.543\pm 0.020$
				& %
				$\ \ -0.427 \pm 0.013$
                & \\
    S1 %
				&$\ \ 0.772\pm 0.116$ 
				& %
				$\ \ 0.231 \pm 0.035$
				& %
				$\ \ 0.725\pm 0.109$ 
				& %
				$\ \ -0.131\pm 0.001$
                &  %
                $\ \ -0.229\pm 0.006$ 
                & %
				$\ \ 0.425 \pm 0.020$ 
                & %
                $\ \ 0.575\pm 0.020$
				& %
				$\ \ -0.175 \pm 0.002$ 
				&   \\
    S2 %
				&$\ \ 0.542\pm 0.081$
				& %
				$\ \ 0.163 \pm 0.024$
				& %
				$\ \ 0.509 \pm 0.076$ 
                & %
				$\ \ -0.152 \pm 0.002$
                & %
                $\ \ -0.270 \pm 0.007$ 
                & %
				$\ \ 0.436 \pm 0.020$ 
                & %
                $\ \ 0.564 \pm 0.020$ 
                & %
				$\ \ -0.203 \pm 0.002$
                & \\
    S3 %
				&$\ \ 0.526\pm 0.078$
				& %
				$\ \ 0.158 \pm 0.023$
				& %
				$\ \ 0.495 \pm 0.073$
				& %
				$\ \ -0.185 \pm 0.003$
                & %
                $\ \ -0.332 \pm 0.007$
                & %
				$\ \ 0.443 \pm 0.020$
                &  %
				$\ \ 0.557 \pm 0.020$ 
                & %
				$\ \ -0.246 \pm 0.004$
			    & \\
    S4 %
				&$\ \ 0.638\pm 0.094$
				& %
				$\ \ 0.191 \pm 0.028$
				& %
				$\ \ 0.600 \pm 0.088$
				& %
				$\ \ -0.241\pm 0.006$
                & %
                $\ \ -0.444 \pm 0.005$
                & %
				$\ \ 0.456\pm 0.020$
                & %
                $\ \ 0.544 \pm 0.020$
				& %
				$\ \  -0.322 \pm 0.008$
				& \\     
                 \hline\hline
				& $\hat
                    {I}_{1s,\parallel}^{K_1 (1400)} $ &$\hat
					{I}_{1c,\parallel}^{K_1 (1400)} $ & $ \hat{I}_{2s,\parallel}^{K_1 (1400)} $
					& $\hat{I}_{2c,\parallel}^{K_1 (1400)} $ & $\hat
					{I}_{3,\parallel}^{K_1 (1400)} $ &$\hat
					{I}_{4,\parallel}^{K_1 (1400)} $ &$\hat
					{I}_{5,\parallel}^{K_1 (1400)} $  & $\hat {I}_{6s,\parallel}^{K_1 (1400)} $ &
			  \\ 
				\hline
    SM %
				& $\ \ 0.407 \pm 0.015$
                & %
				$\ \ 0.458 \pm 0.020$
                & %
			    $\ \ 0.135 \pm 0.005$
				& %
				$\ \ -0.453\pm 0.020$
                & %
				$\ \ -0.116\pm 0.005$
				& %
				$\ \ 0.292 \pm 0.001$
                & %
				$\ \ -0.349 \pm 0.002$
                & %
				$\ \ -0.427 \pm 0.013$
                & %
			  \\
    S1 %
				&$\ \ 0.431 \pm 0.015$
                & %
				$\ \ 0.426 \pm 0.020$
                & %
				$\ \ 0.144 \pm 0.005$
				& %
				$\ \ -0.421\pm 0.020$
                & %
				$\ \ -0.111\pm 0.004$
				& %
				$\ \ 0.286 \pm 0.002$
                & %
				$\ \ -0.184\pm 0.003$
                & %
				$\ \ -0.175 \pm 0.002$
                & %
			  \\ 
    S2 %
                &$\ \ 0.423 \pm 0.015$ 
                & %
				$\ \ 0.438 \pm 0.020$
                & %
				$\ \ 0.141 \pm 0.005$ 
                & %
				$\ \ -0.433 \pm 0.020$
                & %
				$\ \ -0.111 \pm 0.004$ 
                & %
				$\ \ 0.286 \pm 0.002$ 
                & %
				$\ \ -0.214 \pm 0.003$
                & %
				$\ \ -0.203 \pm 0.002$ 
                & %
                \\
    S3 %
                &$\ \ 0.418 \pm 0.015$
                & %
			    $\ \ 0.444\pm 0.020$
                & %
				$\ \ 0.139 \pm 0.005$ 
                & %
				$\ \ -0.439 \pm 0.020$ 
                & %
				$\ \ -0.112 \pm 0.004$ 
			    & %
				$\ \ 0.287 \pm 0.001$
                & %
				$\ \ -0.243 \pm 0.003$
                & %
				$\ \ -0.246 \pm 0.004$
                & %
                \\
    S4 %
                &$\ \ 0.408 \pm 0.015$
                & %
				$\ \ 0.457 \pm 0.020$
                & %
				$\ \ 0.136 \pm 0.005$
                & %
				$\ \ -0.452 \pm 0.020$
                & %
				$\ \ -0.126 \pm 0.005$ 
				& %
				$\ \ 0.295\pm 0.001$
                & %
				$\ \  -0.282 \pm 0.003$
                & %
				$\ \  -0.322 \pm 0.008$
                & %
				\\     
                 \hline\hline
				& $\hat
                    {I}_{1s,\perp}^{K_1 (1400)} $ &$\hat
					{I}_{1c,\perp}^{K_1 (1400)} $ & $\hat{I}_{2s,\perp}^{K_1 (1400)} $
					& $\hat{I}_{2c,\perp}^{K_1 (1400)} $ & $\hat
					{I}_{3,\perp}^{K_1 (1400)} $ &$\hat
					{I}_{4,\perp}^{K_1 (1400)} $ &$\hat
					{I}_{5,\perp}^{K_1 (1400)} $  & $\hat {I}_{6s,\perp}^{K_1 (1400)} $ & 
			  \\ 
				\hline
    SM %
				&$\ \ 0.433 \pm 0.003$
                & %
				$\ \ 0.407 \pm 0.015$
                & %
			    $\ \ -0.159 \pm 0.012$
				& %
				$\ \ 0.135 \pm 0.005$
                & %
				$\ \ 0.058\pm 0.002$
				& %
				$\ \ -0.146 \pm 0.000$
                & %
				$\ \ 0.174 \pm 0.001$
                & %
				$\ \ -0.214 \pm 0.007$
                & %
               
			  \\
    S1 %
				&$\ \ 0.429 \pm 0.003$
                & %
				$\ \ 0.431 \pm 0.015$
                & %
				$\ \ -0.139 \pm 0.013$
				& %
				$\ \ 0.144 \pm 0.005$
                & %
				$\ \ 0.056\pm 0.002$
				& %
				$\ \ -0.143 \pm 0.001$
                & %
				$\ \ 0.092 \pm 0.002$
                & %
                $\ \ -0.088 \pm 0.001$
                & %
				         
			  \\ 
    S2 %
                &$\ \ 0.430 \pm 0.003$ 
                & %
				$\ \ 0.423 \pm 0.015$
                & %
				$\ \ -0.146 \pm 0.013$ 
                & %
				$\ \ 0.141 \pm 0.005$ 
                & %
				$\ \ 0.056\pm 0.002$ 
                & %
				$\ \ -0.143 \pm 0.001$ 
                & %
				$\ \ 0.107 \pm 0.002$
                & %
                $\ \ -0.101 \pm 0.001$
                & %
				  
                \\
    S3 %
                &$\ \ 0.431 \pm 0.003$
                & %
			    $\ \ 0.418 \pm 0.015$
                & %
				$\ \ -0.150 \pm 0.013$ 
                & %
				$\ \ 0.139 \pm 0.005$ 
                & %
				$\ \  0.056 \pm 0.002$ 
			    & %
				$\ \ -0.144 \pm 0.001$
                & %
				$\ \ 0.122 \pm 0.002$
                & %
                $\ \ -0.123 \pm 0.002$
                & %
				
                \\
    S4 %
                &$\ \ 0.433 \pm 0.003$
                & %
				$\ \ 0.408 \pm 0.015$
                & %
				$\ \ -0.158 \pm 0.012$
                & %
				$\ \ 0.136 \pm 0.005$
                & %
				$\ \ 0.063 \pm 0.002$ 
				& %
				$\ \ -0.148 \pm 0.001$
                & %
				$\ \ 0.141 \pm 0.001$
                & %
                $\ \ -0.161 \pm 0.004$
                & %
				
				\\                    
               \hline \hline
			\end{tabular}
   }%
	\end{table}

\begin{table}
		\caption{\small Same as in Table \ref{Bin11400-analysis}, but for the $q^2=[14.0-15.0]$ GeV$^2$ bin.}\label{Bin51400-analysis}
		\scalebox{0.87}{
    \begin{tabular}{|c|ccccccccc|}
    \hline\hline
        \multicolumn{10}{|c|}{$q^2=[14.0-15.0]$ GeV$^2$}\\
        \hline
				\hline 
				\ \ \  & $\left\langle d\mathcal{B}/dq^2\right\rangle 
    $  & $\left\langle d\mathcal{B}/dq^2\right\rangle_{\rho K} 
    $ & $\ \ \left\langle d\mathcal{B}/dq^2\right\rangle_{K^*\pi} 
    $ & $ \ \ \ \
				 \mathcal{A}_{\text{FB}}^{K_1 (1400)}$  & $ \
				\ \ \ \ \mathcal{A}^{K_{1T} (1400)}_{\text{FB}} $ & $\ \ \ \ \ \  f_{L}^{K_1 (1400)} $  & $\ \ \ \ \ \  f_T^{K_1 (1400)} $ & $\hat {I}_{6c,\perp}^{K_1 (1400)} $  & \\ 
    \hline
    SM %
				&$\ \ 0.370\pm 0.063$ 
				& %
				$\ \ 0.111\pm 0.019$
				& %
				$\ \ 0.348\pm 0.059$ 
				& %
				$\ \ -0.174\pm 0.001$
                &  %
                $\ \ -0.263 \pm 0.002$ 
                & %
				$\ \ 0.337 \pm 0.001$ 
                & %
                $\ \ 0.663\pm 0.001$
				& %
				$\ \ -0.233 \pm 0.002$
                & \\
    S1 %
				&$\ \ 0.358\pm 0.060$ 
				& %
				$\ \ 0.107 \pm 0.018$
				& %
				$\ \ 0.336\pm 0.057$ 
				& %
				$\ \ -0.122\pm 0.001$
                &  %
                $\ \ -0.184\pm 0.002$ 
                & %
				$\ \ 0.335 \pm 0.001$ 
                & %
                $\ \ 0.665\pm 0.001$
				& %
				$\ \ -0.163 \pm 0.002$
				&   \\
    S2 %
				&$\ \ 0.260\pm 0.044$
				& %
				$\ \ 0.078 \pm 0.013$
				& %
				$\ \ 0.244 \pm 0.041$ 
                & %
				$\ \ -0.138 \pm 0.001$
                & %
                $\ \ -0.207 \pm 0.002$ 
                & %
				$\ \ 0.336 \pm 0.001$ 
                & %
                $\ \ 0.664 \pm 0.001$ 
                & %
				$\ \ -0.183 \pm 0.002$ 
                & \\
    S3 %
				&$\ \ 0.257\pm 0.044$
				& %
				$\ \ 0.077 \pm 0.013$
				& %
				$\ \ 0.241 \pm 0.041$
				& %
				$\ \ -0.148 \pm 0.001$
                & %
                $\ \ -0.222 \pm 0.002$
                & %
				$\ \ 0.336 \pm 0.001$
                &  %
				$\ \ 0.664 \pm 0.001$ 
                & %
				$\ \ -0.197 \pm 0.002$ 
			    & \\
    S4 %
				&$\ \ 0.319\pm 0.054$
				& %
				$\ \ 0.096 \pm 0.016$
				& %
				$\ \ 0.300 \pm 0.051$
				& %
				$\ \ -0.155 \pm 0.001$
                & %
                $\ \ -0.233 \pm 0.002$
                & %
				$\ \ 0.337 \pm 0.001$
                & %
                $\ \ 0.663 \pm 0.001$
				& %
				$\ \  -0.206 \pm 0.002$ 
				& \\     
                 \hline\hline
				& $\hat
                    {I}_{1s,\parallel}^{K_1 (1400)} $ &$\hat
					{I}_{1c,\parallel}^{K_1 (1400)} $ & $ \hat{I}_{2s,\parallel}^{K_1 (1400)} $
					& $\hat{I}_{2c,\parallel}^{K_1 (1400)} $ & $\hat
					{I}_{3,\parallel}^{K_1 (1400)} $ &$\hat
					{I}_{4,\parallel}^{K_1 (1400)} $ &$\hat
					{I}_{5,\parallel}^{K_1 (1400)} $  & $\hat {I}_{6s,\parallel}^{K_1 (1400)} $ &
			  \\ 
				\hline
    SM %
				& $\ \ 0.497 \pm 0.001$
                & %
				$\ \ 0.337 \pm 0.001$
                & %
			    $\ \ 0.166 \pm 0.000$
				& %
				$\ \ -0.336\pm 0.001$
                & %
				$\ \ -0.307\pm 0.001$
				& %
				$\ \ 0.327 \pm 0.000$
                & %
				$\ \ -0.120 \pm 0.001$
                & %
				$\ \ -0.233 \pm 0.002$
                & %
			  \\
    S1 %
				&$\ \ 0.498 \pm 0.001$
                & %
				$\ \ 0.336\pm 0.001$
                & %
				$\ \ 0.166 \pm 0.000$
				& %
				$\ \ -0.335\pm 0.001$
                & %
				$\ \ -0.307\pm 0.001$
				& %
				$\ \ 0.327 \pm 0.000$
                & %
				$\ \ -0.084 \pm 0.001$
                & %
				$\ \ -0.163 \pm 0.002$
                & %
			  \\ 
    S2 %
                &$\ \ 0.498 \pm 0.001$ 
                & %
				$\ \ 0.336 \pm 0.001$
                & %
				$\ \ 0.166 \pm 0.000$ 
                & %
				$\ \ -0.336 \pm 0.001$
                & %
				$\ \ -0.307 \pm 0.001$ 
                & %
				$\ \ 0.327 \pm 0.000$ 
                & %
				$\ \ -0.095 \pm 0.001$
                & %
				$\ \ -0.183 \pm 0.002$ 
                & %
                \\
    S3 %
                &$\ \ 0.498 \pm 0.001$
                & %
			    $\ \ 0.336 \pm 0.001$
                & %
				$\ \ 0.166 \pm 0.000$ 
                & %
				$\ \ -0.336 \pm 0.001$ 
                & %
				$\ \ -0.307 \pm 0.001$ 
			    & %
				$\ \ 0.327\pm 0.000$
                & %
				$\ \ -0.102 \pm 0.001$
                & %
				$\ \ -0.197 \pm 0.002$
                & %
                \\
    S4 %
                &$\ \ 0.497 \pm 0.001$
                & %
				$\ \ 0.337 \pm 0.001$
                & %
				$\ \ 0.165\pm 0.000$
                & %
				$\ \ -0.336 \pm 0.001$
                & %
				$\ \ -0.309 \pm 0.001$ 
				& %
				$\ \ 0.328 \pm 0.000$
                & %
				$\ \ -0.107 \pm 0.001$
                & %
				$\ \  -0.206 \pm 0.002$
                & %
				\\     
                 \hline\hline
				& $\hat
                    {I}_{1s,\perp}^{K_1 (1400)} $ &$\hat
					{I}_{1c,\perp}^{K_1 (1400)} $ & $\hat{I}_{2s,\perp}^{K_1 (1400)} $
					& $\hat{I}_{2c,\perp}^{K_1 (1400)} $ & $\hat
					{I}_{3,\perp}^{K_1 (1400)} $ &$\hat
					{I}_{4,\perp}^{K_1 (1400)} $ &$\hat
					{I}_{5,\perp}^{K_1 (1400)} $  & $\hat {I}_{6s,\perp}^{K_1 (1400)} $ & 
			  \\ 
				\hline
    SM %
				&$\ \ 0.417 \pm 0.000$
                & %
				$\ \ 0.497 \pm 0.001$
                & %
			    $\ \ -0.085 \pm 0.001$
				& %
				$\ \ 0.166 \pm 0.000$
                & %
				$\ \ 0.153\pm 0.000$
				& %
				$\ \ -0.164 \pm 0.000$
                & %
				$\ \ 0.060 \pm 0.001$
                & %
				$\ \ -0.116 \pm 0.001$
                & %
                
			  \\
    S1 %
				&$\ \ 0.417 \pm 0.000$
                & %
				$\ \ 0.498 \pm 0.001$
                & %
				$\ \ -0.085 \pm 0.001$
				& %
				$\ \ 0.166 \pm 0.000$
                & %
				$\ \ 0.153\pm 0.000 $
				& %
				$\ \ -0.164 \pm 0.000$
                & %
				$\ \ 0.042 \pm 0.001$
                & %
                $\ \ -0.082 \pm 0.001$
                & %
				             
			  \\ 
    S2 %
                &$\ \ 0.417 \pm 0.000$ 
                & %
				$\ \ 0.498 \pm 0.001$
                & %
				$\ \ -0.085 \pm 0.001$ 
                & %
				$\ \ 0.166 \pm 0.000$
                & %
				$\ \ 0.153 \pm 0.000$ 
                & %
				$\ \ -0.164 \pm 0.000$ 
                & %
				$\ \ 0.048 \pm 0.001$
                & %
                $\ \ -0.092 \pm 0.001$
                & %
				   
                \\
    S3 %
                &$\ \ 0.417 \pm 0.000$
                & %
			    $\ \ 0.498 \pm 0.001$
                & %
				$\ \ -0.085 \pm 0.001$ 
                & %
				$\ \ 0.166 \pm 0.000$
                & %
				$\ \ 0.153 \pm 0.000$ 
			    & %
				$\ \ -0.164 \pm 0.000$
                & %
				$\ \ 0.051 \pm 0.001$
                & %
                $\ \ -0.098 \pm 0.001$
                & %
				
                \\
    S4 %
                &$\ \ 0.417 \pm 0.000$
                & %
				$\ \ 0.497 \pm 0.001$
                & %
				$\ \ -0.085 \pm 0.001$
                & %
				$\ \ 0.165 \pm 0.000$
                & %
				$\ \ 0.154 \pm 0.000$ 
				& %
				$\ \ -0.164 \pm 0.000$
                & %
				$\ \ 0.053 \pm 0.001$
                & %
                $\ \ -0.103 \pm 0.001$
                & %
				
				\\                    
               \hline \hline
			\end{tabular}
   }%
	\end{table}

\end{document}